\documentclass[showpacs, pra,twocolumn,preprintnumbers ,amsmath, amssymb, superscriptaddress, aps]{revtex4-2}
\usepackage{color}
\usepackage{amsmath,amssymb}
\usepackage{pifont}
\usepackage{amssymb}  
\usepackage{bbold}
\usepackage{float}
\usepackage{subfloat} 

\usepackage[caption=false]{subfig}
\usepackage{tikz}
\usepackage{makecell}
\usepackage{subfig}
\usepackage{pifont}   
\usepackage{graphicx} 
\graphicspath{{Figures/}}
\usepackage{dcolumn}  
\usepackage{bm}       
\usepackage{multirow} 
\usepackage{placeins}
\usepackage[colorlinks]{hyperref}
\usepackage{mathtools}

\begin{document}

	\title{ 
		Quantum tunneling in graphene Corbino disk in
a solenoid magnetic potential with wedge disclination
 }
	\date{\today}
	\author{Ahmed Bouhlal }
	\email{bouhlal.a@ucd.ac.ma}
	\affiliation{Laboratory of Theoretical Physics, Faculty of Sciences, Choua\"ib Doukkali University, PO Box 20, 24000 El Jadida, Morocco}
		\author{Ahmed Jellal}
	\email{a.jellal@ucd.ac.ma}
	\affiliation{Laboratory of Theoretical Physics, Faculty of Sciences, Choua\"ib Doukkali University, PO Box 20, 24000 El Jadida, Morocco}
	\affiliation{Canadian Quantum Research Center, 204-3002 32 Ave Vernon,  BC V1T 2L7, Canada}
		\author{Mohamed Mansouri}
	\affiliation{Laboratory LAMSAD, National School of Applied Sciences, Hassan First  University,  Berrechid}

	\pacs{ 
	}

		\begin{abstract}			
				
We investigate  the wedge disclination effect on   
quantum  tunneling  of a Corbino disk in  gapped-graphene  of  inner $R_1$ and outer  $R_2$ radii in the presence of 
magnetic flux $\Phi_i$. We solve Dirac equation for different regions and  obtain the solutions  of energy spectrum in terms of Hankel functions. The asymptotic behaviors for large arguments allow us  to determine the transmission,  Fano factor and conductance. We establish the case where the crystal symmetry is modified locally by replacing a hexagon by  pentagon, square, heptagon or octagon. We show that the wedge disclination $n$
modifies the amplitude of  transmission oscillations.
We find that   the period of Fano factor oscillations
is of the Aharonov-Bohm type, 
which strongly depends on  $n$ where intense peaks  are observed.
As another result,  $n$ 
changes  the minimum and period of conductance oscillations  of the Aharonov-Bohm type. 
We show that $n$ 
  minimizes the effect of resonance and decreases the amplitude of    conductance magnitude $\Delta G$ 
  oscillations.

\end{abstract}		
	
\maketitle

	\section{Introduction}

The physics of metal/semiconductor nanostructures led to the discovery of low-dimensional structures called quantum rings \cite{Webb, Mailly, Foldi}.
In such structures,
the confinement of charge carriers associated with the phase coherence of the electronic wave function allows the observation of Aharonov-Bohm (AB) effect \cite{Aharonov}. The AB oscillations manifest by periodic oscillation in the energy and conductance spectrum of the electronic system 
\cite{Webb,Schelter}. As for graphene, 
different quantum rings have already shown AB-conductance oscillations \cite{777,888,999}. 
Graphene-based quantum rings have experimentally be achieved by
lithographic technique in which graphene nanoribbons or ring structures are carved out of a defect-free graphene surface \cite{Zarenia09}. 

In the context of graphene, various transport properties of Corbino disks were recently studied experimentally \cite{Zhao, Peters}  and theoretically \cite{Rycerz10, Khatibi, Moomivand}. 
 Recently A. Rycerz and D. Suszalski \cite{b8} showed that a Corbino graphene ballistic disc pierced by a solenoid of a magnetic potential vector can present Aharonov-Bohm type conductance oscillations when the current flows through only one conductive element. In our former paper  \cite{babe} we considered the  system used in \cite{b8} but by  adding a mass term. Our results showed that the energy difference removes the tunnel effect by creating zero transmission singularity points. It is found that  when the ratio of radii $ R_2 / R_1 $ varies the transmission presents an oscillatory behavior 
 with a decrease in periods and amplitudes. In addition the appearance of the minimum conductance is observed at the points $ k_F R_1 = R_1 \delta $, with Fermi wave vector $k_F$ and rescaled energy gap $\delta$. It is demonstrated that  the conductance as a function of the magnetic flux passing through the disk shows periodic oscillations of the Aharonov-Bohm type and becomes very clear in the presence of energy gap. 
	
In this paper we follow a similar set of ideas and consider a graphene quantum ring  with a wedge disclination $n=0, \pm1, \pm2$ that can be understood from Volterra construction \cite{cludio94}. After a general description of the model, we solve it by taking into account the radial and angular degrees of freedom. We find the eigenspinors  in terms of Hankel functions. With the  use of pairing conditions and the asymptotic behaviors of Hankel functions for large arguments, we compute the transmission, Fano factor and conductance.
It is found that the index 
$n$ is responsible for the appearance of intense peaks in 
the Fano factor oscillations and changes of its period. Also
 our results show that the index   $n$  modifies the frequency of the oscillations of  conductance
and  
minimizes the effect of the resonance induced by  the conductance magnitude $\Delta G$.

The manuscript is organized as follows. In section \ref{TTMM}, we present our theoretical model based on the Dirac Hamiltonian to describe the new geometry obtained via Volterra construction. We establish the solutions of energy spectrum in the three regions. We use the  matching conditions to end up with the transmission, conductance and Fano factor in section \ref{Trans}. In section \ref{RRDD}, we numerically discuss our results under suitable choices of the physical parameters. Our conclusions are summarized in the final section.

	\section{Theory and methods}\label{TTMM}
\begin{figure}[H]\centering
	\includegraphics[width=0.47\linewidth]{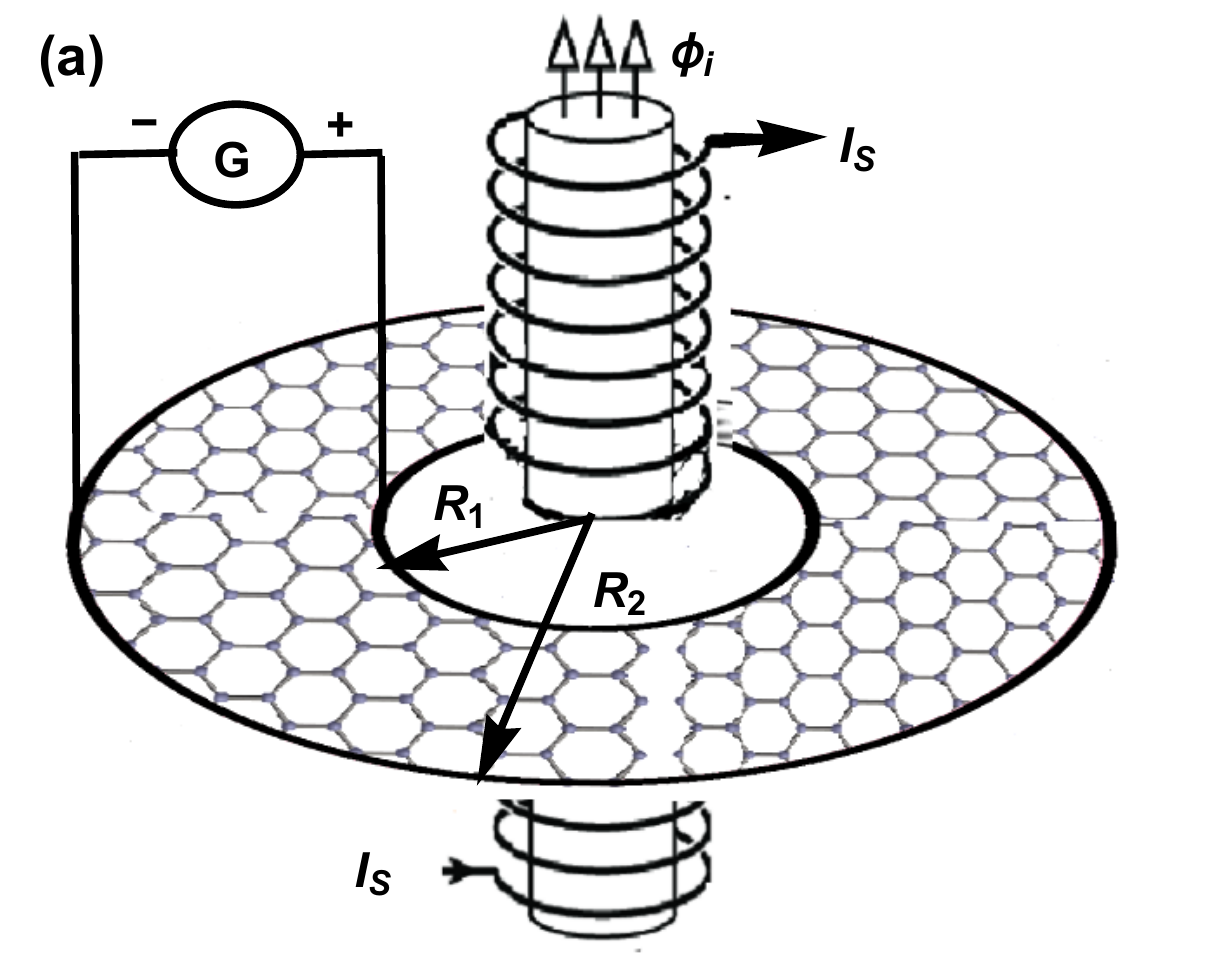} \includegraphics[width=0.51\linewidth]{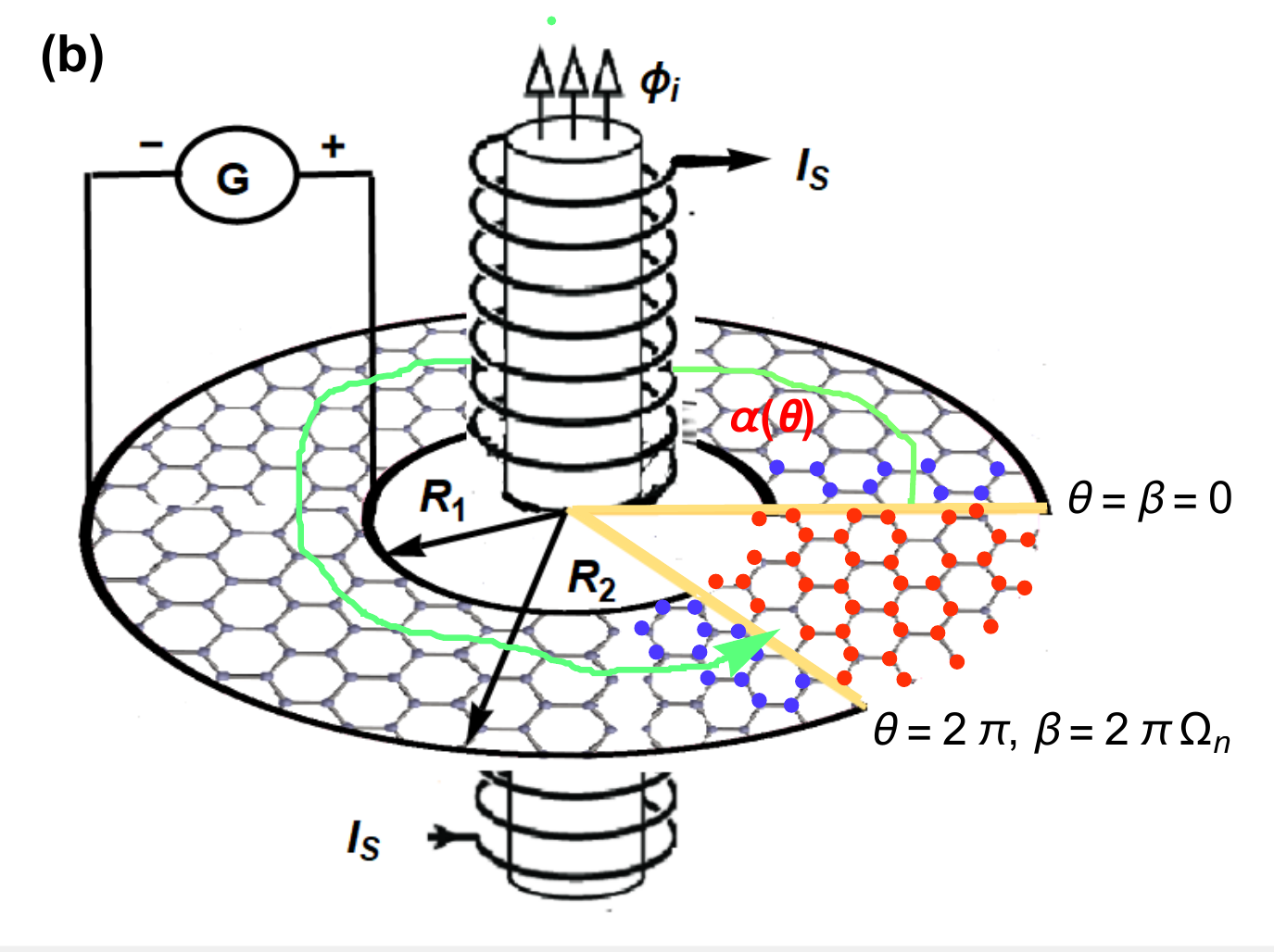}
	\caption{(color online) (a): 
		Graphene quantum rings 
		after Volterra construction of the inner radius $R_1$ and the outer radius $R_2$ contacted by two electrodes (thick black circles). (b):  Unfolded plane of lattice where a wedge of angle $n \pi /3$ is removed, here $n=1$.  $\alpha (\theta)$ is a closed path around the cone.  We rescaled the angle $\beta$ of the unfolded plane to   $\theta = \frac{\beta}{\Omega_n}$ with the wedge disclination $\Omega_n=1-\frac{n}{6}$. The carbon atoms of the removed sector are indicated by red balls, those which remain after
		the cups are represented by blue balls.} \label{syst}
\end{figure}

Let us start by considering quantum rings characterized by the inner radius $R_1$ and the outer radius $R_2$,
surrounded by metallic contacts modeled by heavily-doped graphene areas, see Fig. \ref{syst}\textcolor{red}{a}. We use the Volterra construction \cite{cludio94} to
model the disclination defect 
 by the regularized rings of radius $R_1$ and $R_2$ around the
apex and the removed wedge disclination as presented in Fig. \ref{syst}\textcolor{red}{b}.

To study the above system we introduce the following Hamiltonian 
\begin{equation}
\label{eq:Dirac}
 H_{\tau}=v_F ( \tau \pi_x \sigma_x +\pi_y \sigma_y ) + U(r) \mathbb{I}+\tau \Delta(r) \sigma_z
\end{equation}
where $v_F = 10^6$ m/s is the Fermi velocity,  $\tau=1(-1)$ refers to the valley  $K(K^{'})$, $\sigma_{i}$ ($i$= $x$, $y$, $z$)  are Pauli  matrices in the basis of the two sublattices of $A$ and $B$ atoms,  $\pi_{i}=p_i + e A_{i}$ are the conjugate    momentum components.
The involved vector potential
of a solenoid is chosen in the symmetric gauge 
\begin{equation}
\vec A = 
 \frac{\hbar \Phi}{\Omega_n e \Phi_0 r^2} (-y,x)
\end{equation}
such that the wedge disclination is described by
 $\Omega_n=1-\frac{n}{6}$ \cite{bohm} and $n=0,\pm1,\pm2$ is its associated index, with $\Phi$ is the solenoid flux and   its unit $\Phi_0=\frac{h}{e}$. It is convenient for our task to fix 
 the potential and  
  gap 
  as
\begin{align}\label{eq2}
&U(r)=
\left\{%
\begin{array}{ll}
0,  & \ \ \hbox{$R_1<r< R_2$} \\
- \hbar v_F V_{\infty}, & \ \  \hbox{otherwise} \\
\end{array}%
\right.\\
&
\Delta(r)=
\left\{%
\begin{array}{ll}
\Delta,  & \ \ \hbox{$R_1<r< R_2$} \\
0, &  \ \ \hbox{otherwise}. \\
\end{array}%
\right.
\end{align}

The geometry of Fig. \ref{syst} suggests to work with 
the polar coordinates $(r,\theta)$. Then, one can 
 map the Hamiltonian \eqref{eq:Dirac} as
\begin{equation}\label{e4}
H= \begin{pmatrix}
V_{+} & \partial_{-}\\ 
\partial_{+} &  V_{-}
\end{pmatrix}
\end{equation}
where we have set the potentials 
\begin{align}
\label{eq5}
V_{\pm}(r)=U(r)\pm \tau \Delta(r)
\end{align}
and introduced the operators 
\begin{align}
\partial_{\pm}=-i\hbar\nu_{F} e^{\pm i \tau \theta}\left(\tau \dfrac{\partial}{\partial r}\pm\frac{i}{r}\dfrac{\partial}{\partial\theta}\mp\frac{\Phi_i }{\Omega_n r}\right)
\end{align}
with the  dimensionless flux
 $\Phi_i=\frac{\Phi}{\Phi_0}$.
The circular symmetry of the quantum rings ensures the conservation of the total angular momentum $J_z =L_z +\tau \hbar \frac{\sigma_z}{2}$, i.e. $ [H, J_z] = 0$. Thus,  the eigenspinors can be expressed in terms of angular and radial
components as
\begin{equation}\label{eq6}
\Psi^{\tau}_m(r,\theta)=\begin{pmatrix} \chi^{\tau}_{A}(r)\psi^{+}_m(\theta) \\ \chi^{\tau}_{B}(r)\psi^{-}_{m \pm {\tau}}(\theta)  \end{pmatrix}
\end{equation}
such that 
the eigenstates  of $J_z$ 
are given by
\begin{equation}\label{eq7}
	\psi_m^{+}(\theta)=\frac{e^{im\theta}}{\sqrt{2\pi}}\begin{pmatrix} 1  \\ 0  \end{pmatrix}, \quad \psi_{m\pm \tau}^-(\theta)=\frac{e^{i(m \pm {\tau})\theta}}{\sqrt{2\pi}}\begin{pmatrix} 0  \\ 1  \end{pmatrix}
\end{equation}
where $m=\pm1,\pm2,\cdots$, is the integer-value angular momentum quantum number, and the subscripts $ A (B) $ labels the upper (lower) spinor element.

Now, we solve the Dirac equation in the three regions shown
in Fig. \ref{syst}\textcolor{red}{b}
\begin{equation}
H_{\tau}\Psi^{\tau}_m(r,\theta)=E_{\tau}\Psi^{\tau}_m(r,\theta)
\end{equation}
which gives rise to  
\begin{align}\label{eq8}
&\left(\tau\dfrac{\partial}{\partial r} +\frac{m+\tau}{r}+\frac{\Phi_{i} }{\Omega_n r} \right)\chi^{\tau}_{B}(r)=i\kappa_{+}^\tau\chi^{\tau}_{A}(r)
\\
\label{eq9}
&\left(\tau\dfrac{\partial}{\partial r} -\frac{m}{r}-\frac{\Phi_{i} }{\Omega_n r} \right)\chi^{\tau}_{A}(r)=i\kappa_{-}^\tau\chi^{\tau}_{B}(r)
\end{align}
where we have defined 
$ \kappa^{\tau}_{\pm} =\epsilon_{\tau}-V \pm\tau \delta $
 and the dimensionless parameters are used $\epsilon_{\tau}=\frac{E_{\tau}}{\hbar v_F}$, $V=\frac{U}{\hbar v_F}$, $\delta=\frac{\Delta}{\hbar v_F}$. We proceed further by  deriving a second differential equation for $\chi^{\tau}_{A}(r)$
\begin{equation}\label{eq10}
\left(r^2\frac{\partial^2 }{\partial r^2}+r \dfrac{\partial}{\partial r} +r^2 k^2-(m+\frac{ \Phi_{i}}{\Omega_n })^2\right)\chi^{\tau}_{A}(r)=0
\end{equation}
with the parameter
\begin{equation}\label{eq11}
	k = \sqrt{\left|(\epsilon_{\tau}-V_i)^2-\delta^2\right|}.
\end{equation}
Under the variable change $\rho=k r$ we obtain 
\begin{equation}\label{eq12}
\left(\rho^2\frac{\partial^2 }{\partial \rho^2}+\rho \dfrac{\partial}{\partial \rho} +\rho^2-(m+\frac{ \Phi_{i}}{\Omega_n })^2\right)\chi^{\tau}_{A}(\rho)=0
\end{equation} 
which has the Hankel functions $ H^{\pm}_{\nu}(k r)$ as general solution  associated to the quantum number
\begin{equation}\label{eq111}
\nu =m+\frac{\Phi_{i}}{\Omega_n}.
\end{equation}
Consequently, the radial components
$\chi^{\tau}_{m}=(\chi^{\tau}_{A},\chi^{\tau}_{B})^T$ for the incoming and outgoing waves are given 
by 
\begin{align}\label{eq14}
	&
\chi^{{\tau}(inc)}_{\nu}(r)=
\begin{pmatrix}
H_{\nu}^{-}(kr)\\
i H^{-}_{\nu+\tau}(kr) 
\end{pmatrix}\\
&\chi^{{\tau}(out)}_{m}(r)=
\begin{pmatrix}
H^{+}_{\nu}(kr)\\
i H^{+}_{\nu+\tau}(kr)  
\end{pmatrix}.
\end{align}

Now we consider the solutions of each region. Indeed, 
in the disk area ($R_1<r<R_2$), we have $U(r)=0$,  $\Delta(r)\neq0$  and the electron-doping case $E_\tau>U(r)$. Then the  solution  can be represented as
\begin{equation}\label{wave2}
\chi^{\tau(2)}_{\nu}(r)= a^{\tau}
\begin{pmatrix}
H^{-}_{\nu}(k r)\\
i H^{-}_{\nu+\tau}(k r) 
\end{pmatrix}
+b^{\tau} 
\begin{pmatrix}
H^{+}_{\nu}(k r)\\
i H^{+}_{\nu+\tau}(k r)  
\end{pmatrix}
\end{equation}
with $a^{\tau}$, $b^{\tau}$ being arbitrary constants and from \eqref{eq11}
we derive
\begin{align}
 k=\sqrt{\left| \epsilon_{\tau}^2-\delta^2\right|}
\end{align}

For the region $r<R_1$ (the inner disk) and $r>R_2$ (the  outer disk), we have $U(r)=U_\infty, \Delta(r)=0$. As a result, we get   the solutions 
for $r<R_1$
\begin{equation} \label{eq18}
\chi^{\tau(1)}_{\nu}(r)=
\begin{pmatrix}
H^{-}_{\nu}(k_\infty r)\\
i H^{-}_{\nu+\tau}(k_\infty r)
\end{pmatrix}
+r^{\tau}_\nu
\begin{pmatrix}
H^{+}_{\nu}(k_\infty r)\\
i H^{+}_{\nu+\tau}(k_\infty r) 
\end{pmatrix}
\end{equation}
and  $r>R_2$
\begin{equation} \label{eq19}
\chi^{\tau(3)}_{\nu}(r)= t^{\tau}_\nu 
\begin{pmatrix}
H^{-}_{\nu}(k_\infty r)\\
i H^{-}_{\nu+\tau}(k_\infty r)
\end{pmatrix}
\end{equation}
with \eqref{eq11} goes to 
\begin{align}
	k_\infty=\epsilon_{\tau}+V_{\infty}\longrightarrow \pm\infty.
\end{align}
Here $ r^{\tau}_\nu $ and $ t^{\tau}_\nu $ are the reflection and transmission coefficients, respectively.
For the need
we recall the useful formula of  the Hankel functions 
\begin{equation}
H^{\pm}_{\nu+\tau}(rk)= \pm i H^{\pm}_{\nu}(rk), \quad H^\pm_{\nu}(rk)=[H^\mp_{\nu}(rk)]^*.
\end{equation}

\section{Transport properties}\label{Trans}

We compute the transmission probability, conductance and Fano factor
associated to our system. For this, 
we consider the limit of a highly doped lead $k_{\infty} r\gg 1$ to approximate the asymptotic behavior of the Hankel functions for large arguments as
\begin{equation}
H^{(\pm)}_\nu(\rho)\approx (2/\pi \rho)^{\frac{1}{2}} e^{\pm i(\rho-\nu\frac{\pi}{2}-\frac{\pi}{4})}.
\end{equation}
As consequence, 
 \eqref{eq18}  reduce to 
\begin{align}
\chi^{\tau(1)}_{\nu}=\frac{e^{+ik_{\infty} r}}{\sqrt{r}}
\begin{pmatrix}
1\\
1
\end{pmatrix}
+r^{\tau}_{\nu} \frac{e^{-ik_{\infty} r}}{\sqrt{r}} 
\begin{pmatrix}
1\\
-1  
\end{pmatrix}
\end{align}
as well as  \eqref{eq19}
\begin{align}
\chi^{\tau(3)}_{\nu}=t^{\tau}_{\nu}\frac{e^{+ik_{\infty} r}}{\sqrt{r}}
\begin{pmatrix}
1\\
1
\end{pmatrix}. 
\end{align}

Now we use the continuity of wave functions at the edges of  three regions 
 \begin{align}
 &	 \chi^{\tau(1)}_{\nu}(R_1)= \chi^{\tau(2)}_{\nu}(R_1)\\ & \chi^{\tau(2)}_{\nu}(R_2)= \chi^{\tau(3)}_{\nu}(R_2)
 \end{align}
 to determine the transmission coefficient for the $\nu^{th}$ mode 
\begin{equation}
t^{\tau}_{\nu}=\frac{4}{\pi k \sqrt{R_1 R_2}}  \frac{e^{+i k_{\infty}(R_1-R_2)}}{\Gamma_{\nu}^{-\tau}+i \Gamma_{\nu}^{+\tau}}.
\end{equation}
Therefore,  the transmission probability can be obtained from 
$ T^{\tau}_\nu= |t^{\tau}_{\nu}|^2 $ as
\begin{equation}
T^{\tau}_\nu= \frac{16}{\pi^2 (k R_1) (k R_2)}  \frac{1}{(\Gamma_{\nu}^{-{\tau}})^{2}+(\Gamma_{\nu}^{+{\tau}})^{2}} \label{eqtrans}
\end{equation}
where we have defined  the quantities 
\begin{align}
\Gamma_{\nu}^{+{\tau}}&= \\
&\operatorname{Im} \left[H_{\nu}^{-}(k R_1) H_{\nu}^{+}(k R_2) + H_{\nu+{\tau}}^{-}(k R_1)H_{\nu+{\tau}}^{+}(k R_2)\right]\nonumber
\\
\Gamma_{\nu}^{-{\tau}}&= \\
&\operatorname{Im} \left[H_{\nu}^{-}(k R_1) H_{\nu+{\tau}}^{+}(k R_2) - H_{\nu+{\tau}}^{-}(k R_1)H_{\nu}^{+}(k R_2)\right]\nonumber.
\end{align}
Recall that  the wedge index $n$ is included in 
the quantum number
$\nu$ given in \eqref{eq111}. As a result, the wedge will affect the transmission associated to our system. 

At this level we compute two interesting physical quantities related the transmission. Indeed, the conductance  can be calculated within the Landauer–Büttiker formalism in the linear-response regime \cite{RolfL,Buttiker1992}. It is provided by the relation
 \begin{equation}\label{eq22}
G=g_0\sum_{\tau=\pm1, \nu} 
T^{\tau}_\nu
\end{equation} 
where  $ g_0=\frac{4 e^2}{h} $, with 4 accounting for spin and valley degeneracy in the graphene. Here summation is over the valley index $\tau$ and quantum number $\nu$.

As for  the Fano factor quantifying the power of  shot noise for graphene, it is given by  the summation over all modes. This is 
\begin{equation}\label{eq23}
\mathcal{F}^{\tau}=\frac{ \sum_{\nu} \left[T^{\tau}_\nu\left(1-T^{\tau}_\nu\right)\right]}{ \sum_{\nu} T^{\tau}_\nu}.
\end{equation}
The results obtained so far will numerically be analyzed under
suitable choices of the physical parameters. With this we can 
characterize the influence of wedge disclination  on
the transport properties of our system described schematically in Fig. \ref{syst}.

\section{Results and discussions}\label{RRDD}

 \begin{figure}[H]\centering
 	\includegraphics[width=0.49\linewidth]{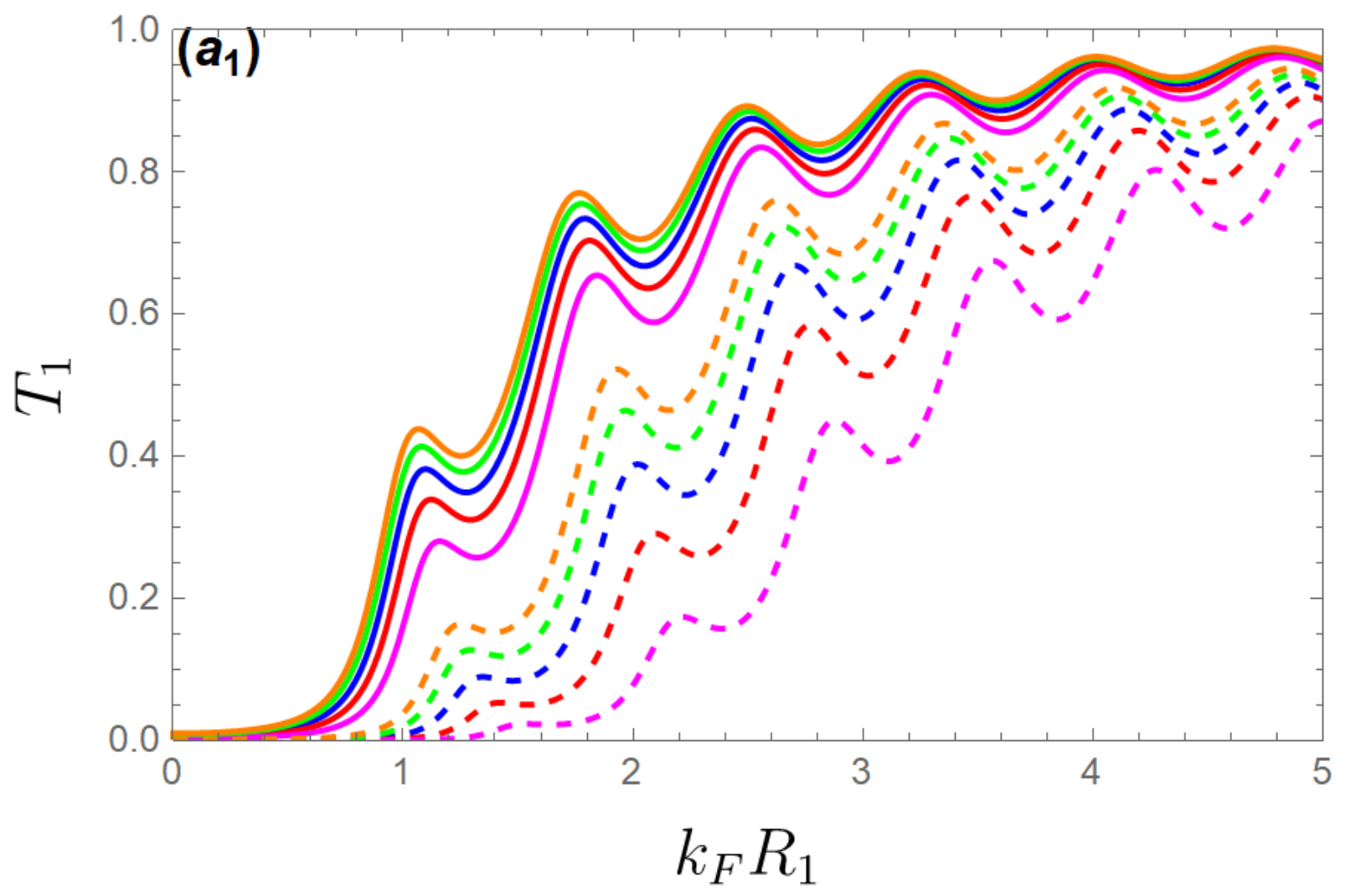}
 	\includegraphics[width=0.49\linewidth]{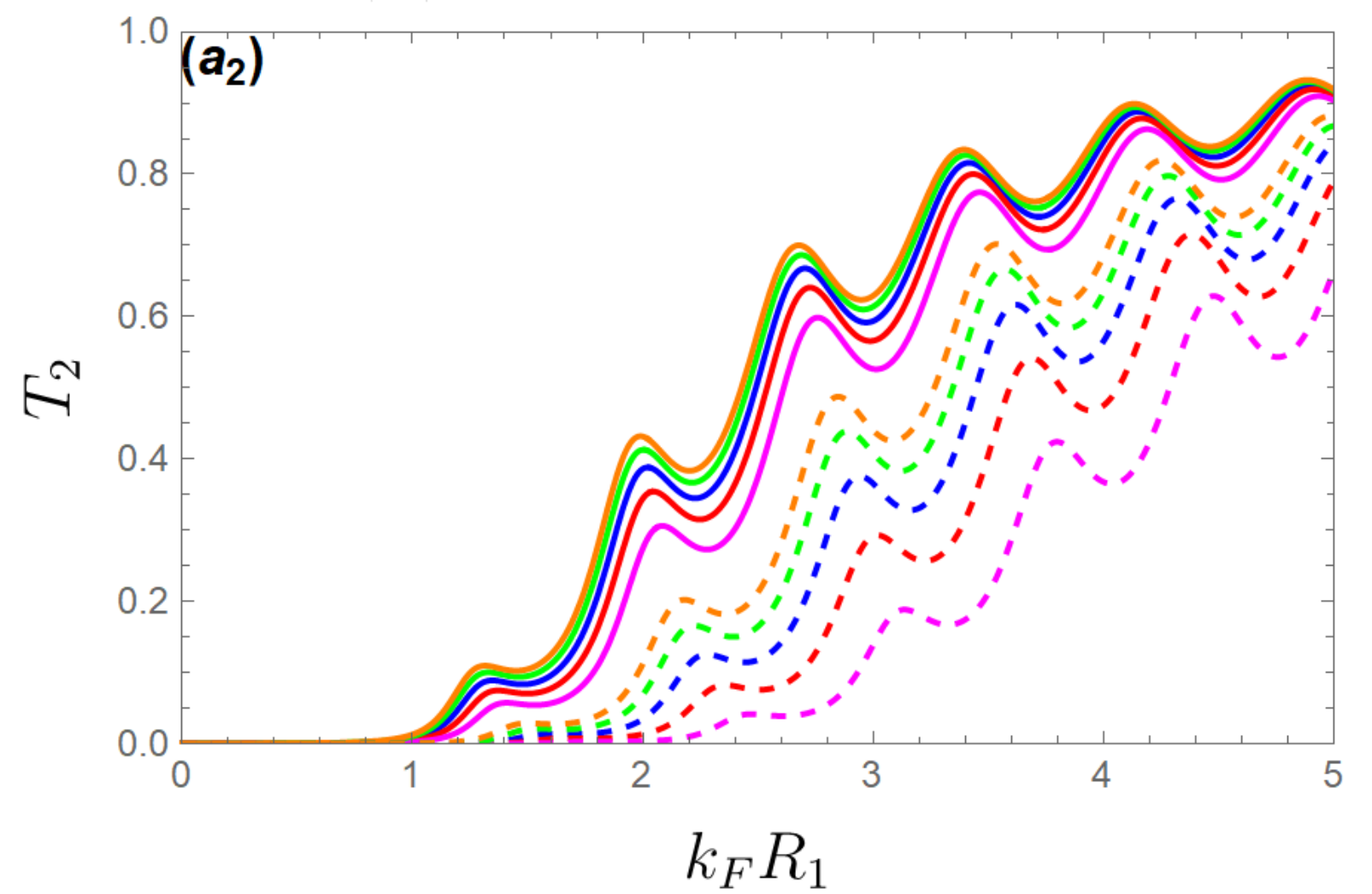}
 	\includegraphics[width=0.49\linewidth]{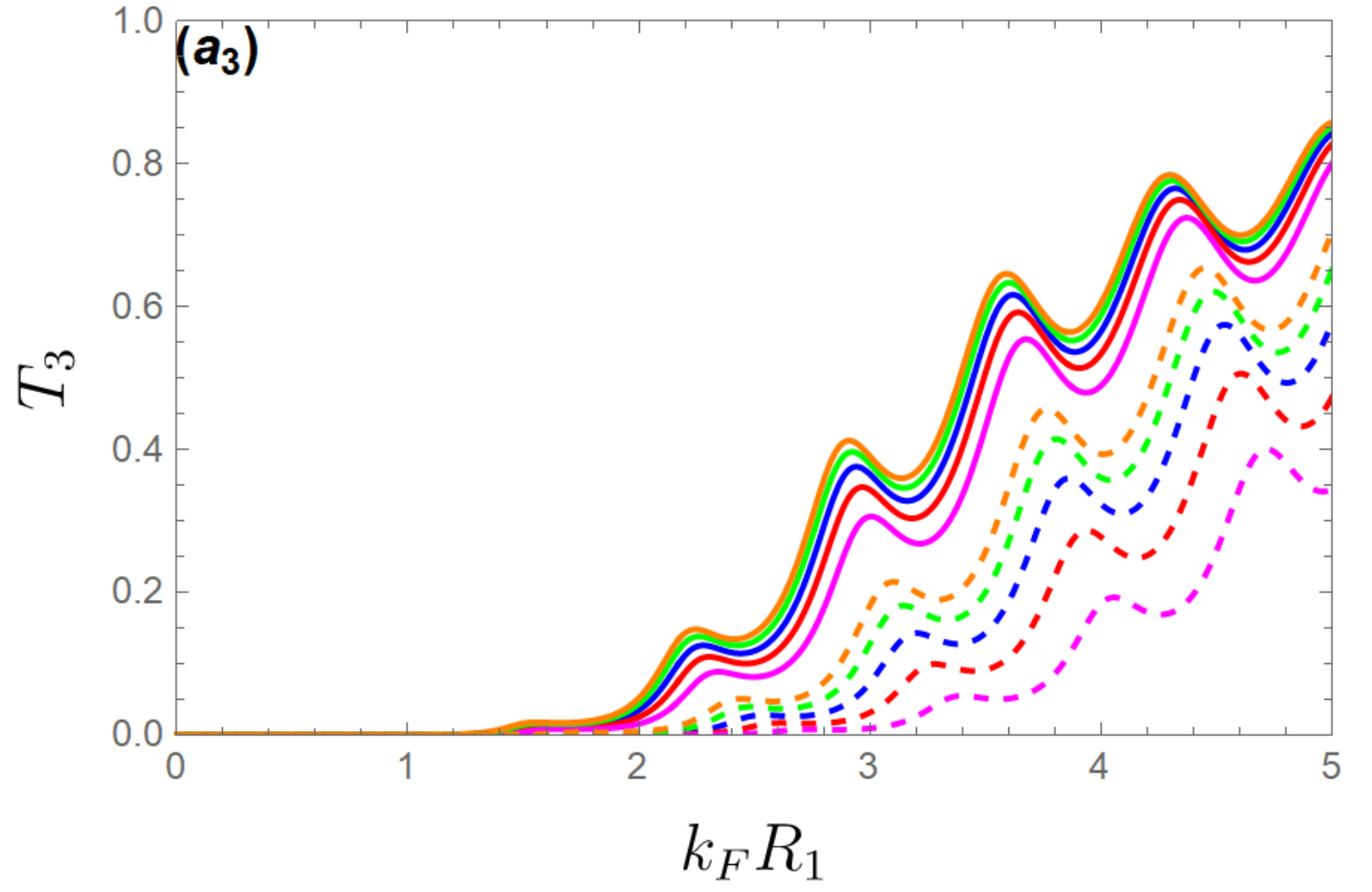}
 	\includegraphics[width=0.49\linewidth]{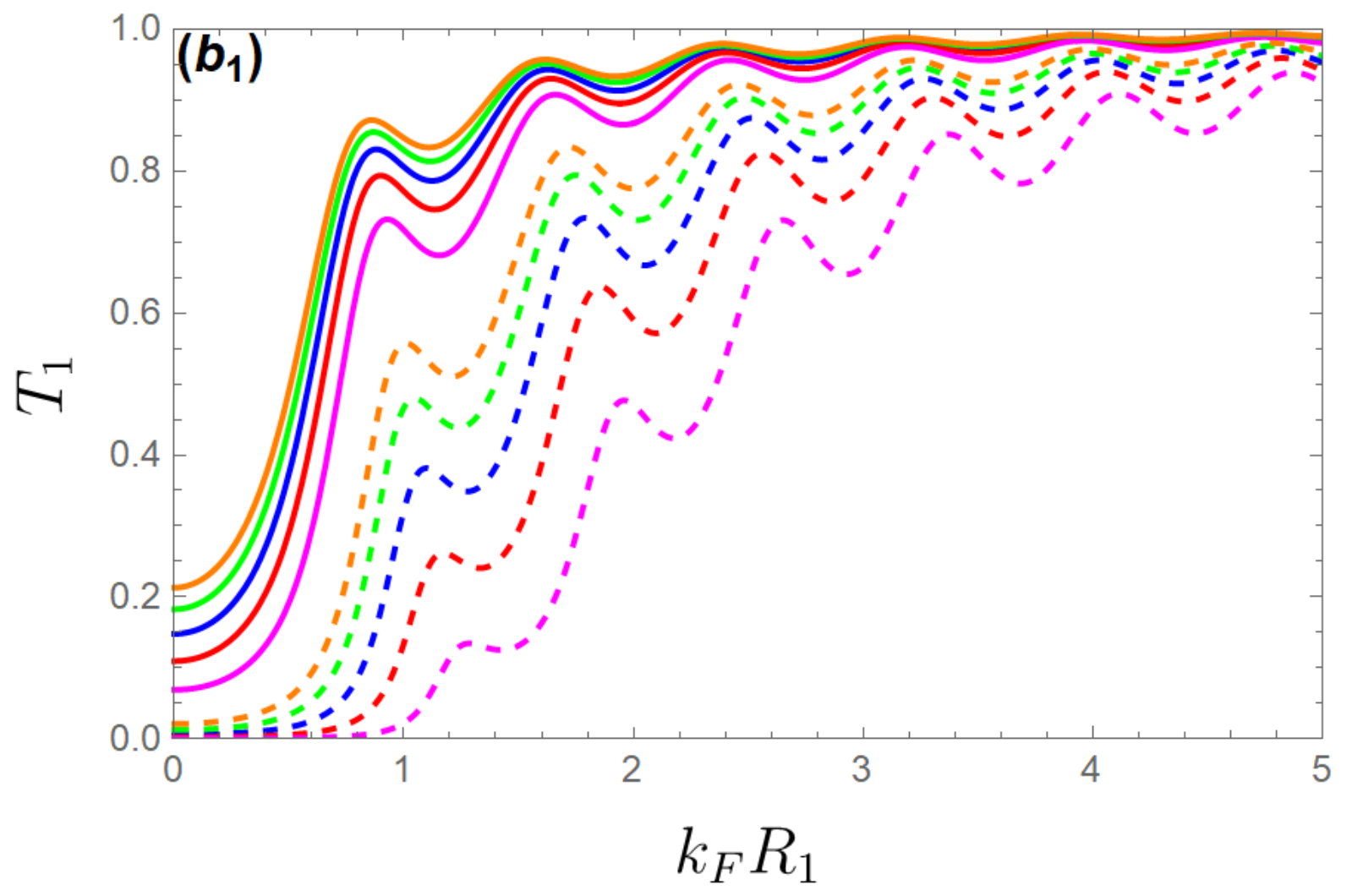}
 	\includegraphics[width=0.49\linewidth]{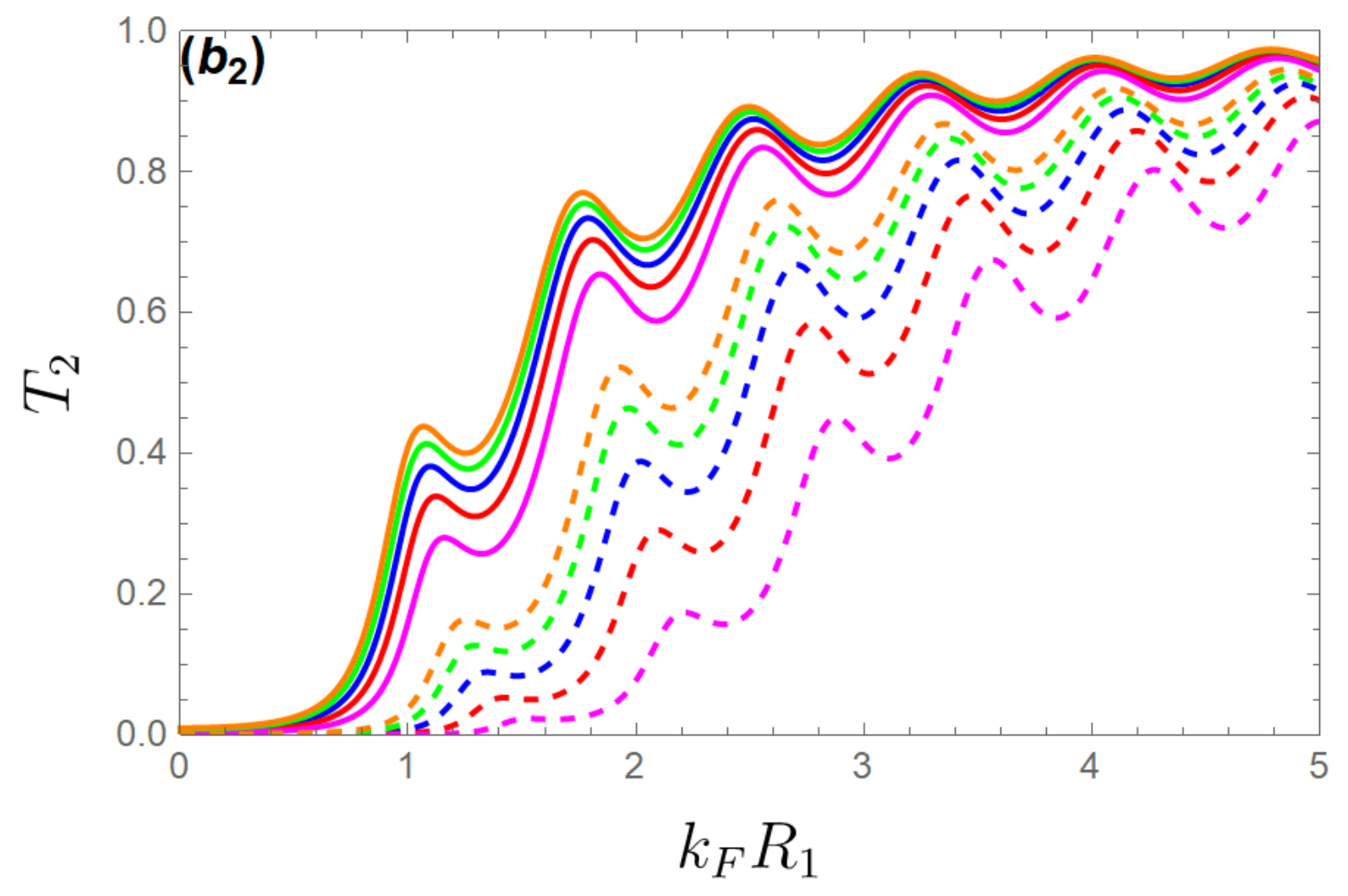}
 	\includegraphics[width=0.49\linewidth]{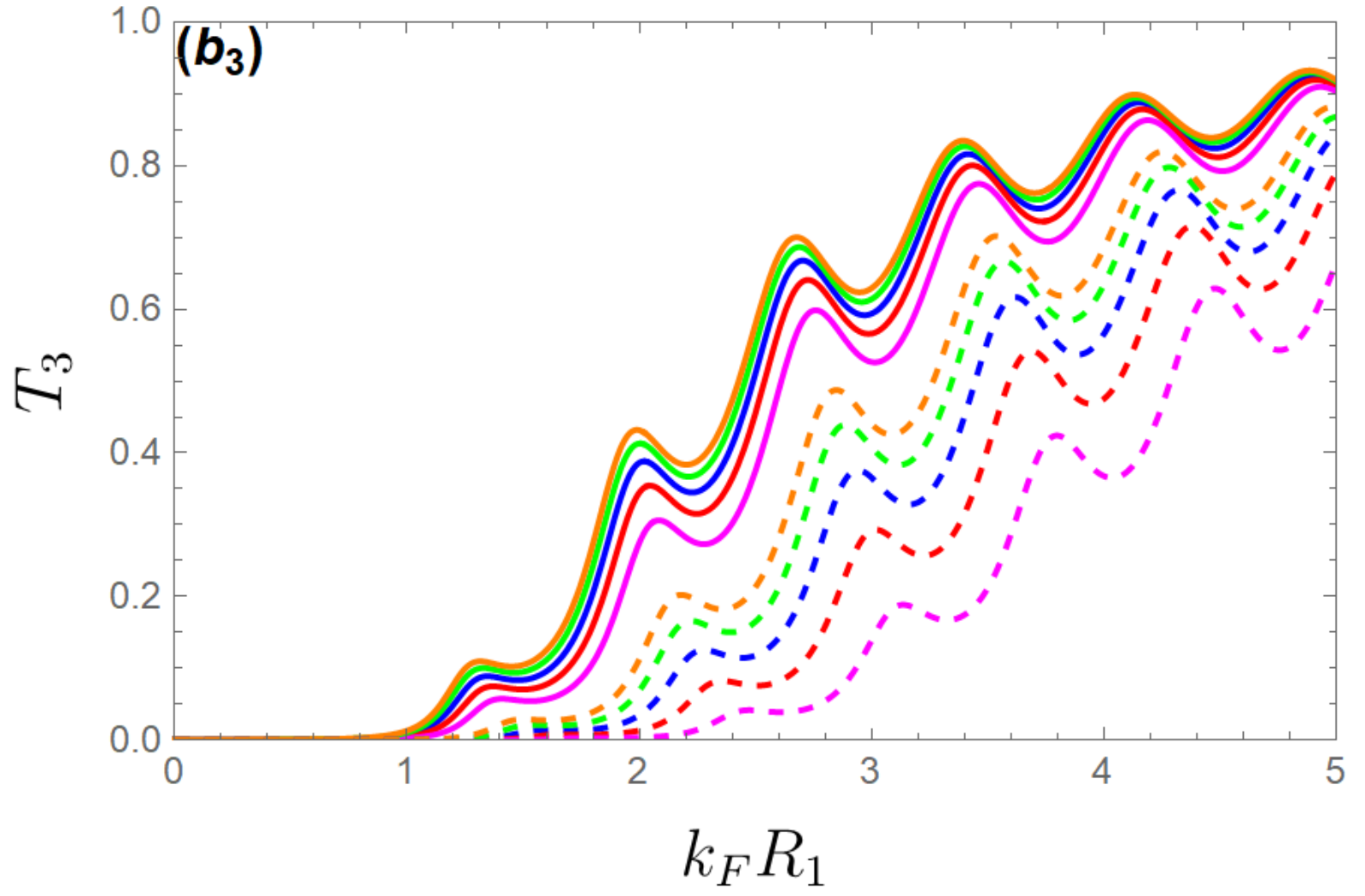}
 	\includegraphics[width=0.49\linewidth]{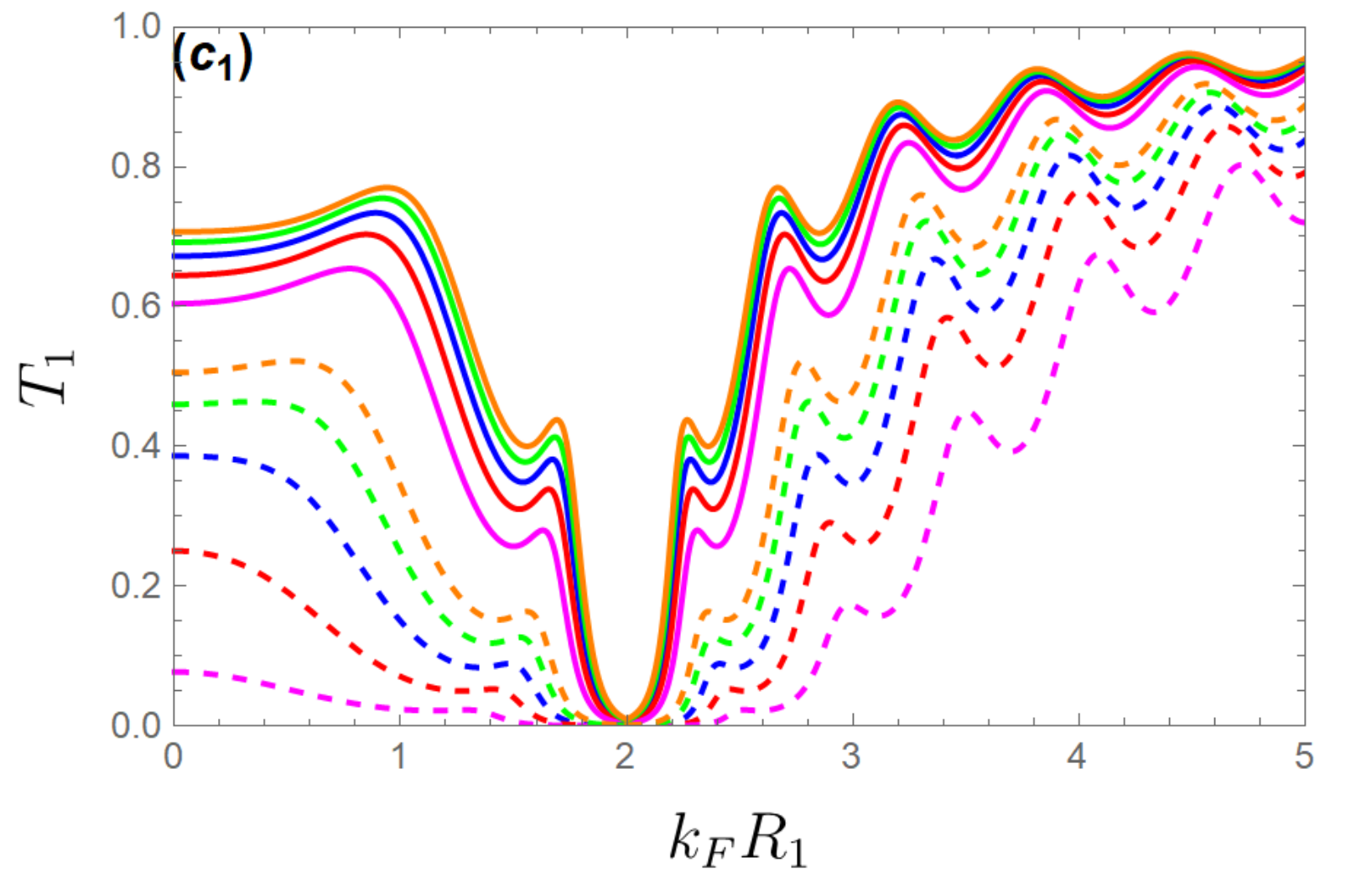}
 	\includegraphics[width=0.49\linewidth]{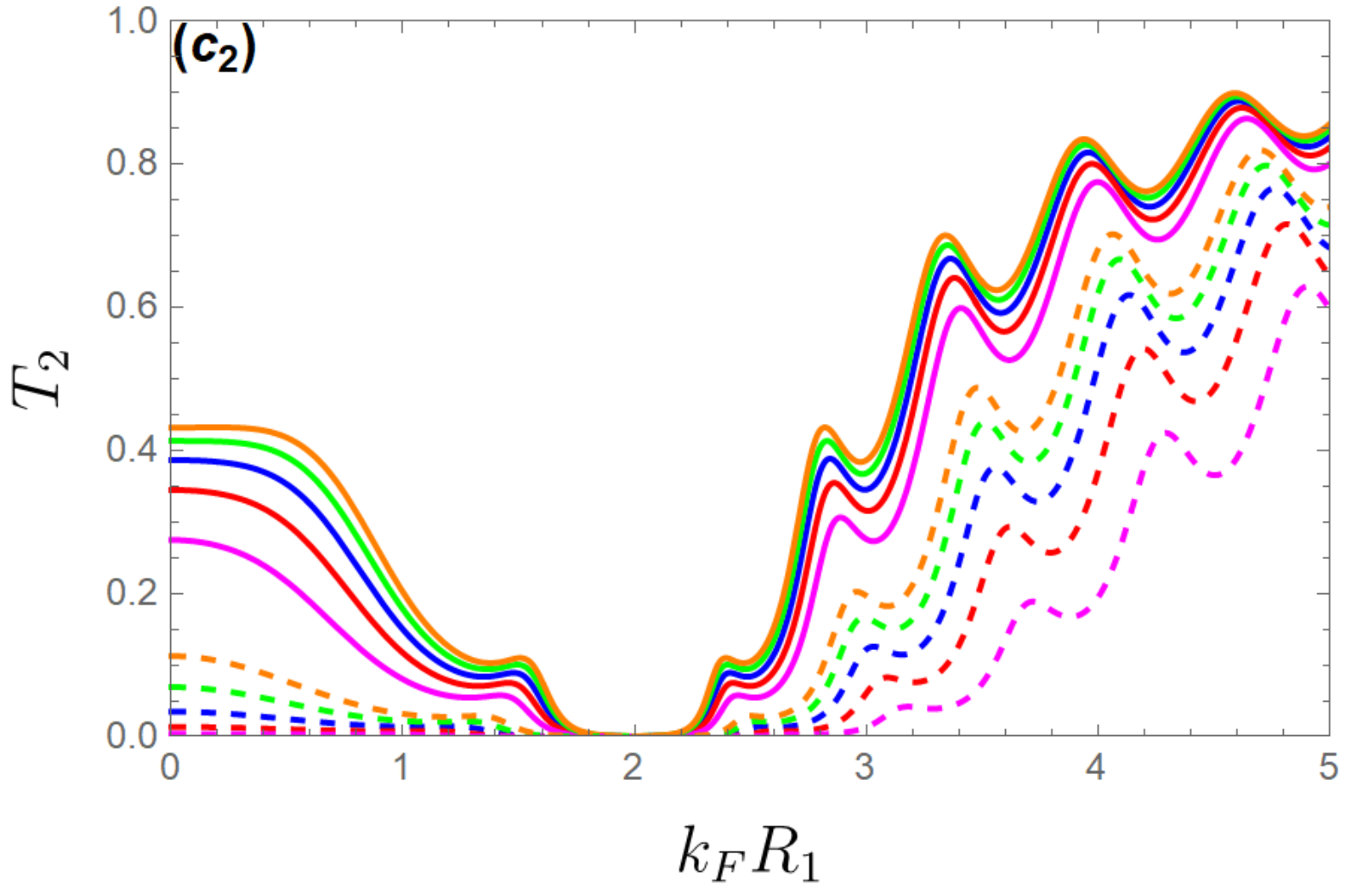}
 	\includegraphics[width=0.49\linewidth]{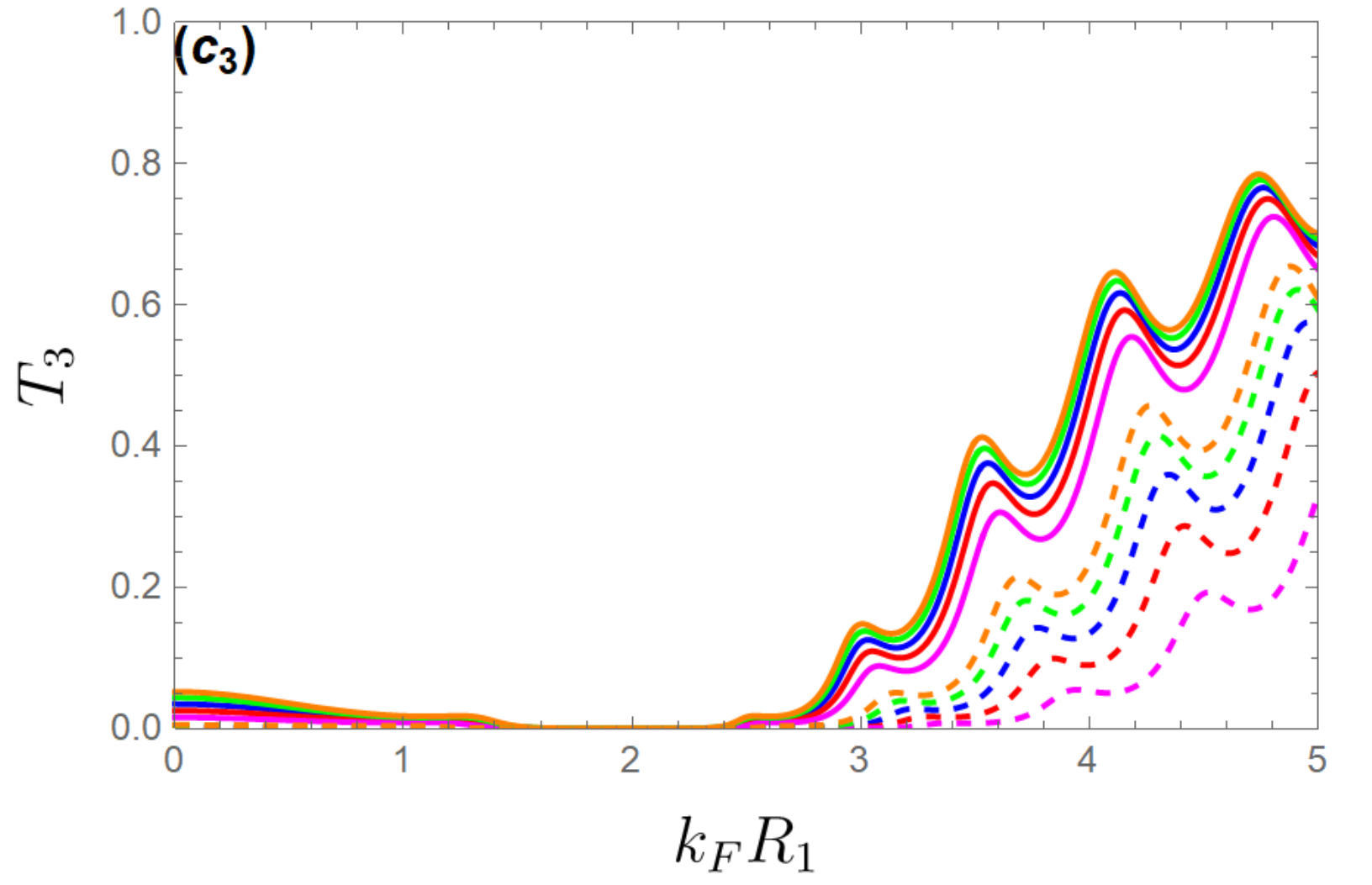}
 	\includegraphics[width=0.49\linewidth]{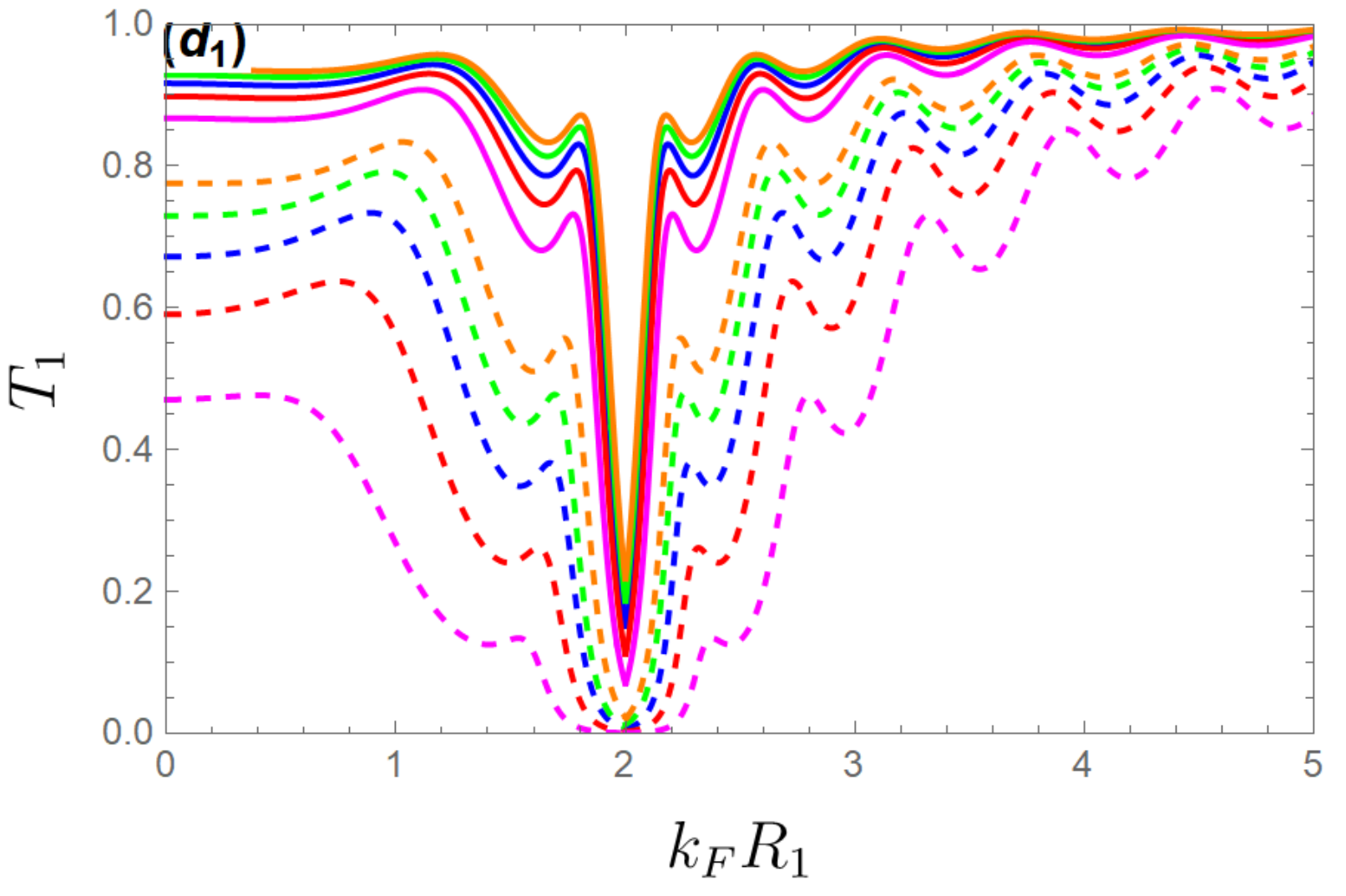}
 	\includegraphics[width=0.49\linewidth]{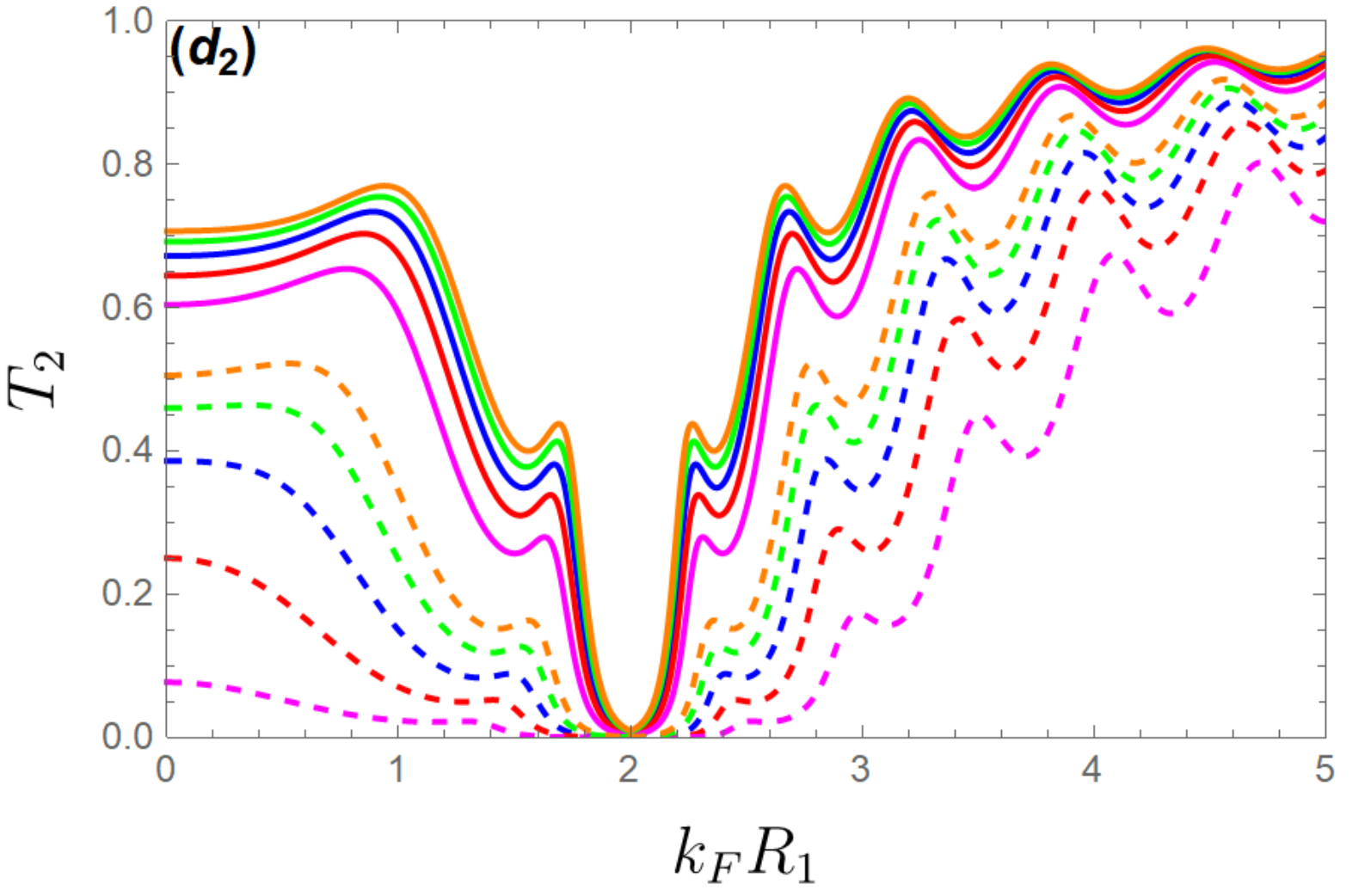}
 	\includegraphics[width=0.49\linewidth]{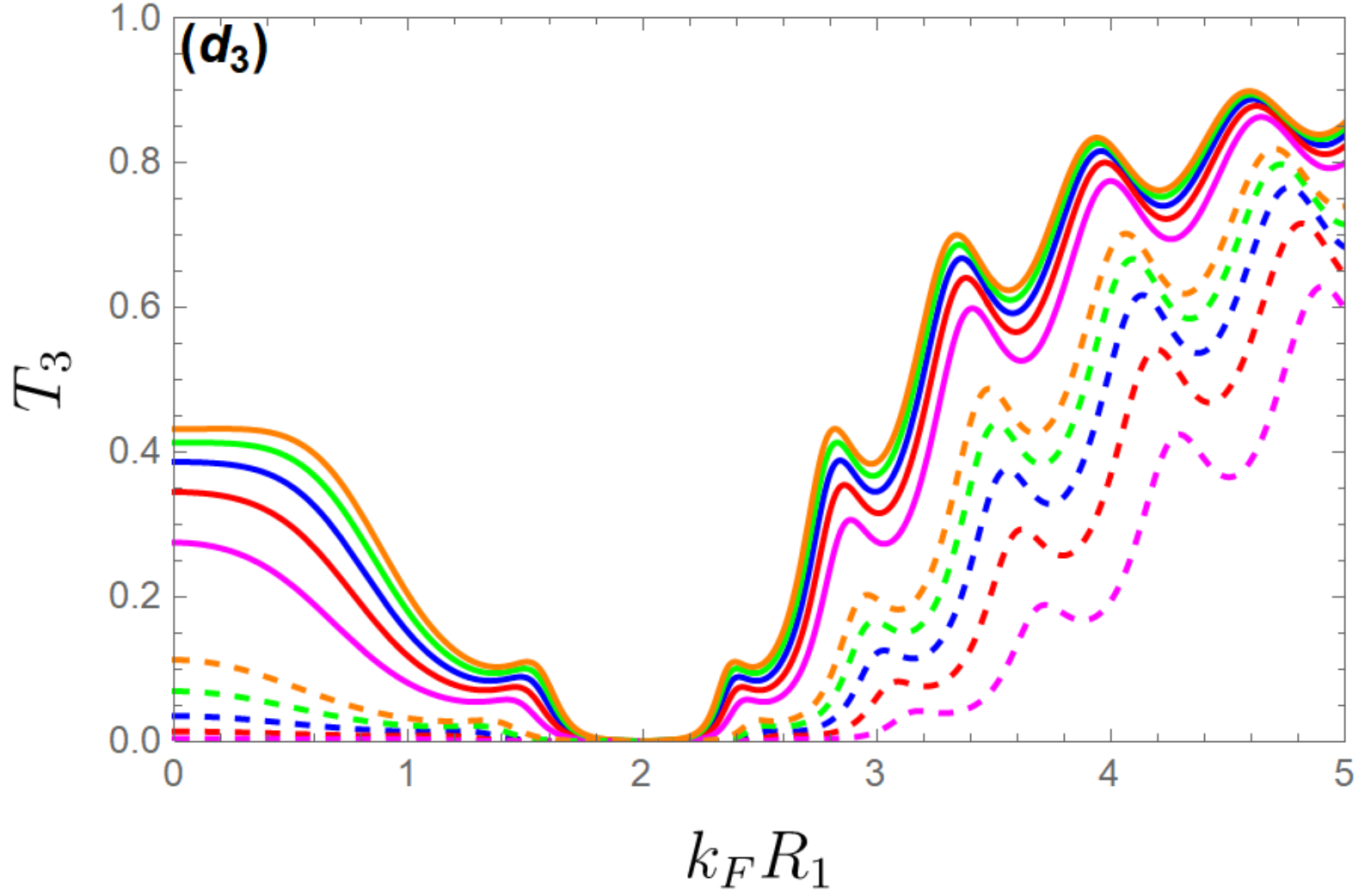}		
 	\caption{(color online)  The transmission
 		$ T_\nu$  $(m=1,2,3)$ as a function of the doping $k_F R_1$  for a radii ratio  $R_2/R_1 =
 		5$,   $ \Phi_i=\frac{1}{2} $ (solid line) and $ \Phi_i=\frac{3}{2} $ (dashed line), with $n = 0$ (blue line), $ n=1 $ (red line), $ n=2 $ (magenta line), $ n=-1 $ (green line), $ n=-2 $ (orange line).  (a$_{\text m}$): $K(\tau=+1)$ and $R_1 \delta=0$,  (b$_{\text m}$):  $K^{\prime}(\tau=-1)$ and $R_1 \delta=0$,  (c$_{\text m}$): $K(\tau=+1)$ and $R_1 \delta=2$,  (d$_{\text m}$): $K^{\prime}(\tau=-1)$ and $R_1 \delta=2$.}\label{trans1}
 \end{figure}
 We numerically study  the transmission $T_\nu$  $ (m=1,2,3)$, the conductance $G$
 and 
 Fano factor $\mathcal{F}$   of the Corbino disk in graphene obtained by creating a gap in the disk area ($R_1<r<R_2$). Indeed,  Fig. \ref{trans1} represents the transmission  as function of doping $k_F R_1$ for the two valleys $K(\tau=+1)$ and $K^{\prime}(\tau=-1)$ with two values of  energy gap $R_1 \delta=0,2$, magnetic flux $\Phi_{i}=\frac{1}{2}, \frac{3}{2}$ and 
 the wedge index $n=0, \pm1, \pm2$.
 Knowing that $n=0$ corresponds to  an isolated in the defect-free case, $n=1$ to an isolated pentagon defect, $n=2$ to an  isolated square defect, $(n=-1)$ to an isolated  heptagon defect and $(n=-2)$ corresponds to an isolated  octagon defect. One sees that more the flux magnetic is increased, the more the transmission is decreasing. The transmission of  $K^{\prime}(\tau=-1)$ is greater than that for  $K(\tau=+1)$. We observe a suppression of the tunneling and the electrons are totally reflected for  $k_F R_1= R_1\delta $, which depends on whether the back-scattering phenomenon is canceled in the presence of energy gap $\delta$ or not. This  is in agreement with our previous results obtained in \cite{babe}. 
 The transmission changes compared to the defect free case $n = 0$ knowing that the transmission for a negative value of $n$ is greater than that for a positive value of $n$. We notice that the difference in the transmission between  $n = 0$ and $n\neq 0$ becomes very important in the presence of magnetic flux $\Phi_i$. We observe that the transmission strongly depends on the valley and  magnetic flux knowing that the configuration which can have a full transmission (Klein tunneling) is $(m=1, \tau=-1, \Phi_i=\frac{1}{2})$, see Figs. \ref{trans1}\textcolor{red}{(b$_{\text 1}$,d$_{\text 1}$)}.

\begin{figure}[htp]\centering
	\includegraphics[width=0.49\linewidth]{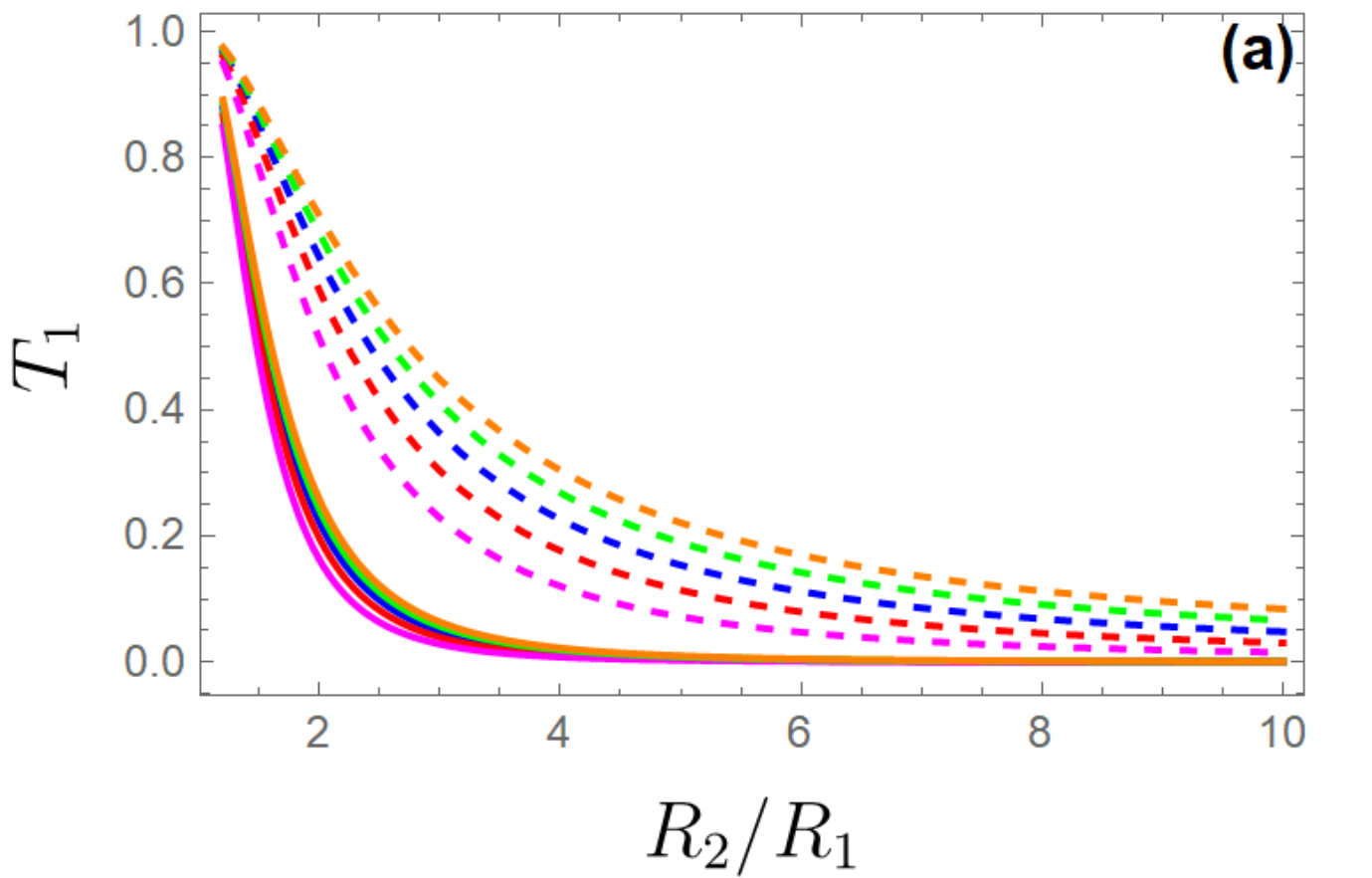}
	\includegraphics[width=0.49\linewidth]{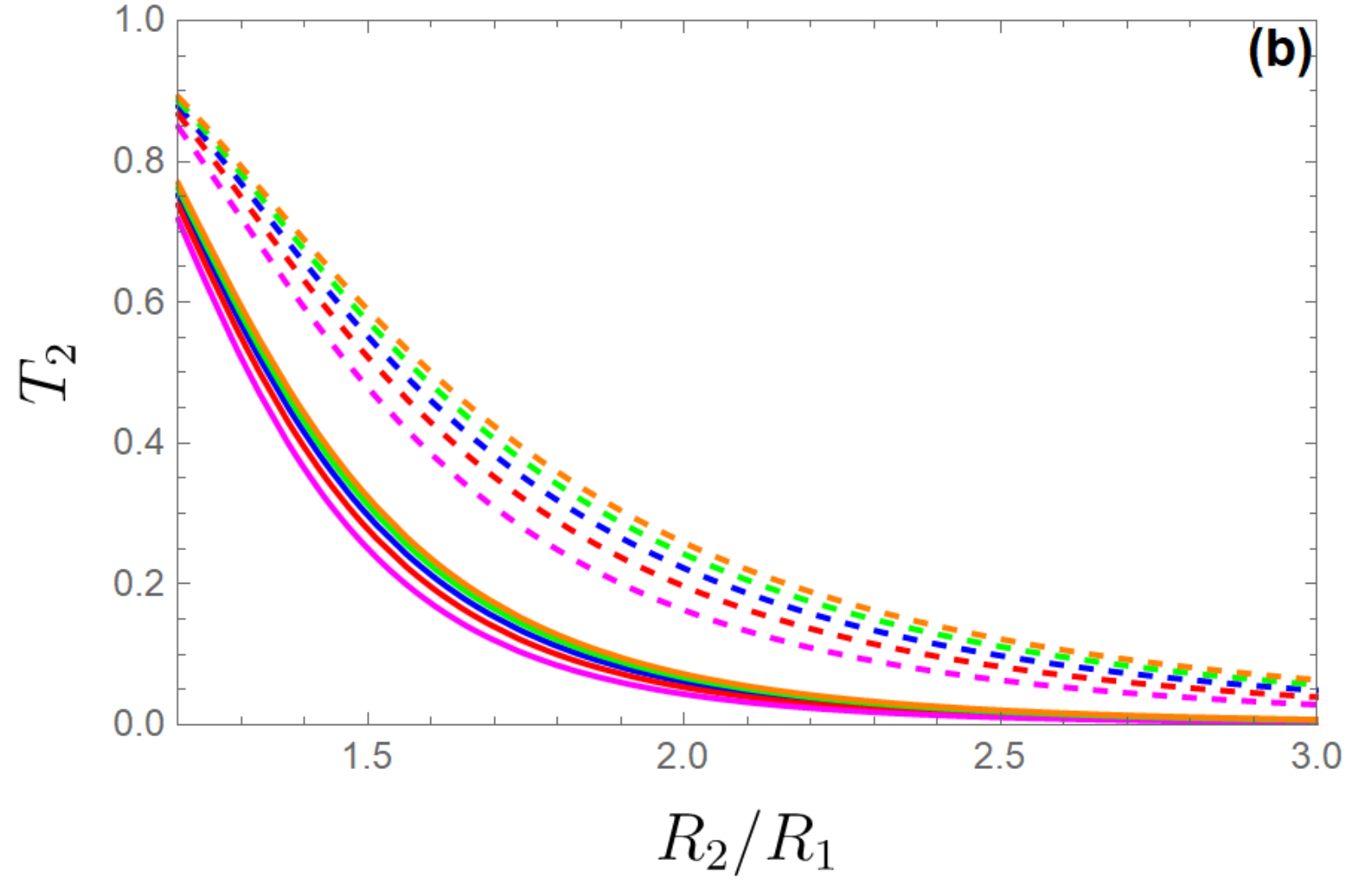}
	\includegraphics[width=0.49\linewidth]{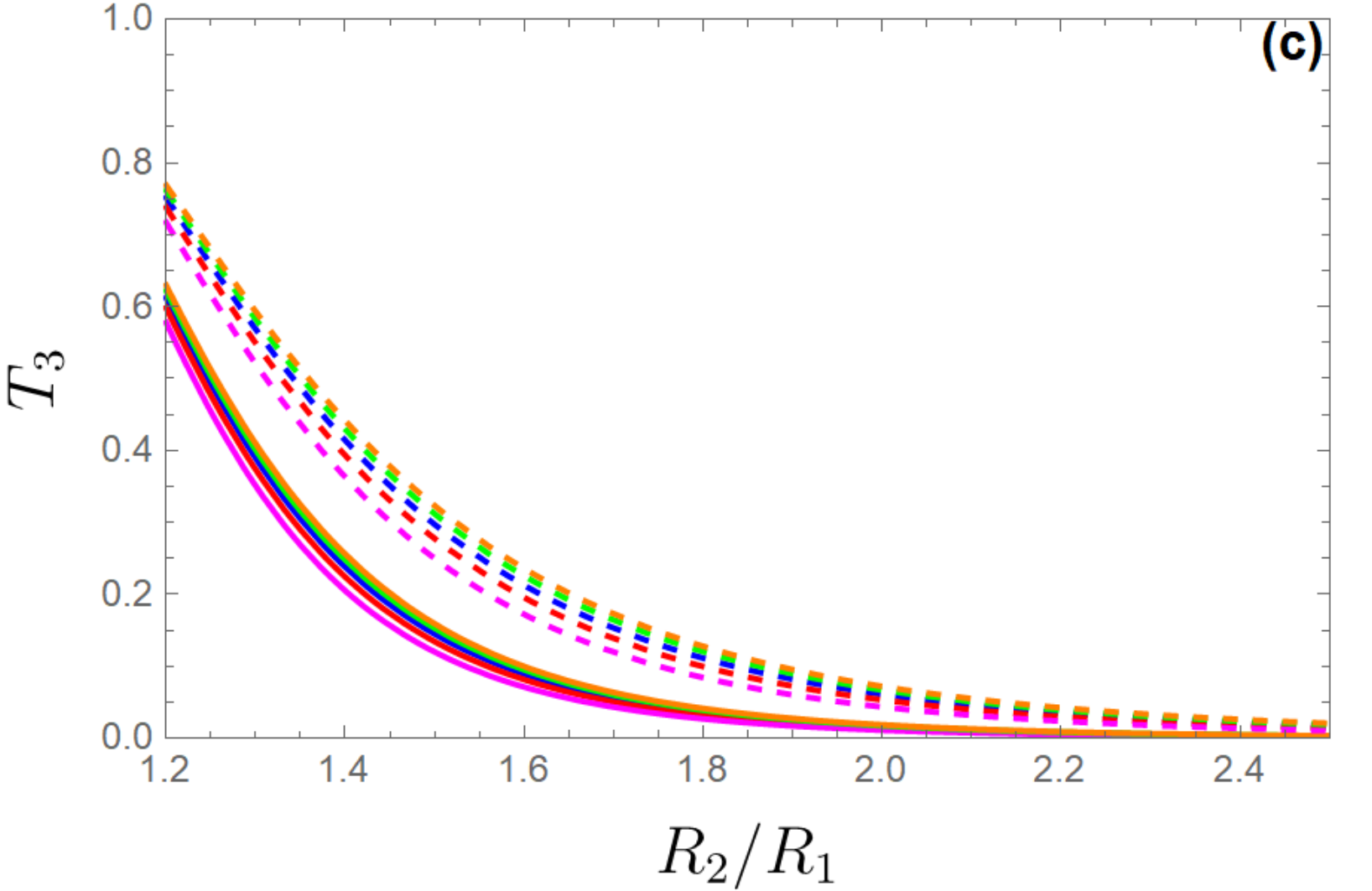}
\includegraphics[width=0.49\linewidth]{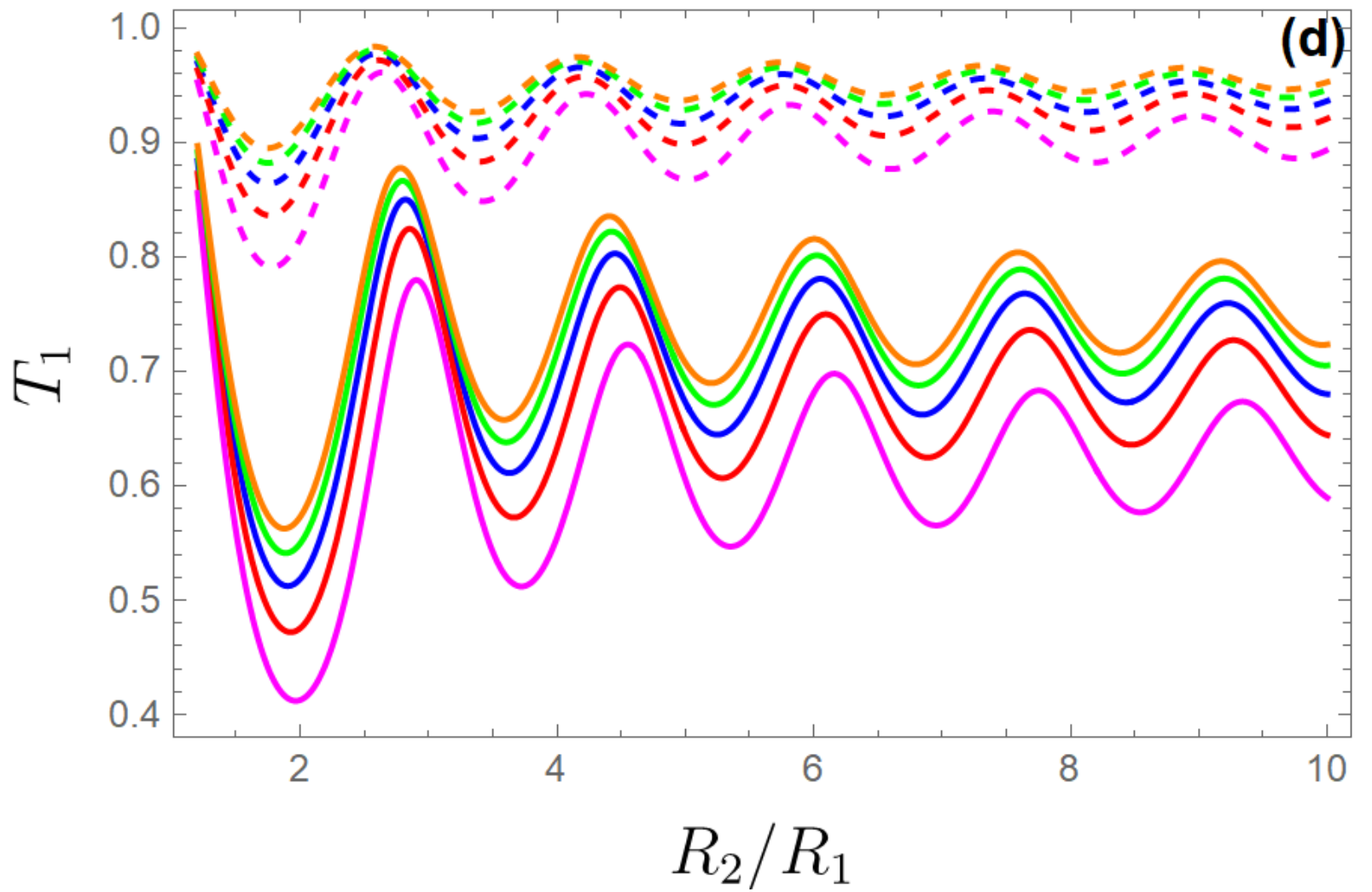}
\includegraphics[width=0.49\linewidth]{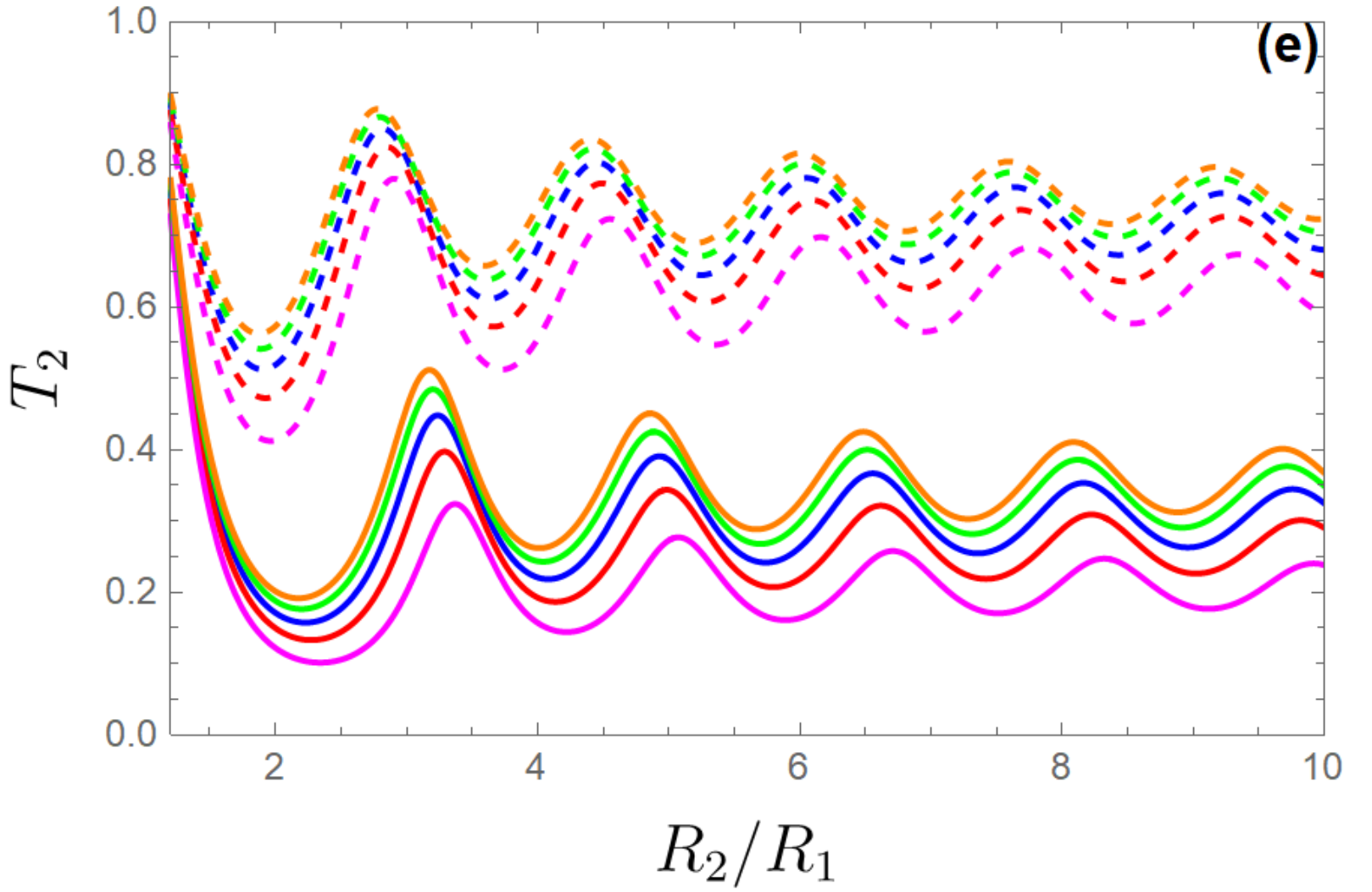}
\includegraphics[width=0.49\linewidth]{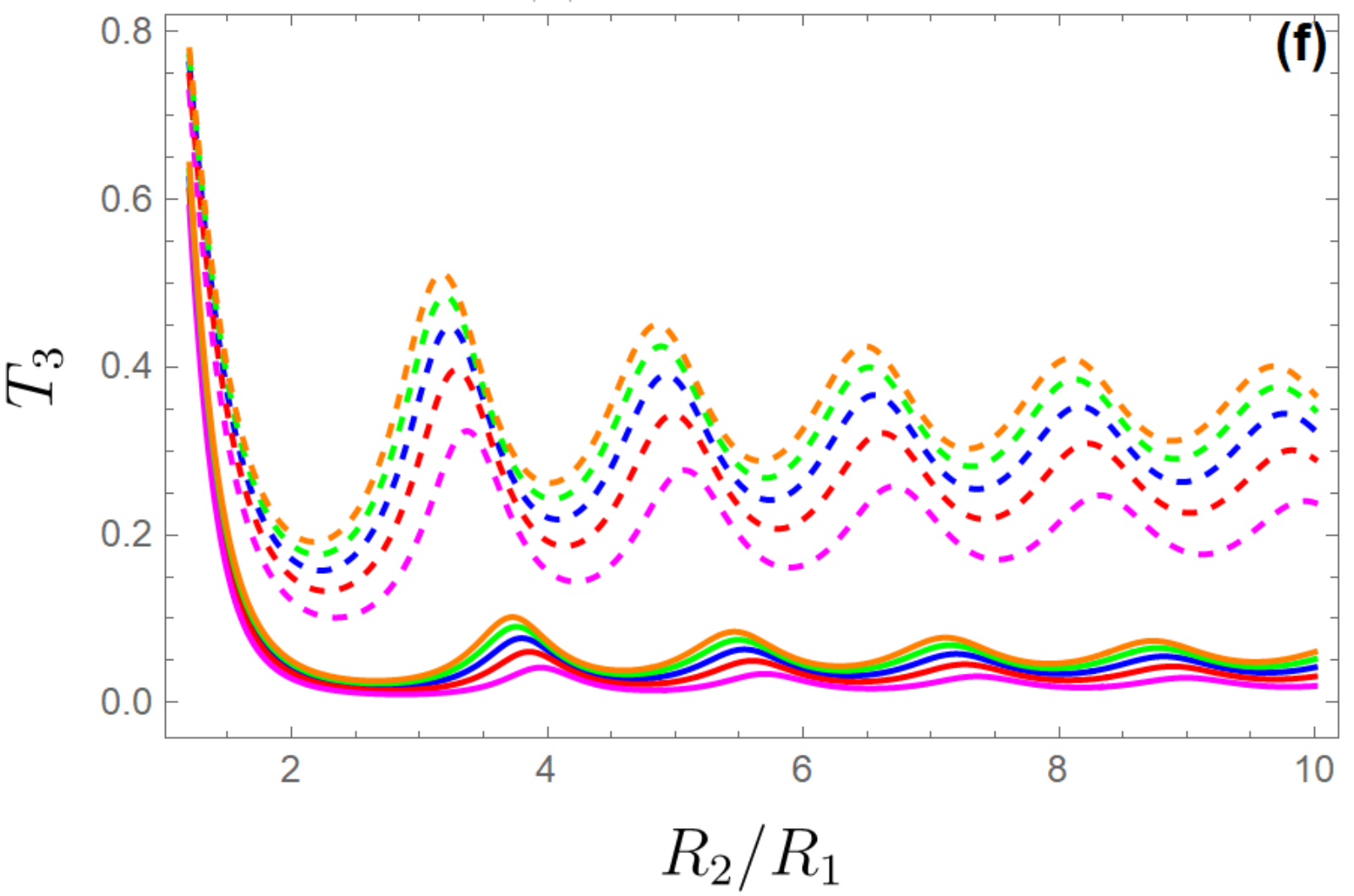}			
	\caption{(color online)  The transmission 
		$ T_\nu$ $ (m=1, 2 ,3) $ as a function of the radii ratio $R_2/R_1$ for the doping  $ k_F R_1 =
		0.1 $, $ \Phi_i=\frac{1}{2} $,    $K (\tau=+1)$ (solid line) and $K^{\prime} (\tau=-1)$ (dashed line).  $n = 0$ (blue line), $ n=1 $ (red line), $ n=2 $ (magenta line), $ n=-1 $ (green line), $ n=-2 $ (orange line). (a,b,c):  $R_1 \delta=0$,   (d,e,f):  $R_1 \delta=2$. } \label{trans2}
\end{figure}

Fig. \ref{trans2} shows the transmission  $ T_\nu$ $ (m=1, 2, 3) $ as a function of the radii ratio $ R_2 / R_1 $ 
for the  valleys $K (\tau=+1)$ and $K^{\prime} (\tau=-1)$, 
two values of energy gap $ R_1 \delta = 0$ (Figs. \ref{trans1}\textcolor{red}{(a,b,c)})   and $ R_1 \delta = 2$ (Figs. \ref{trans2}\textcolor{red}{(d,e,f)}),  magnetic flux $\Phi_i=\frac{1}{2}, \frac{3}{2}$ and  $n=0,\pm1, \pm2$. 
For $R_1 \delta=0$, the transmission  decreases exponentially whatever the value of $n$ and the valley index $\tau=\pm1$. For $R_1 \delta = 2$, we observe  oscillations that become more and more of distinct peaks for  $K^{\prime} (\tau=-1)$ (Figs. \ref{trans2}\textcolor{red}{(d,e,f)}). We have 
a full transmission (Klein tunneling) for $R_1 \delta=0$, $\tau=-1$ and $n=-2$. 
We notice that the amplitude of these oscillations decreases when the parameter of the radii ratio $R_2/R_1$ is increased and depends on the value of the disclination effect $n$ knowing that its smallest value corresponds to the case where $n=-2$ the octagon defect. Both Figs. \ref{trans1} and \ref{trans2} show that the transmission is reduced if the value of the angular momentum $\nu$ is increased.

\begin{figure}[H]\centering
	\includegraphics[width=0.49\linewidth]{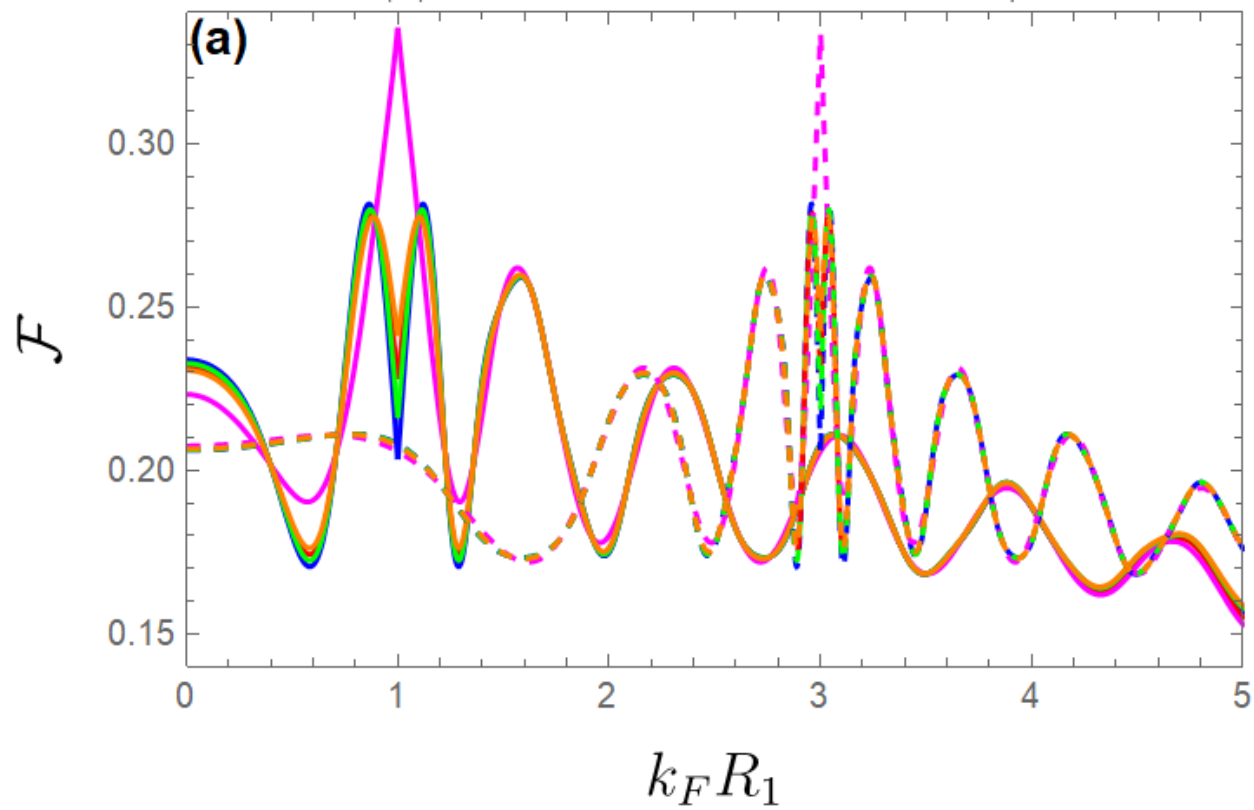}
	\includegraphics[width=0.49\linewidth]{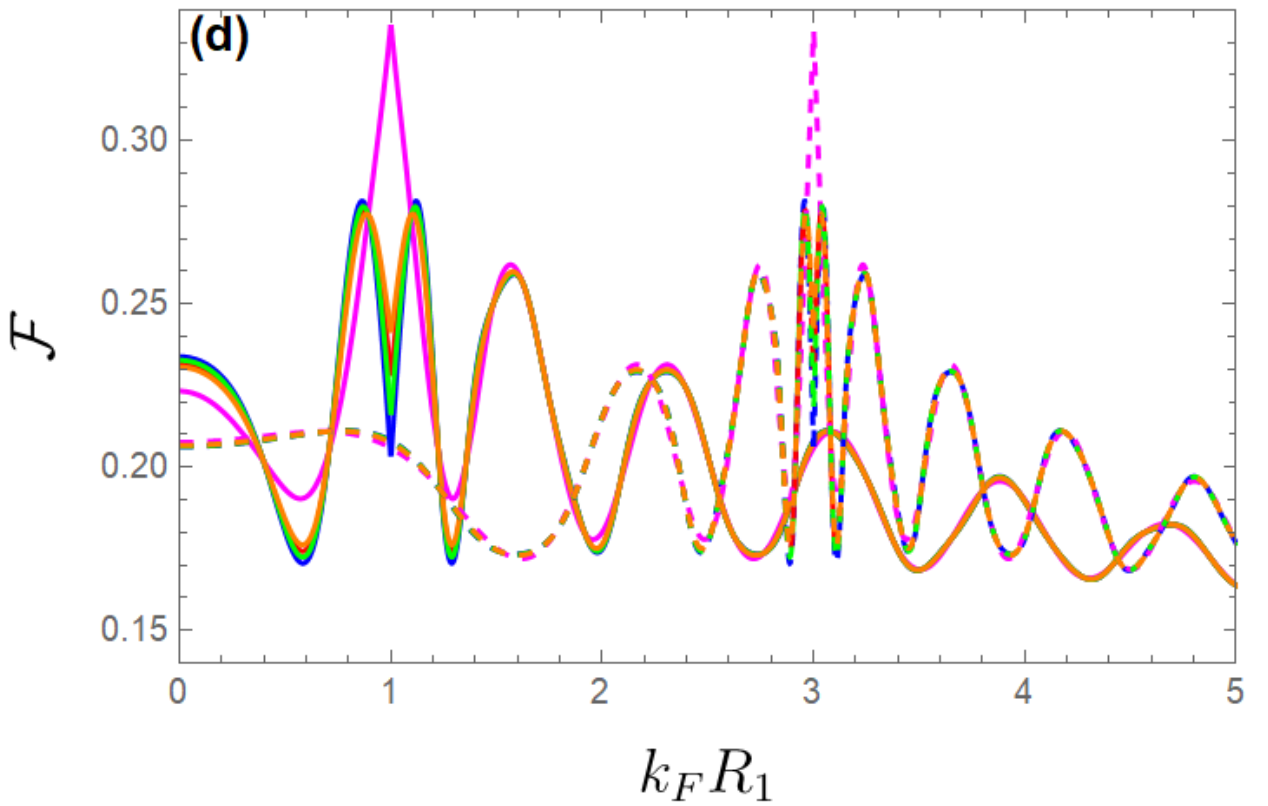}
	\includegraphics[width=0.49\linewidth]{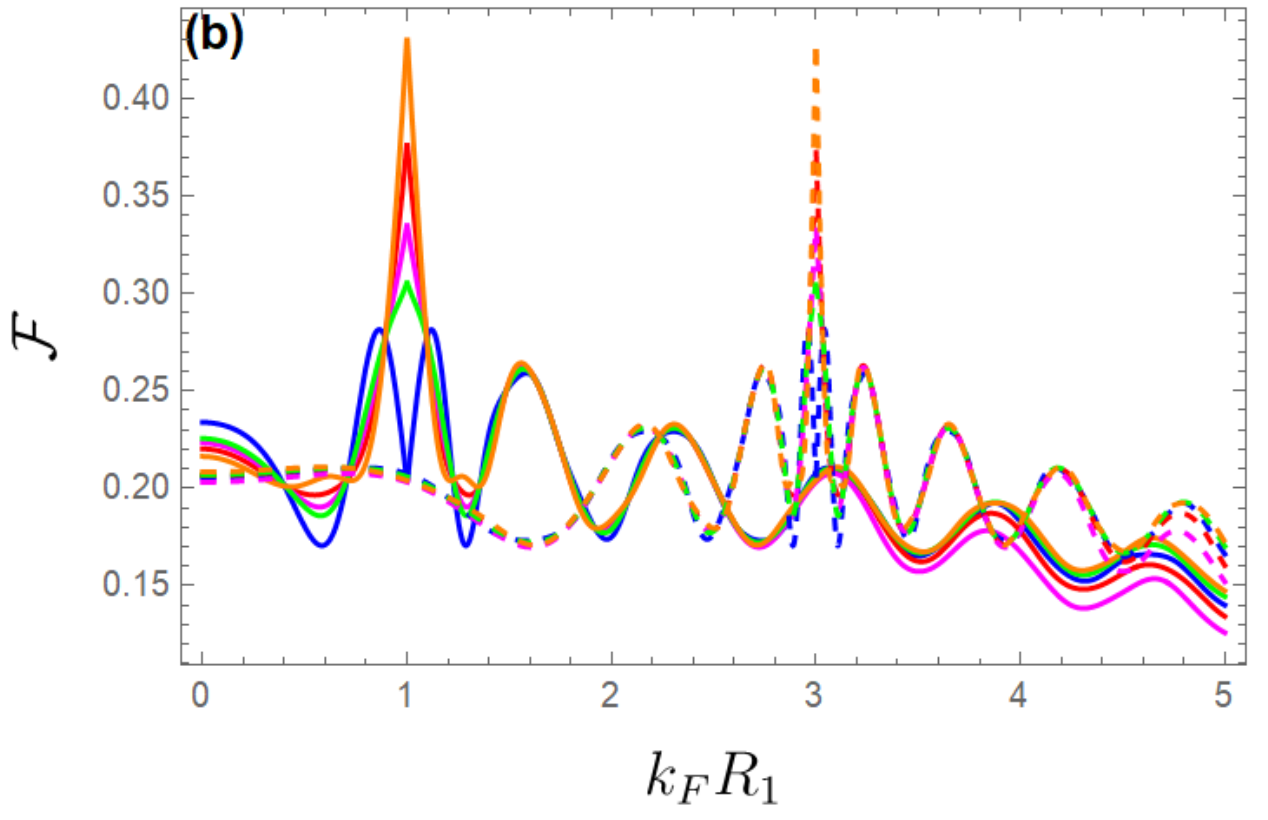}
	\includegraphics[width=0.49\linewidth]{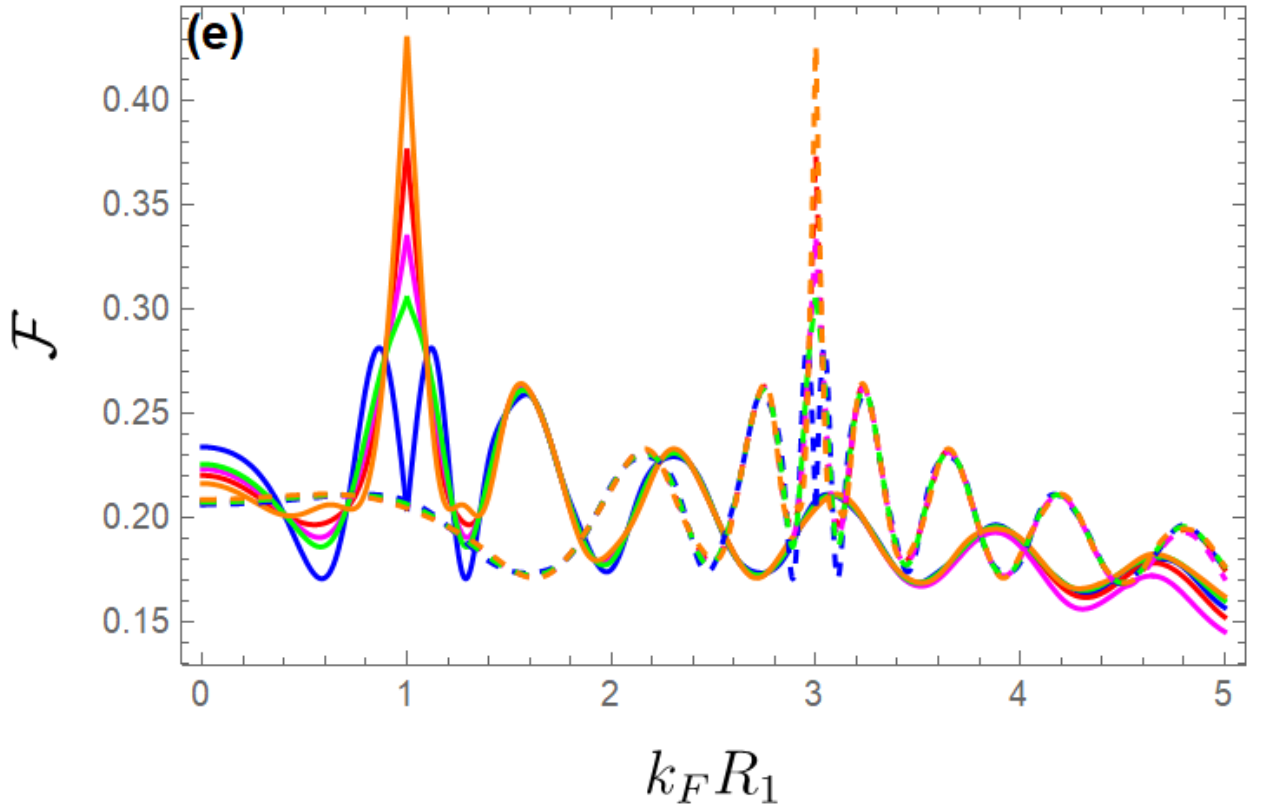}
	\includegraphics[width=0.49\linewidth]{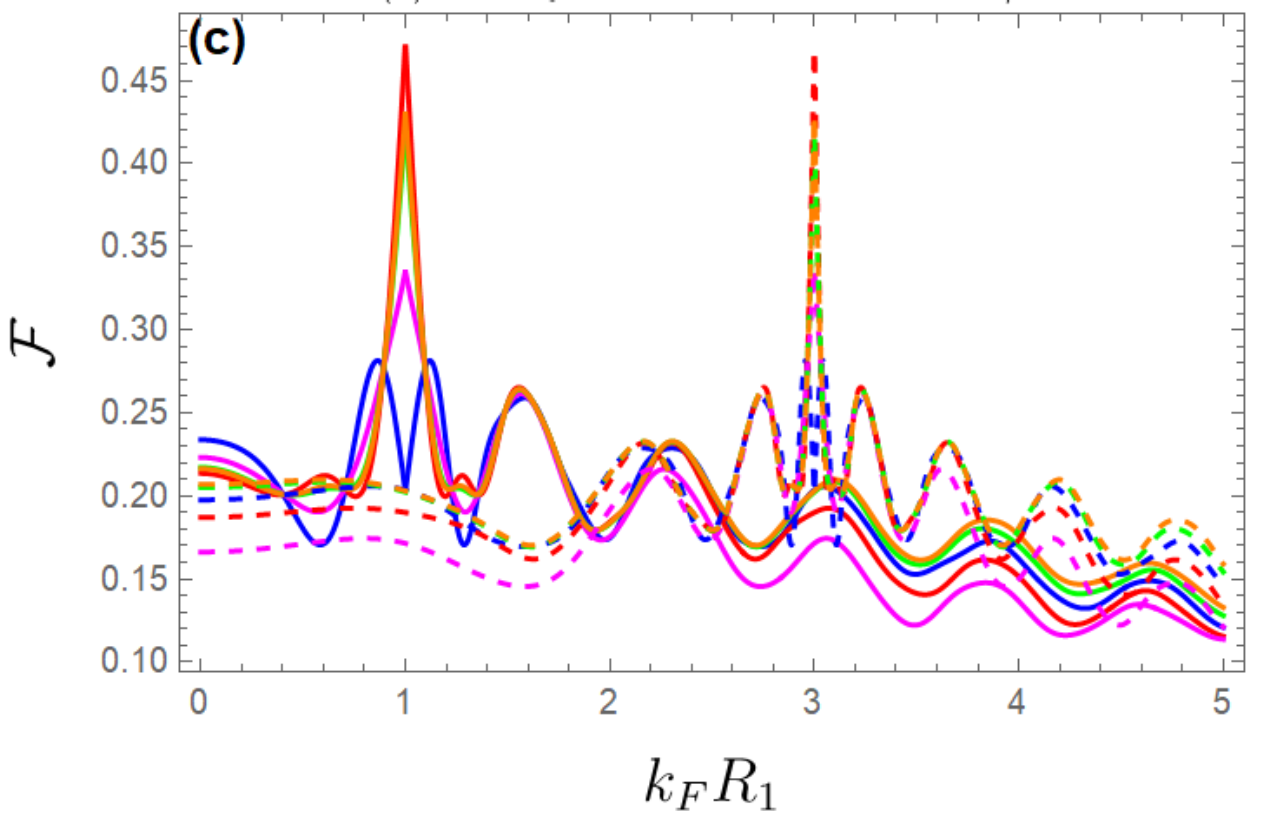}
	\includegraphics[width=0.49\linewidth]{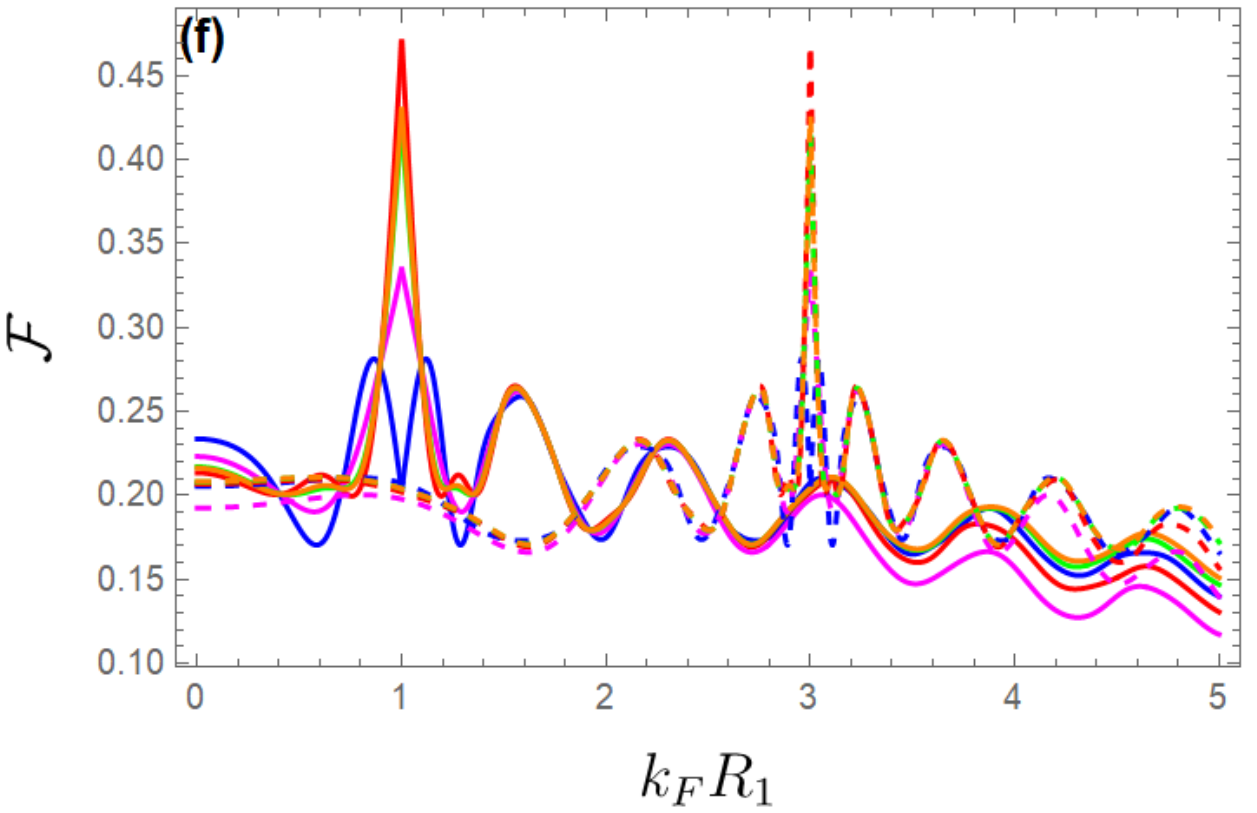}		
	\caption{(color online) The Fano factor $\mathcal{F}$  as a function of the doping $k_F R_1$  for the  radii ratio  $R_2/R_1 =
		5$,  $ R_1\delta=1 $ (solid line) and $ R_1\delta=3 $ (dashed line).  $n = 0$ (blue line), $ n=1 $ (red line), $ n=2 $ (magenta line), $ n=-1 $ (green line), $ n=-2 $ (orange line).  (a,b,c):  $K(\tau=+1)$,  (d,e,f):  $K^{\prime}(\tau=-1)$. (a,d):  $\Phi_i=\frac{1}{2}$,  (b,e):  $\Phi_i=\frac{3}{2}$,  (c,f):  $\Phi_i=\frac{5}{2}$. } \label{fano1}
\end{figure} 
 
In Fig. \ref{fano1}, we report the Fano
factor as a function of the doping $k_F R_1$ for the radii ratio  $R_2/R_1 =5$,  energy gap $ R_1\delta=1 $ (solid line) and $ R_1\delta=3 $ (dashed line),  $n=0, \pm1, \pm2$ and  magnetic flux $\Phi_i=\frac{1}{2}, \frac{3}{2}, \frac{5}{2}$. We first observe that  $\mathcal{F}$ is independent of the valley $K(\tau=+1)$ or $K^{\prime}(\tau=-1)$. Figs. \ref{fano1}\textcolor{red}{(a,d)}  are for $\Phi_i=1/2$ where we observe intense peaks at the point
$k_F R_1 = R_1 \delta$, then the curves follow an oscillatory process for high doping just for  the case of a square defect $n = 2$. For  $n=0, \pm1, -2$ there exists a double identical peaks of average intensity at $\mathcal{F}=0.28$. 
We stress that  
the intensity of peaks is independent of  the  energy gap $R_1 \delta$. Figs. \ref{fano1}\textcolor{red}{(b,c,e,f)}  are for $\Phi_i=\frac{3}{2}, \frac{5}{2}$,  we observe intense peaks at the points $k_F R_1 = R_1 \delta$ for $n\neq 0$ whose maximum intensity value varies from one case to another. Obviously the case where $n = 0$ free defect remains unchanged. 

\begin{figure}[htp]\centering
	\includegraphics[width=0.49\linewidth]{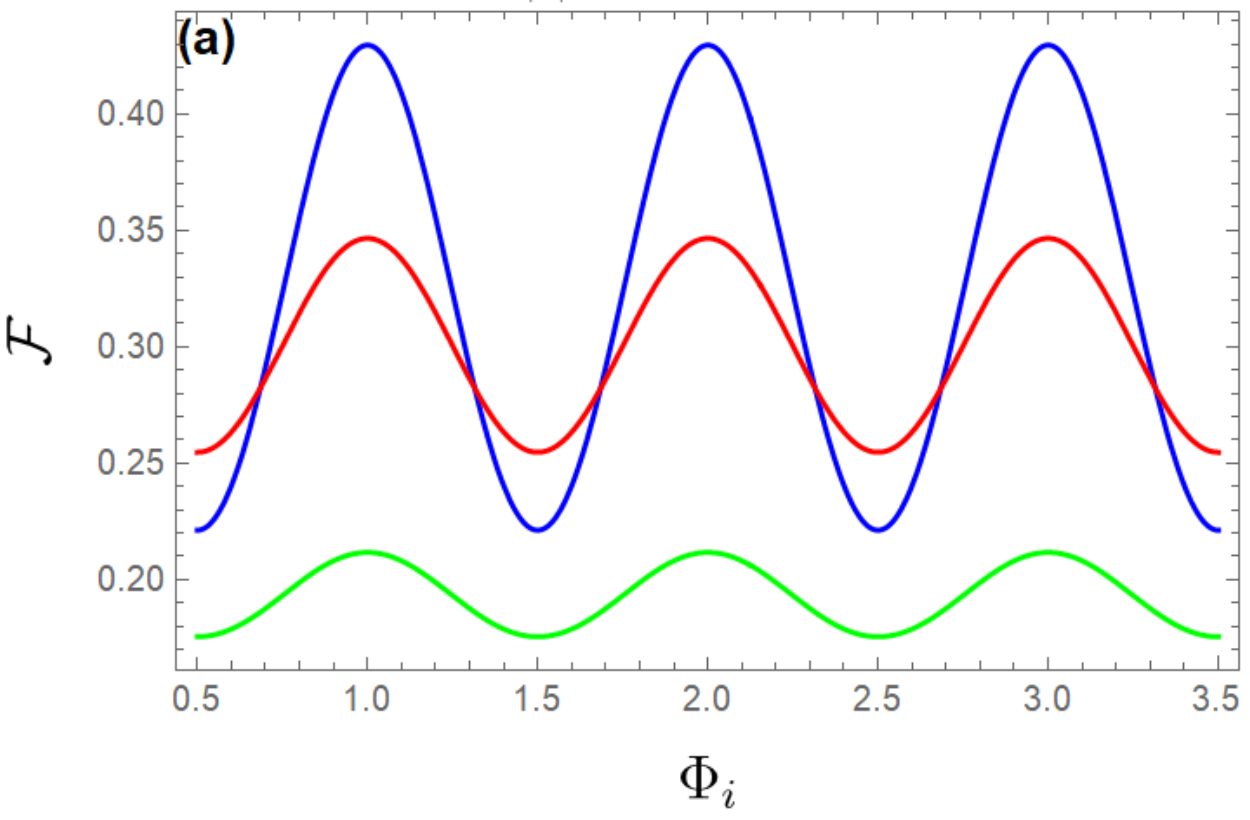}
	\includegraphics[width=0.49\linewidth]{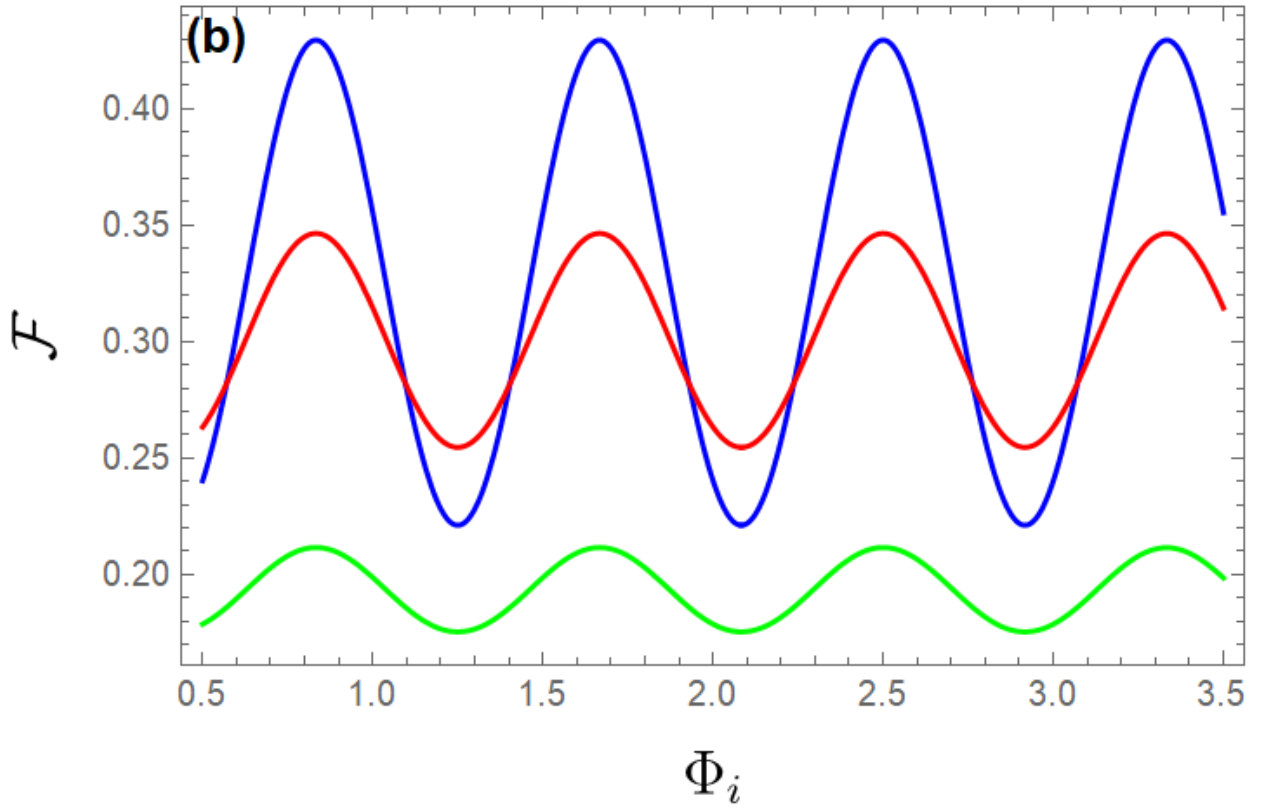}
	\includegraphics[width=0.49\linewidth]{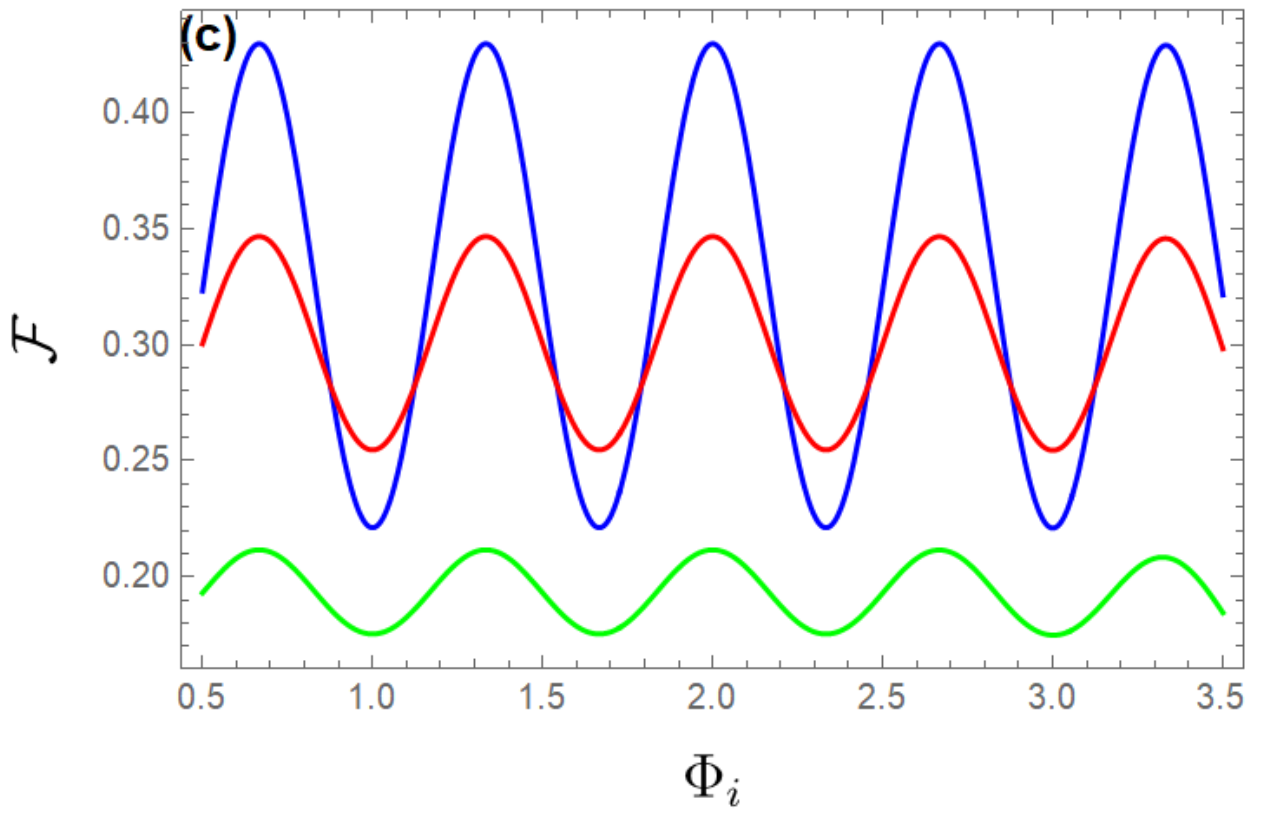}
	\includegraphics[width=0.49\linewidth]{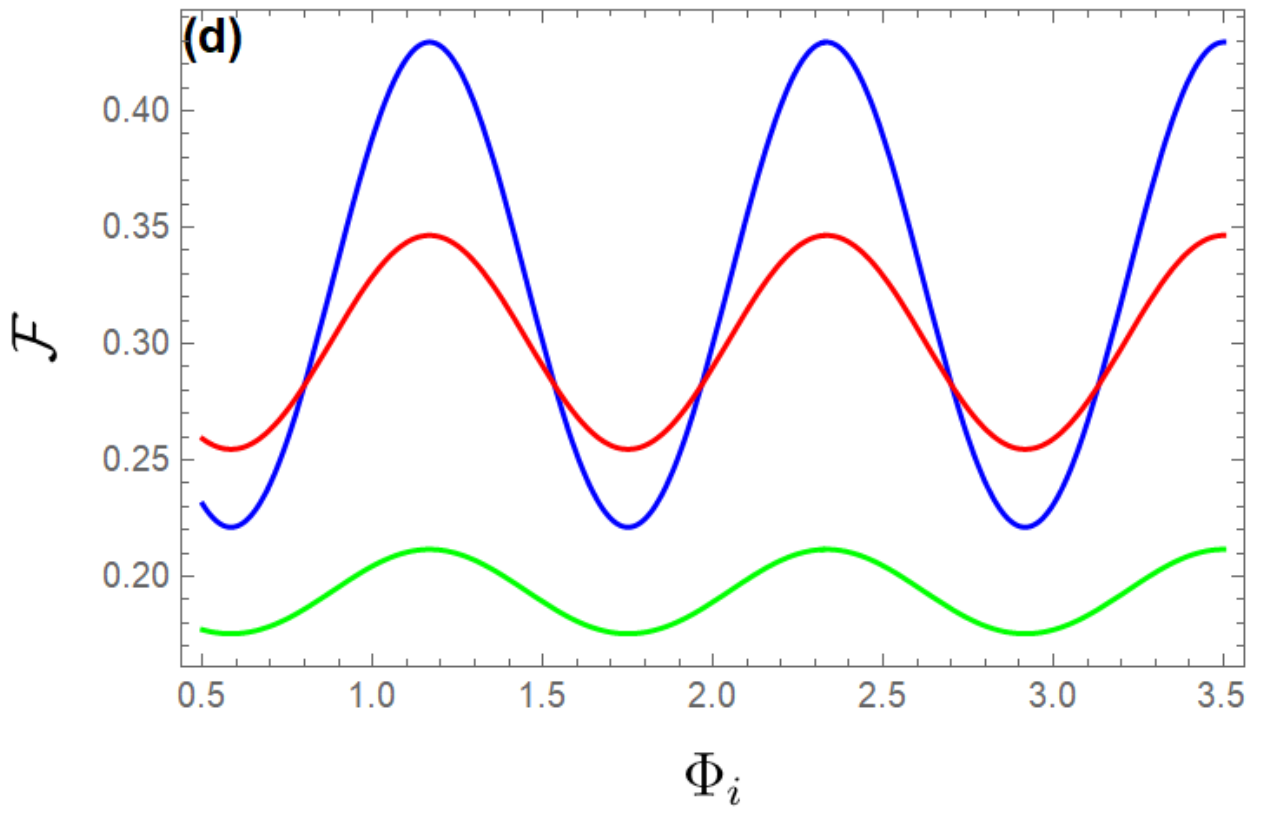}
	\includegraphics[width=0.49\linewidth]{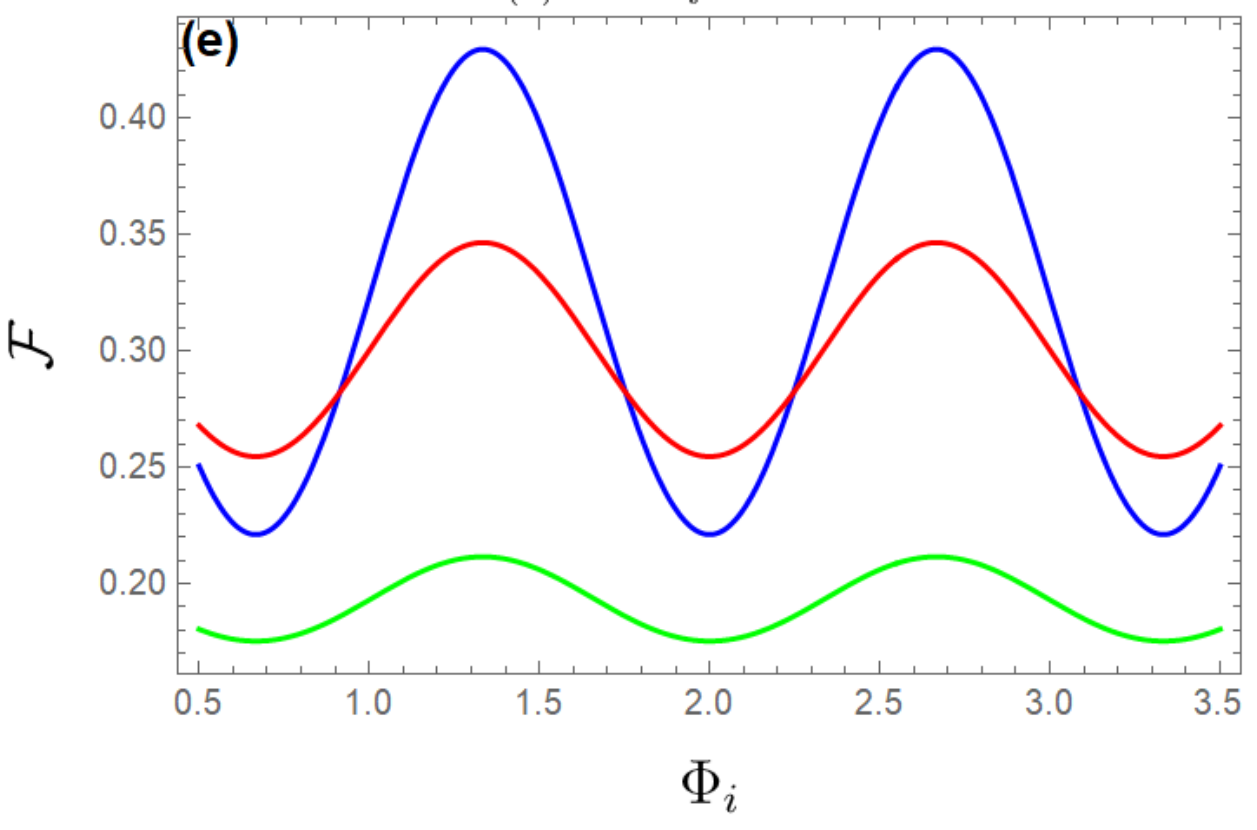}
	\includegraphics[width=0.49\linewidth]{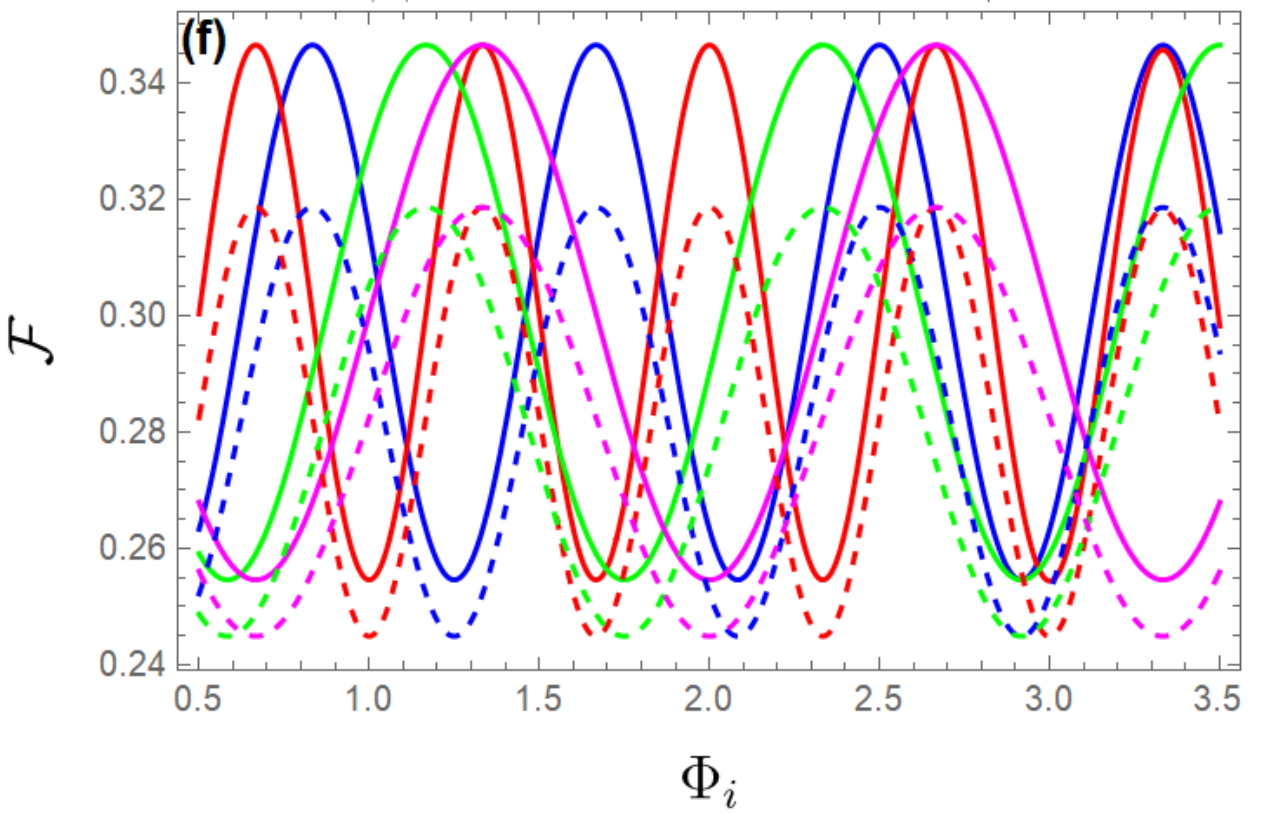}		
	\caption{(color online) The Fano factor $\mathcal{F}$ as a function of the magnetic flux $\Phi_i$ for the  doping  $ k_F R_1 = 0.2 $ and  the valley  $K (\tau=+1)$.   (a,b,c,d,e): $R_2/R_1 =5$, $R_1 \delta = 0$ (blue line), $ R_1 \delta = 0.4 $ (red line), $ 0.8 $ (green line). (a):  $n=0$,  (b):  $n=1$, (c):  $n=2$, (d):  $n=-1$,  (e):  $n=-2$. (f):  $ R_1 \delta = 0.4 $, $R_2/R_1 =
		5$ (solid line ), $R_2/R_1 =
		6$ (dashed line ), $n = 0$ (blue line), $ n=1 $ (red line), $ n=2 $ (magenta line), $ n=-1 $ (green line),  $ n=-2 $ (orange line). } \label{fano2}
\end{figure}

Fig. \ref{fano2} represents the Fano factor $\mathcal{F}$ as function the magnetic flux $\Phi_i$ for a low doping  $ k_F R_1 = 0.2 $.  
We observe that $\mathcal{F}$ shows a periodic oscillation and its amplitude depends on both  $ R_1\delta$ and $R_2/R_1$ because its decreases with the increasing of them.
It is interesting to note that 
 the period of these oscillations depends on   $n$. Indeed, the maximal period  of the oscillations corresponds to the  octagon defect $n=-2$ while  the minimal  one
 corresponds to   square defect $n=2$. Then it is clearly seen that  the wedge index $n$ acts by changing the period of the oscillations of
  $\mathcal{F}$.

\begin{figure}[htp]\centering
	\includegraphics[width=0.5\linewidth]{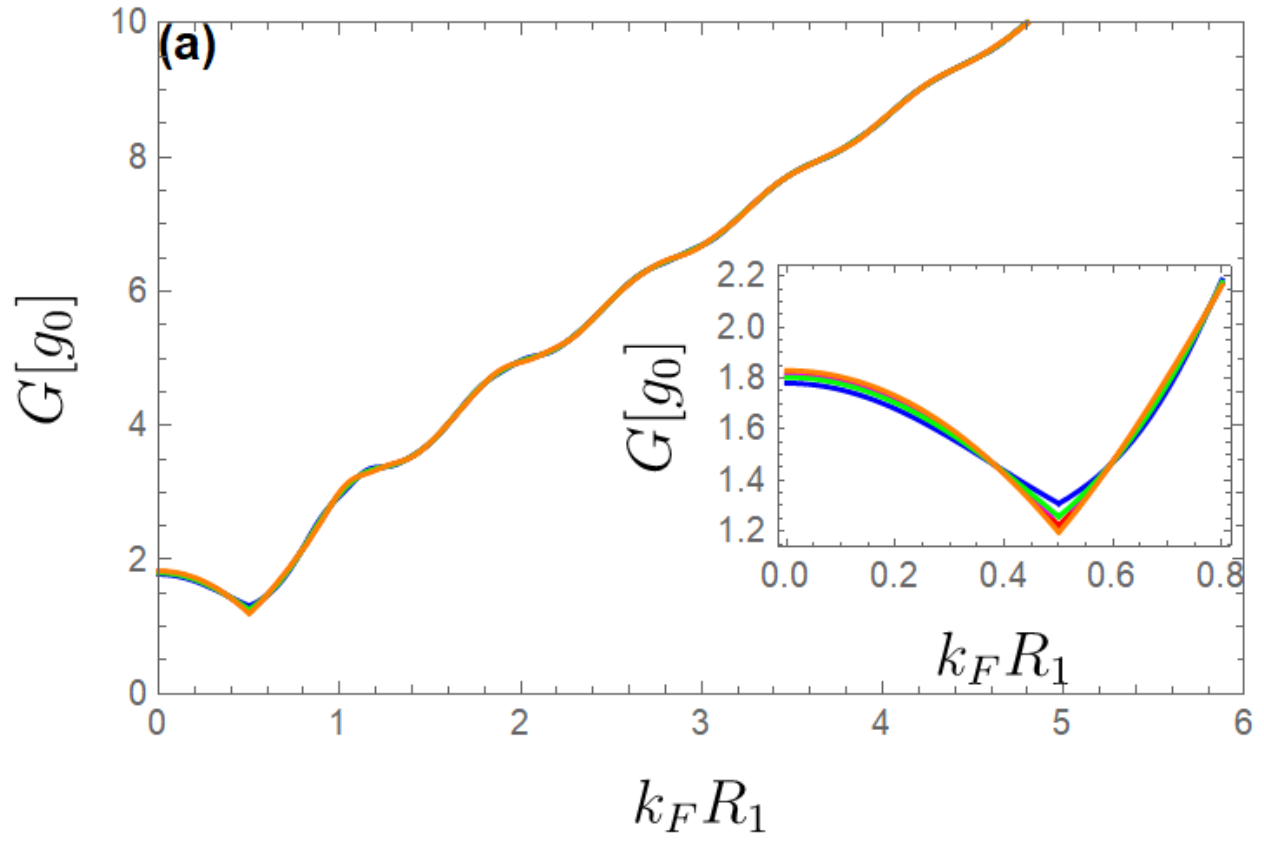}~~\includegraphics[width=0.5\linewidth]{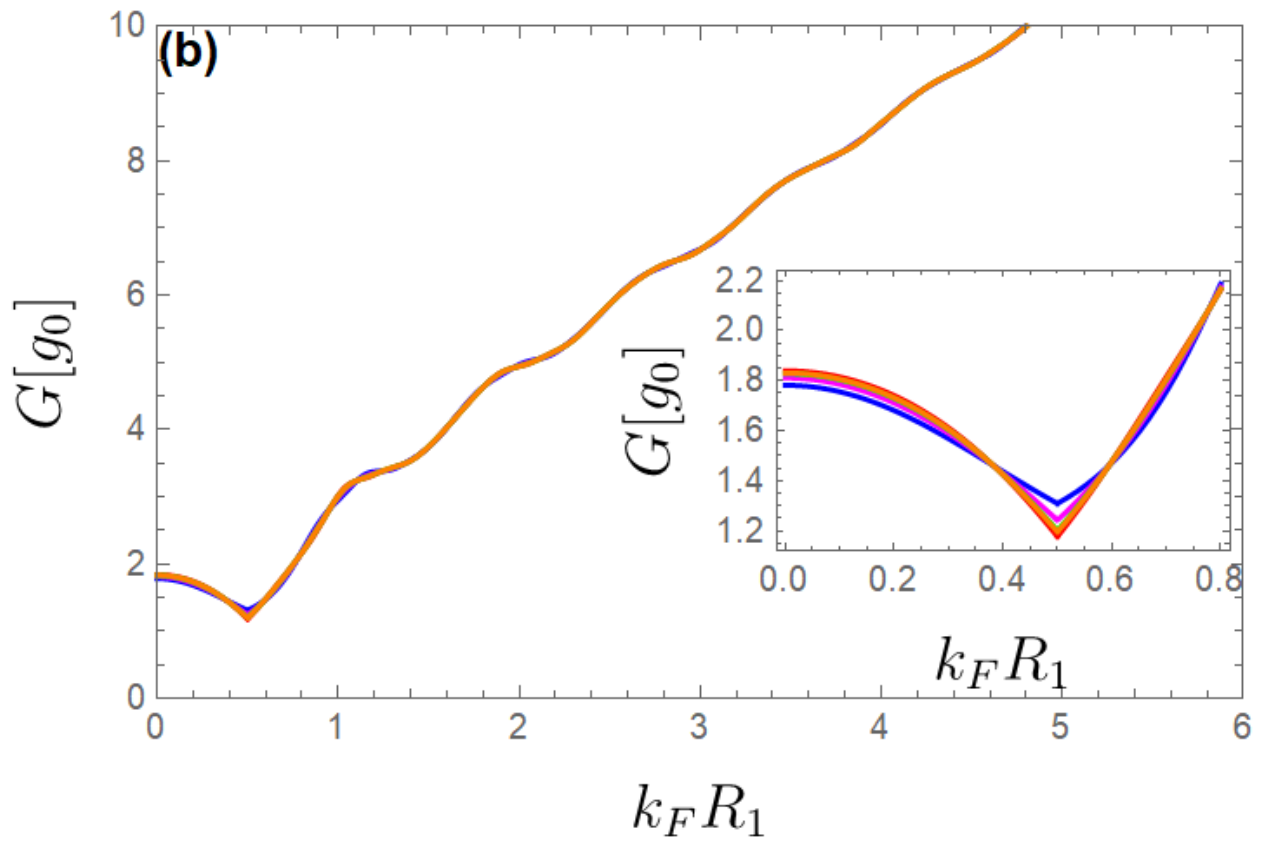}\\
	\includegraphics[width=0.5\linewidth]{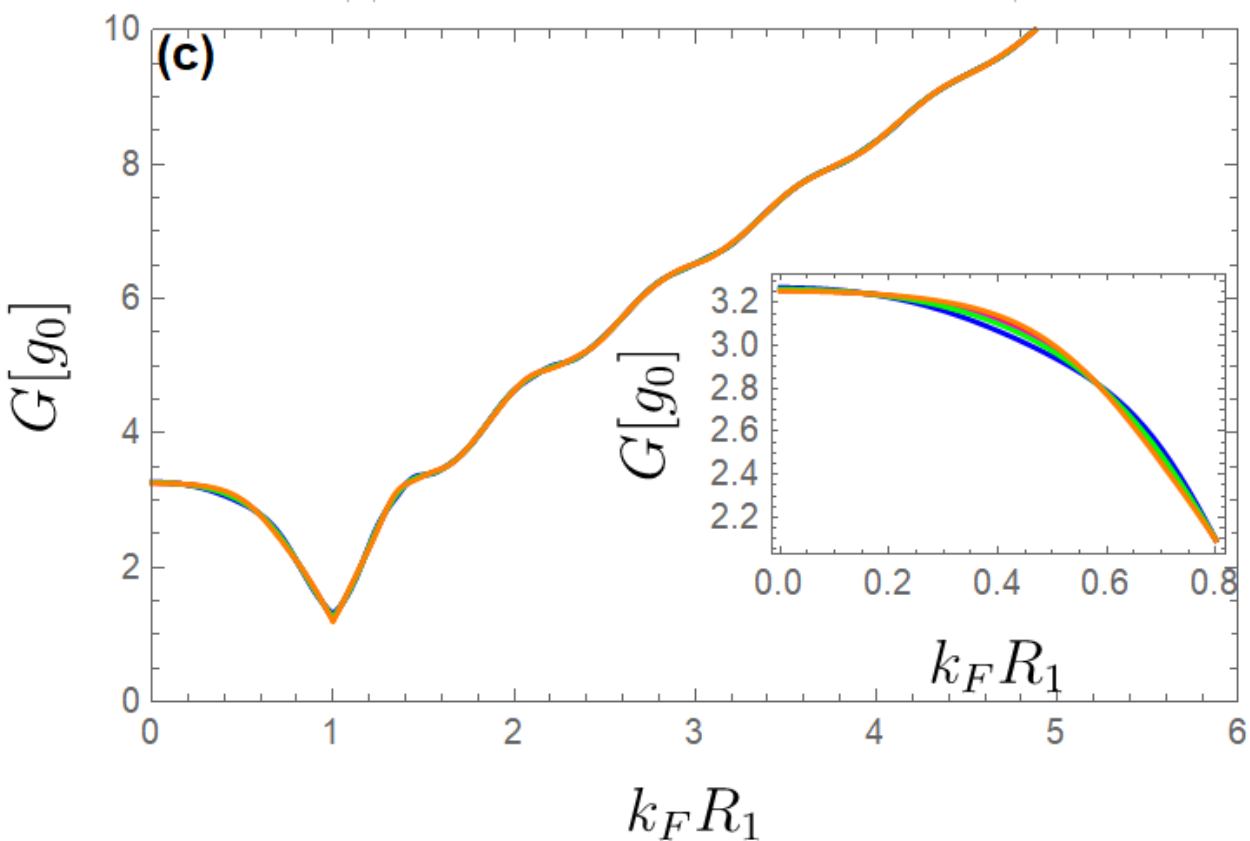}~~\includegraphics[width=0.5\linewidth]{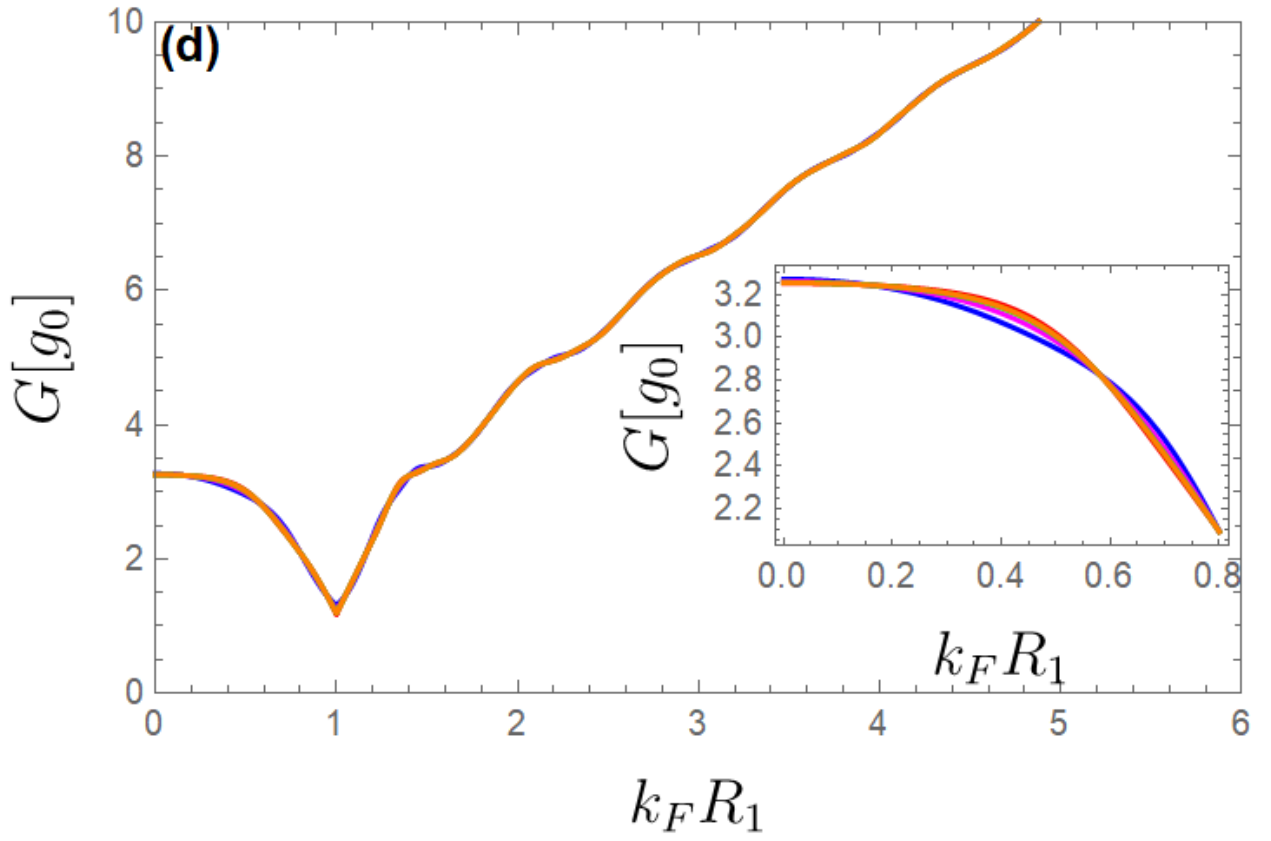}		
	\caption{(color online) The conductance $G[g_0]$ as a function of the doping $k_F R_1$  for the radii ratio  $R_2/R_1 =
		5$ with 
		$n = 0$ (blue line), $ n=1 $ (red line), $ n=2 $ (magenta line), $ n=-1 $ (green line) and $ n=-2 $ (orange line). Inset presents a zoom-in for low doping $k_F R_1<0.8$.  (a,c):  $R_1\delta=0.5$,  (b,d):  $R_1\delta=1$,   (a,b):  $\Phi_i=\frac{1}{2}$,  (c,d):  $\Phi_i=\frac{3}{2}$.} \label{conductance1}
\end{figure}

Fig. \ref{conductance1} shows  the conductance $G[g_0]$ as a function of the doping $ k_F R_1 $ for  
 $\Phi_i=\frac{1}{2}, \frac{3}{2}$. 
It can be written in the approximate form 
\begin{equation}
G\approx2 g_0 R_1 \sqrt{|k_F^2-\delta^2|}+G_{min}(n,\Phi_i)
\label{conducmin} 
\end{equation}
where  $G_{min}(n,\Phi_i)$ is the minimum value of $ G $, which is depending on  $n$ (disclination effect) and $\Phi_i$ (magnetic flux).  \eqref{conducmin} is in agreement with the result that we  previously found \cite{babe}. $ G $ increases when the doping increases and for zero doping ($k_F R_1=0$), it  increases by increasing the energy gap. It is clearly seen that $ G $ takes a minimal value $G_{min}(n,\Phi_i)$ for $k_F R_1=R_1 \delta$.
We observe that adding an energy gap $R_1\delta$ implies an appearance of a singularity where the conductance becomes minimal for the doping case $k_F R_1=R_1\delta$. The conductance for a zero doping $G(k_F R_1=0)$ is increased when the value of $R_1\delta$ increases. 

\begin{figure}[H]\centering
	\includegraphics[width=0.46\linewidth]{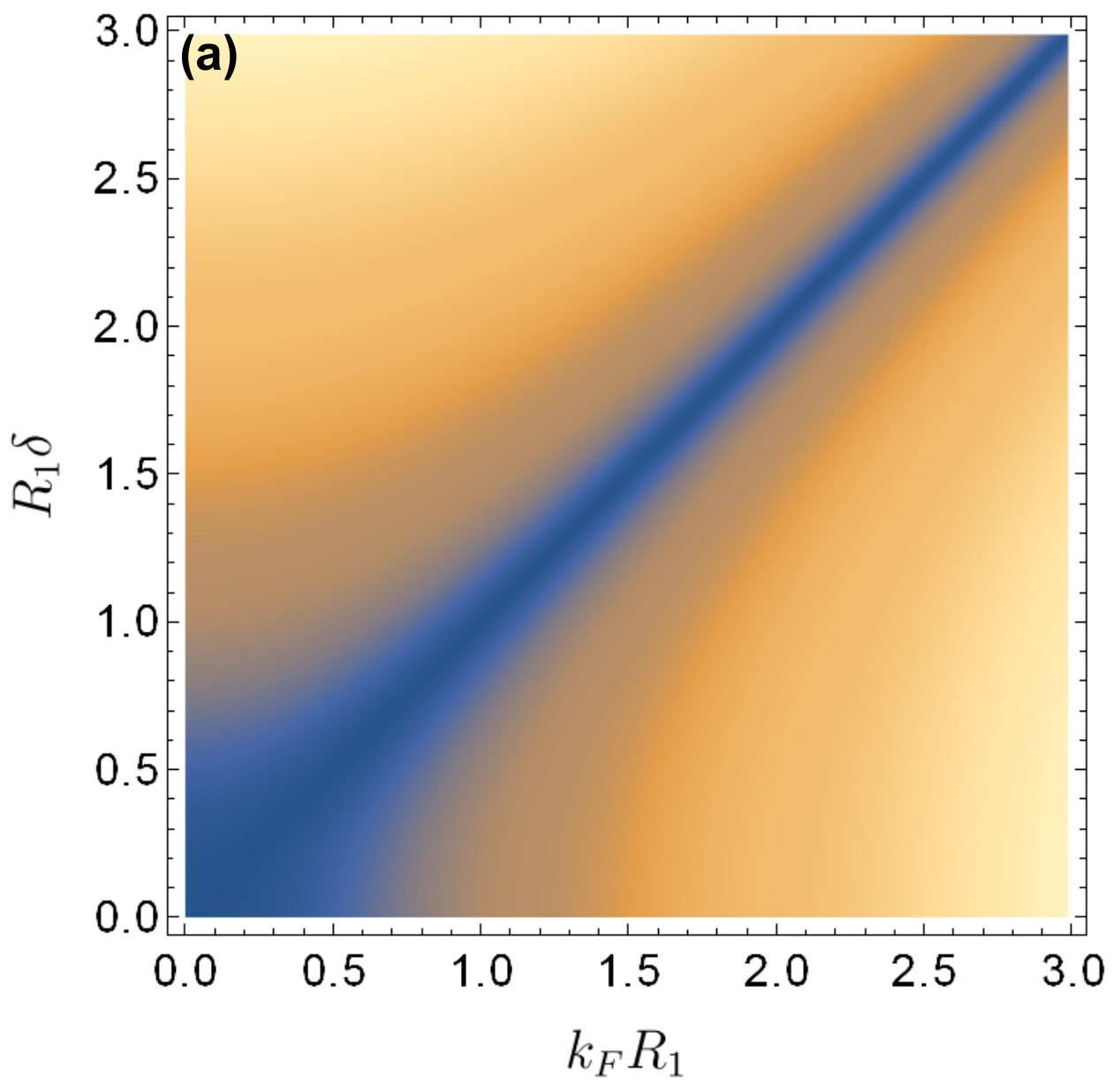}
	\includegraphics[width=0.51\linewidth]{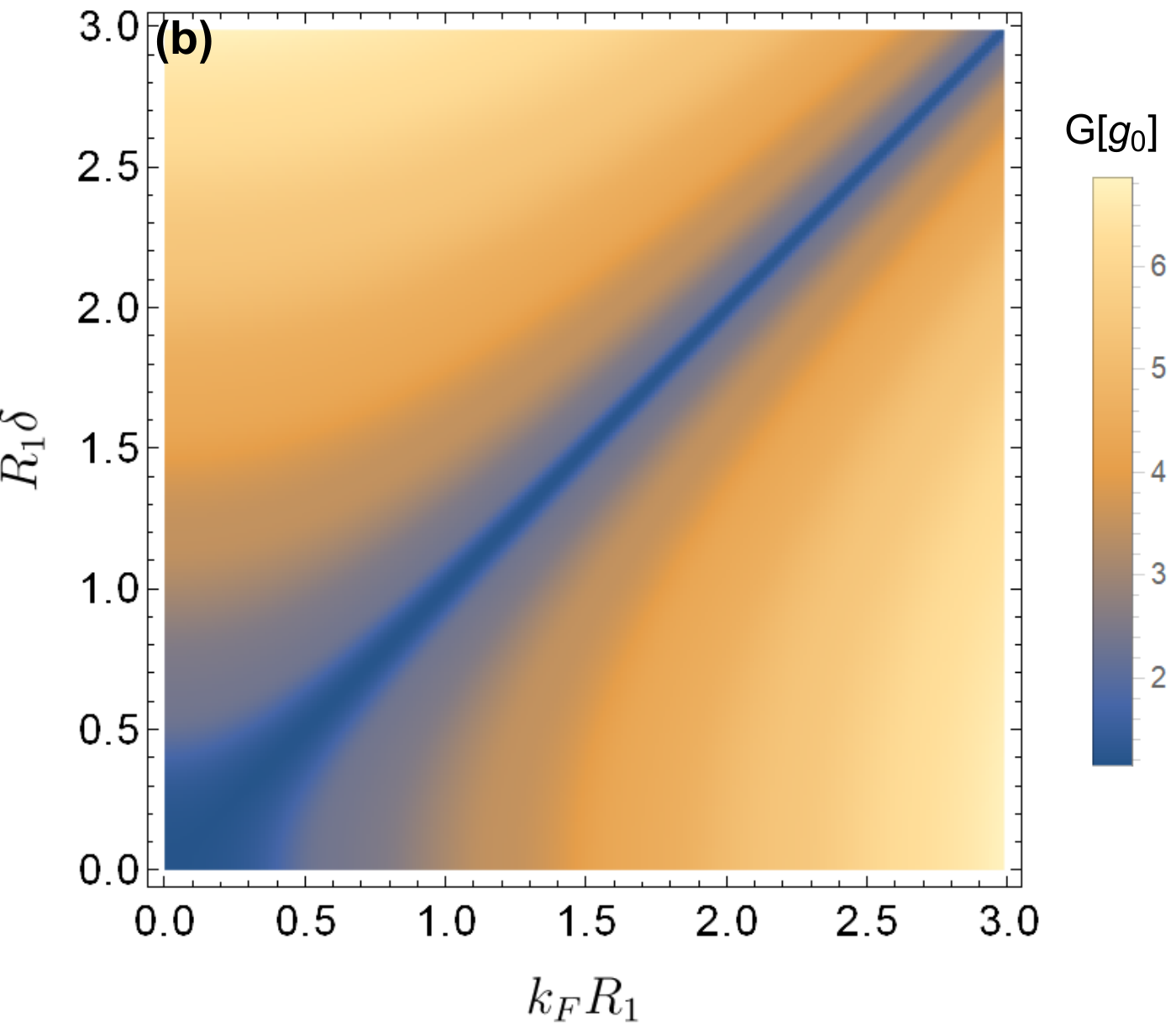}
	\caption{(color online) Contour plot of the conductance $G[g_0] $ as a function of the doping $ k_F R_1 $ and energy gap $ R_1 \delta $ for 
		free defect $n=0$. (a):  $R_2/R_1=5$,  (b):  $R_2/R_1=7.5$.}
	\label{contourplot}
\end{figure}

\begin{figure}[H]\centering
\includegraphics[width=0.5\linewidth]{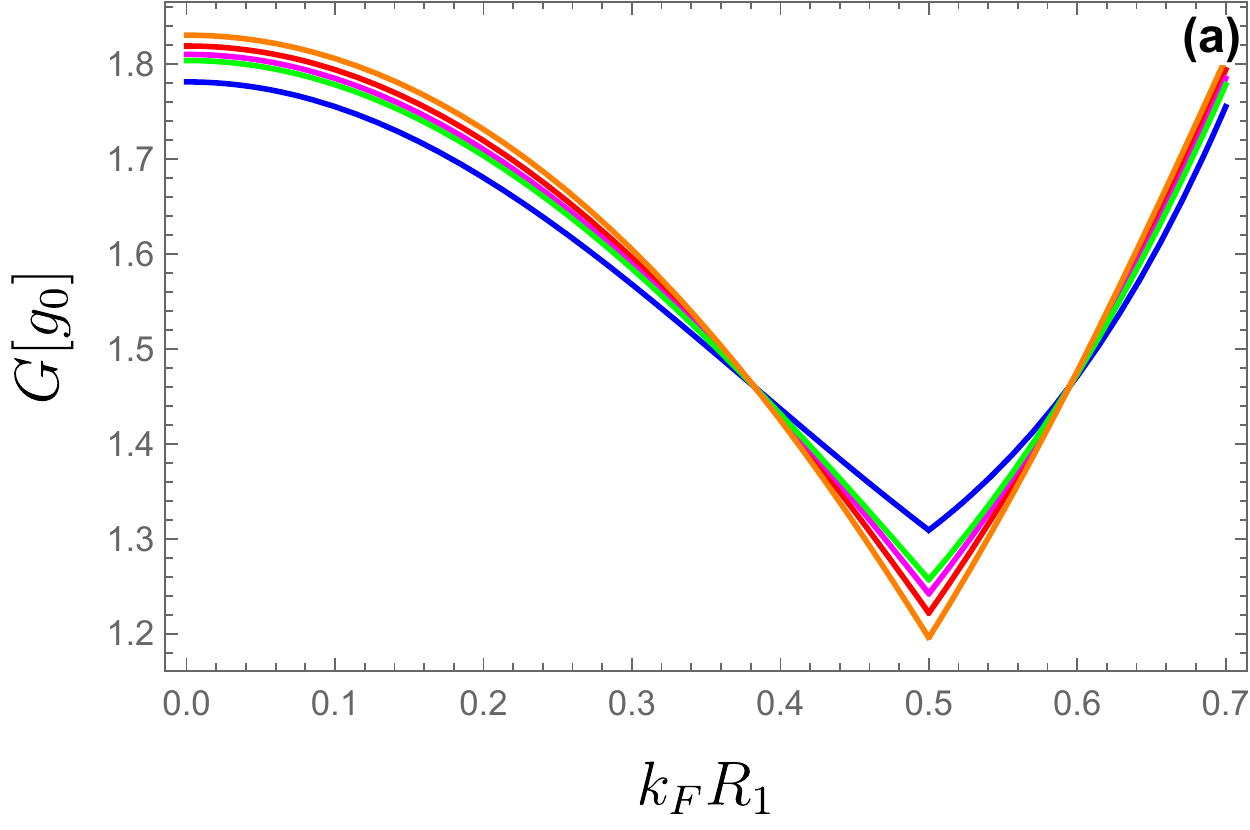}~~\includegraphics[width=0.5\linewidth]{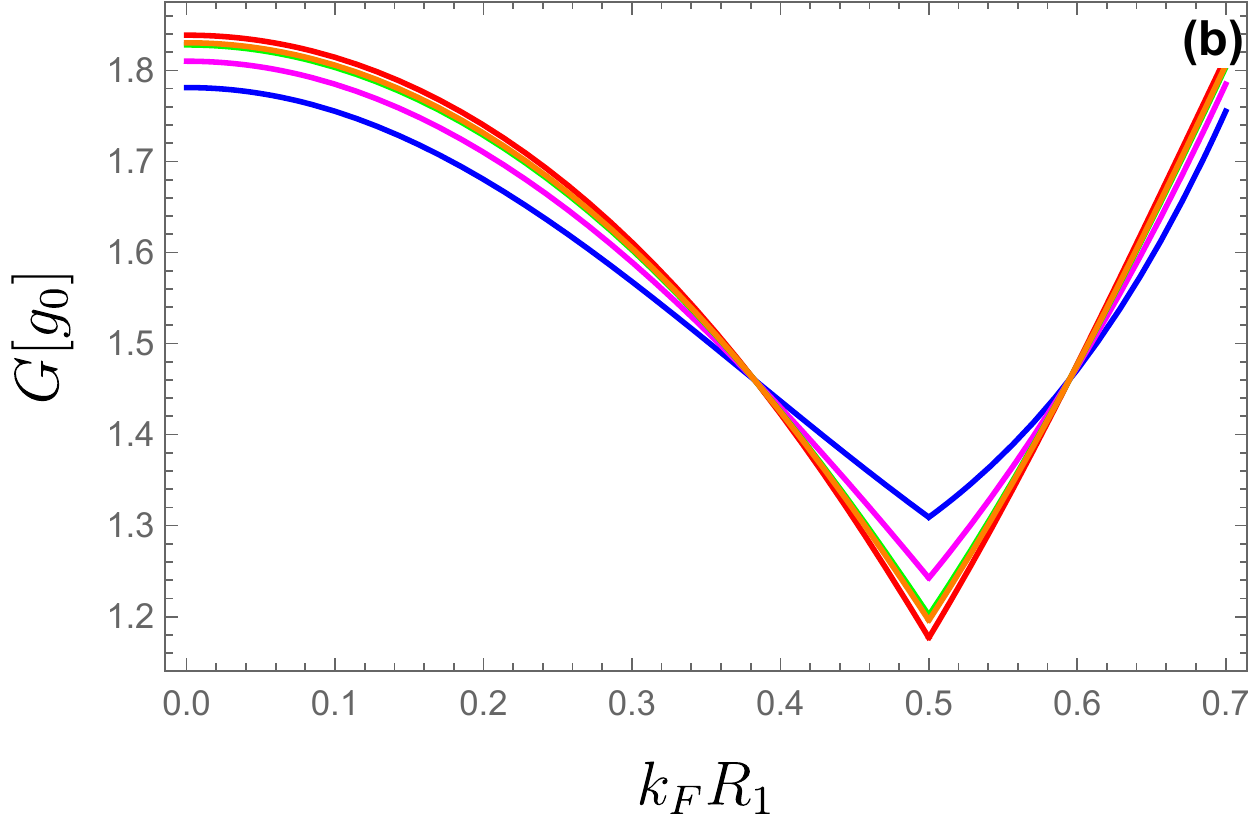}		
	\caption{(color online)  The conductance $G[g_0]$ as a function of the doping $k_F R_1$ for  $R_2/R_1 =
		5$ and 
		$R_1\delta=0.5$ with $n = 0$ (blue line), $ n=1 $ (red line), $ n=2 $ (magenta line), $ n=-1 $ (green line) and $ n=-2 $ (orange line). (a): $\Phi_i=\frac{1}{2}$, (b):  $\Phi_i=\frac{3}{2}$.} \label{conductance2}
\end{figure}

 Fig. \ref{contourplot} shows  contour plot of the conductance $G[g_0]$ as a function of the doping $k_F R_1 $ and the energy gap $R_1 \delta$ with (a): $R_2/R_1=5$  and (b): $R_2/R_1=7.5$ for free defect $n=0$. 
We observe that the conductance for zero doping is proportional to the energy gap  and all points where the conductance is minimum are reduced  for $R_2 \gg R_1$ (surface hatched by the color black).
To study the effect of wedge index $n=0, \pm1, \pm2$ and  magnetic flux  $\Phi_i=\frac{1}{2}, \frac{3}{2}$ on  
 $G_{min}(n,\Phi_i)$, in Fig. \ref{conductance2} we show the conductance as a function of the doping  $k_F R_1$. To analyze such case, let us show
some illustrations under suitable choices. 
\begin{enumerate}
	\item The minimal value of the conductance $G_{min}(n,\Phi_i)$
	 \begin{itemize}
	 	\item $G_{min}(n,\frac{1}{2})$:
	 	\begin{align*}
	 		&G_{min}(0)=1.31 g_0,\ G_{min}(1)=1.22 g_0,\ G_{min}(2)=1.24 g_0\\
	 		&G_{min}(-1)=1.26 g_0,\ G_{min}(-2)=1.20 g_0
	 	\end{align*}
	 \item $G_{min}(n,\frac{3}{2})$:
	 \begin{align*}
	 	&G_{min}(0)=1.31 g_0,\ G_{min}(1)=1.17 g_0,\ G_{min}(2)=1.24 g_0\\
	 	&
	 	G_{min}(-1)=1.21 g_0,\ G_{min}(-2)=1.19 g_0
	 \end{align*}	
	 \end{itemize}
 \item The initial value of the conductance $G_{int}(k_F R_1=0, n, \Phi_i)$ 
 \begin{itemize}
 	\item $G_{int}(k_F R_1=0, n, \frac{1}{2})$
	\begin{align*}
	&
	G_{int}(0)=1.78 g_0,\ G_{int}(1)=1.82 g_0,\ G_{int}(2)=1.81 g_0\\
	&
	G_{int}(-1)=1.80 g_0,\ G_{int}(-2)=1.83 g_0
	\end{align*}
	\item $G_{int}(k_F R_1=0, n, \frac{3}{2})$
	\begin{align*}
	&
	G_{int}(n0)=1.78 g_0,\ G_{int}(1)=1.84 g_0,\ G_{int}(2)=1.81 g_0\\
	&G_{int}(-1)=1.82 g_0,\ G_{int}(-2)=1.83 g_0
	\end{align*}
\end{itemize}
\end{enumerate}

\begin{figure}[H]\centering
	\includegraphics[width=0.49\linewidth]{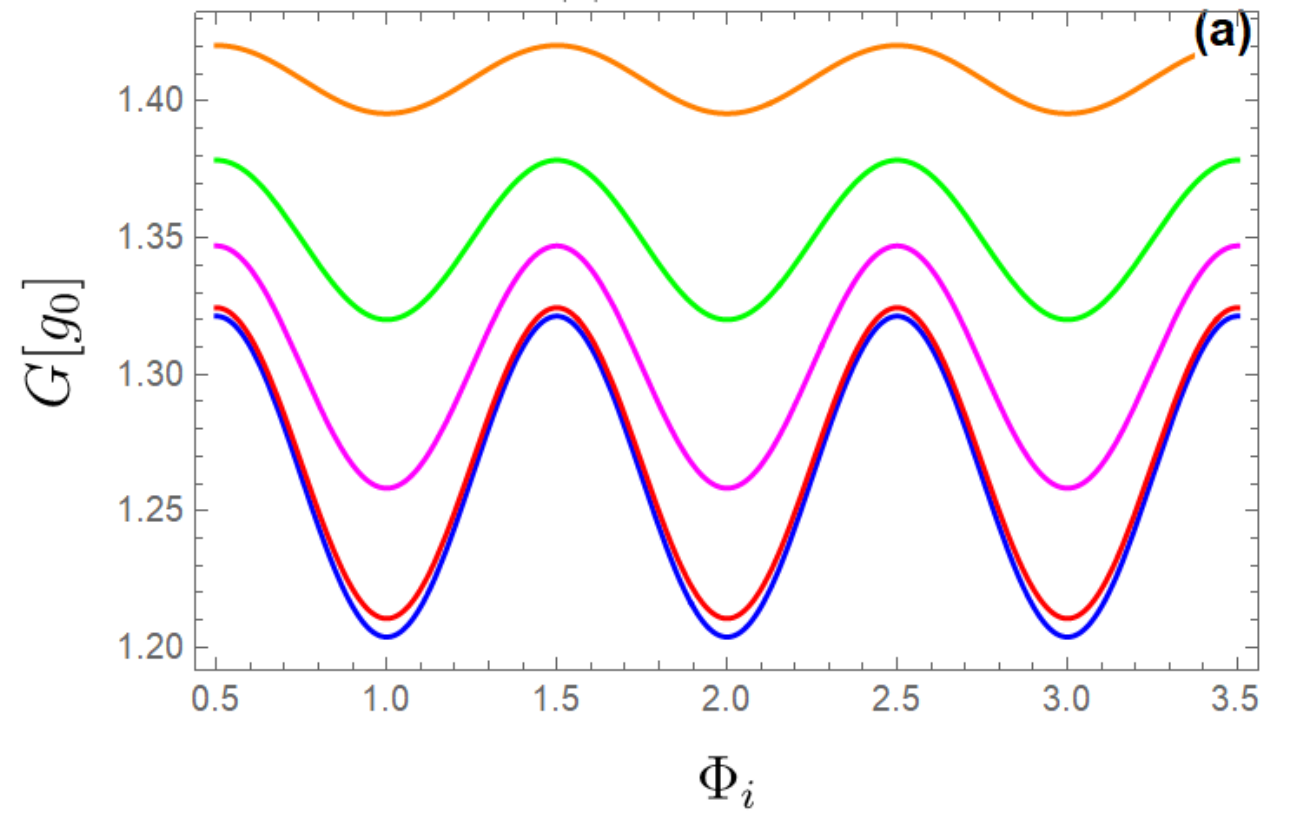}
	\includegraphics[width=0.49\linewidth]{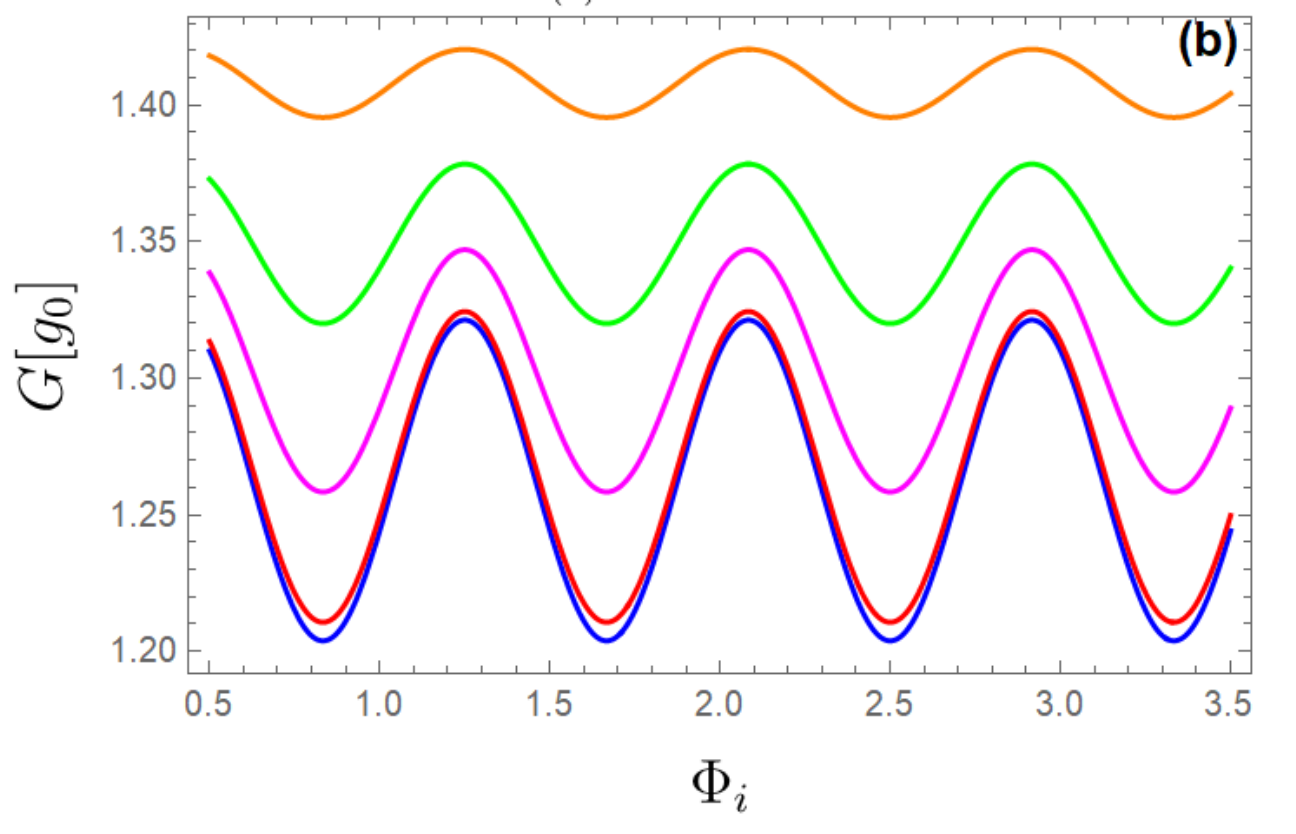}
	\includegraphics[width=0.49\linewidth]{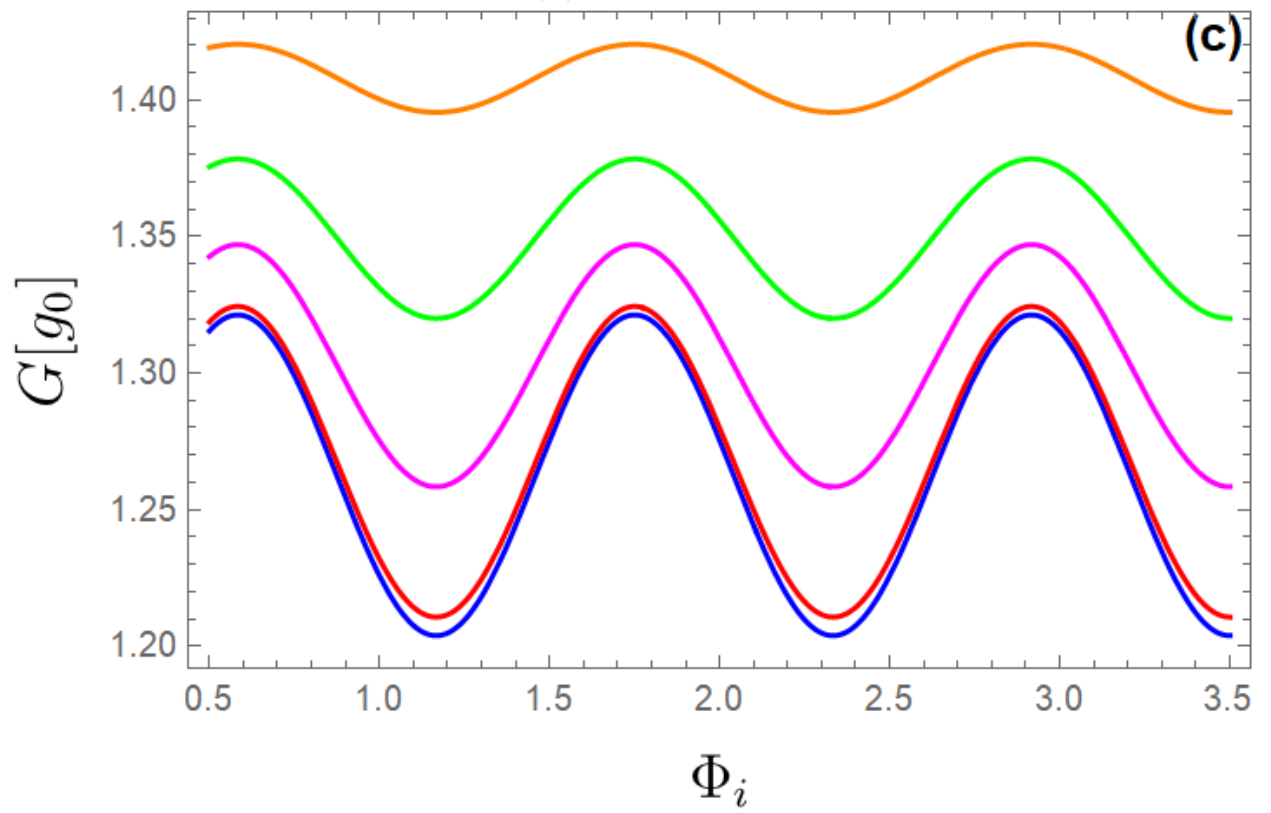}
	\includegraphics[width=0.49\linewidth]{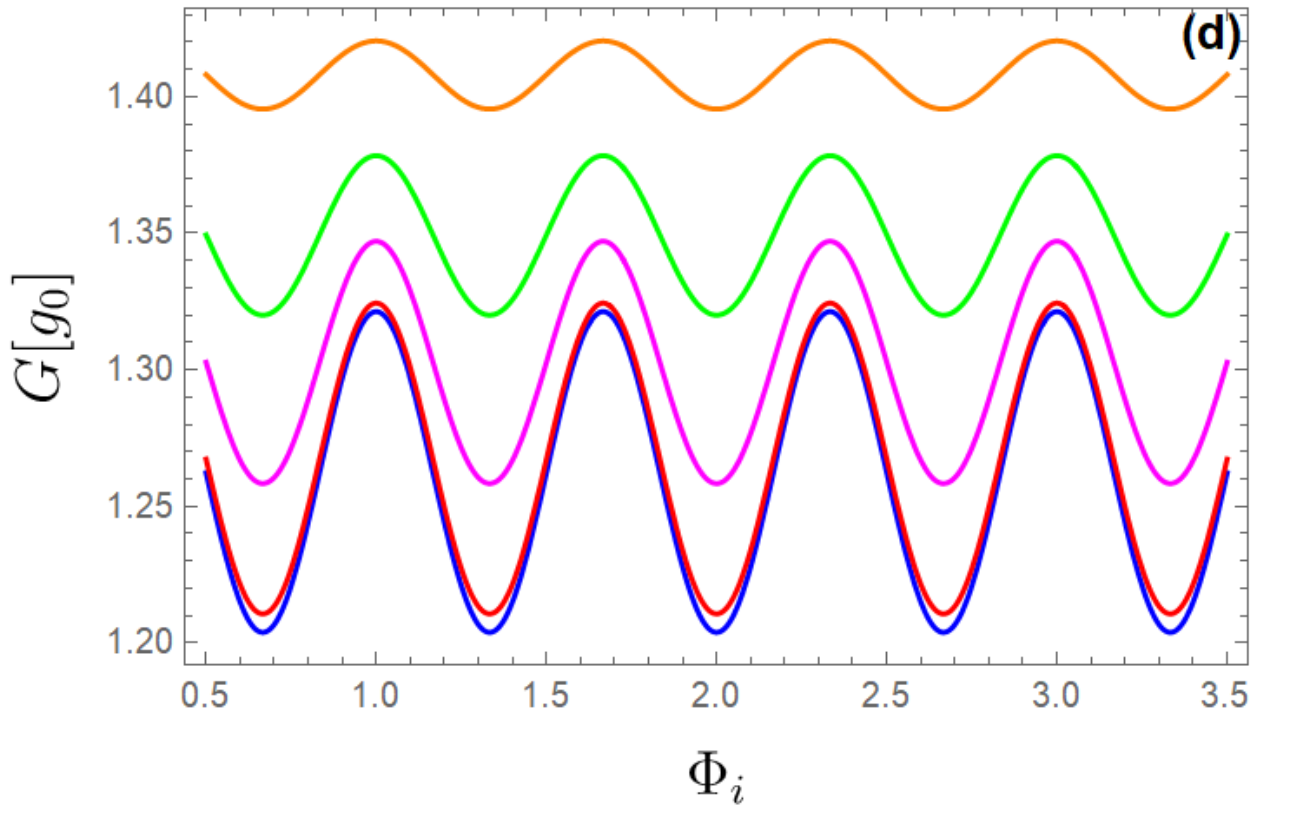}
	\includegraphics[width=0.49\linewidth]{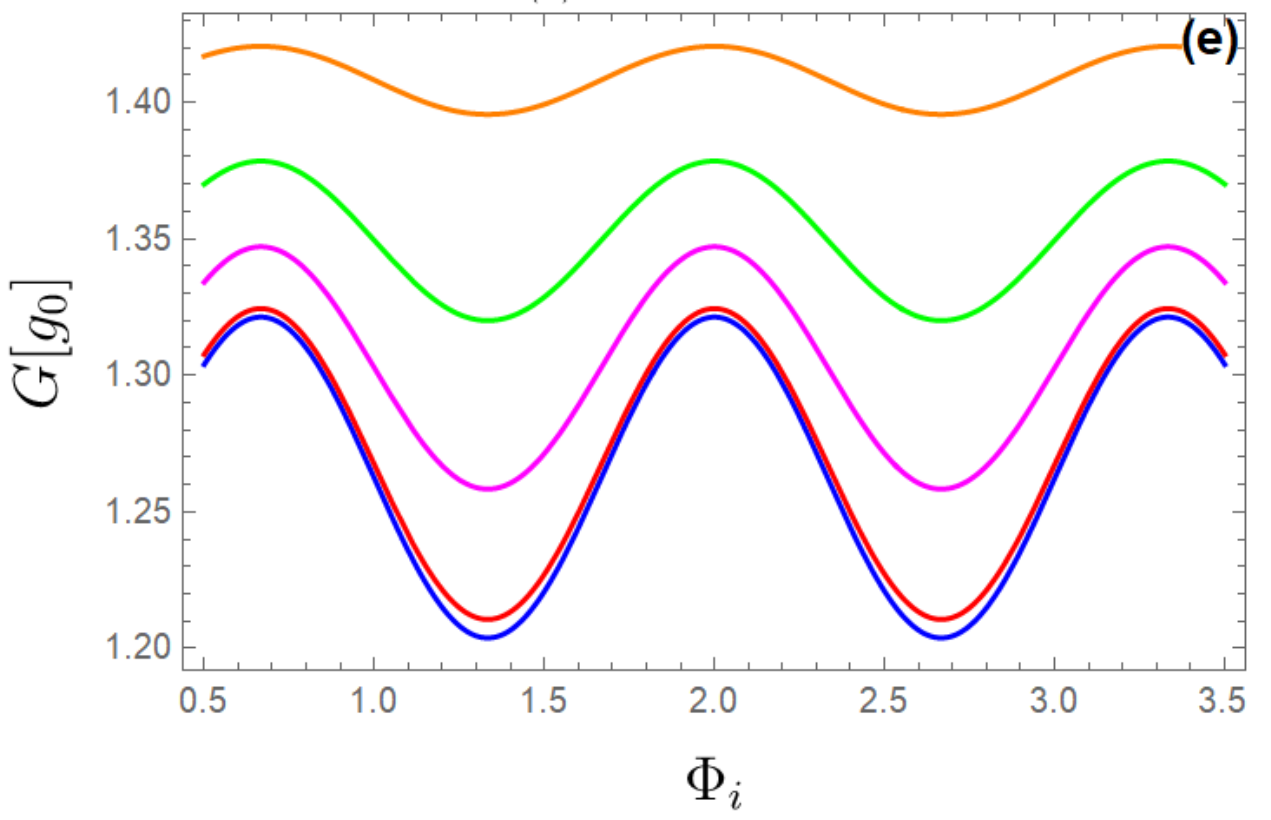}
	\includegraphics[width=0.49\linewidth]{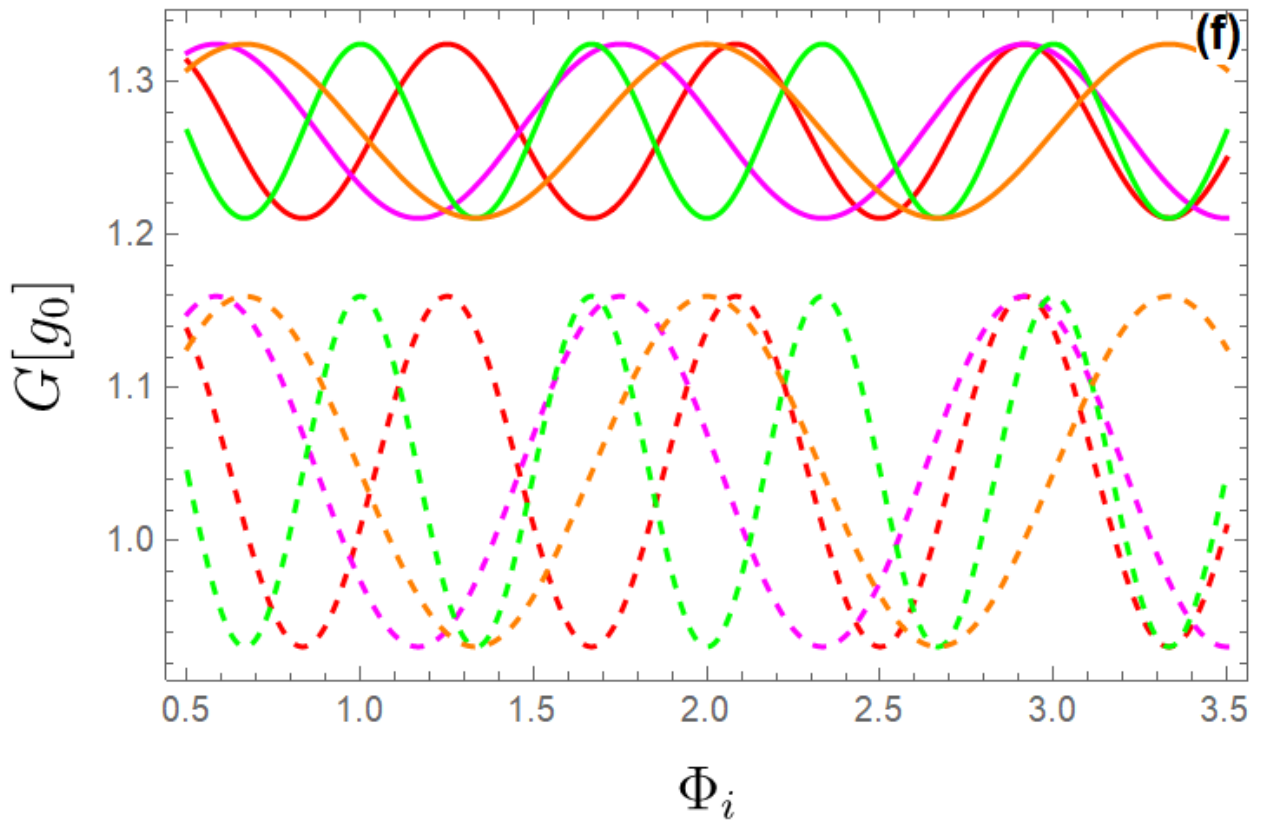}		
	\caption{(color online) The conductance $G[g_0]$ as a function of the magnetic flux $\Phi_i$ for 
		 $k_F R_1=0.1$.  (a,b,c,d,e):  $R_2/R_1 =
		5$ with  $R_1 \delta = 0$ (blue line), $ 0.15 $ (red line), $ 0.2 $ (magenta line), $ 0.25 $ (green line), $ 0.3 $ (orange line).  (a): $n=0$, (b):  $n=1$, (c): $n=-1$, (d): $n=2$,  (e):  $n=-2$.  (f):  $R_1 \delta = 0$, $R_2/R_1 =
		5$ (solid line), $R_2/R_1 =
		7.5$ (dashed line) with $ n=1 $ (red line), $ n=2 $ (magenta line), $ n=-1 $ (green line),  $ n=-2 $ (orange line).} \label{conductance3}
\end{figure}

\begin{figure}[H]\centering
	\includegraphics[width=0.49\linewidth]{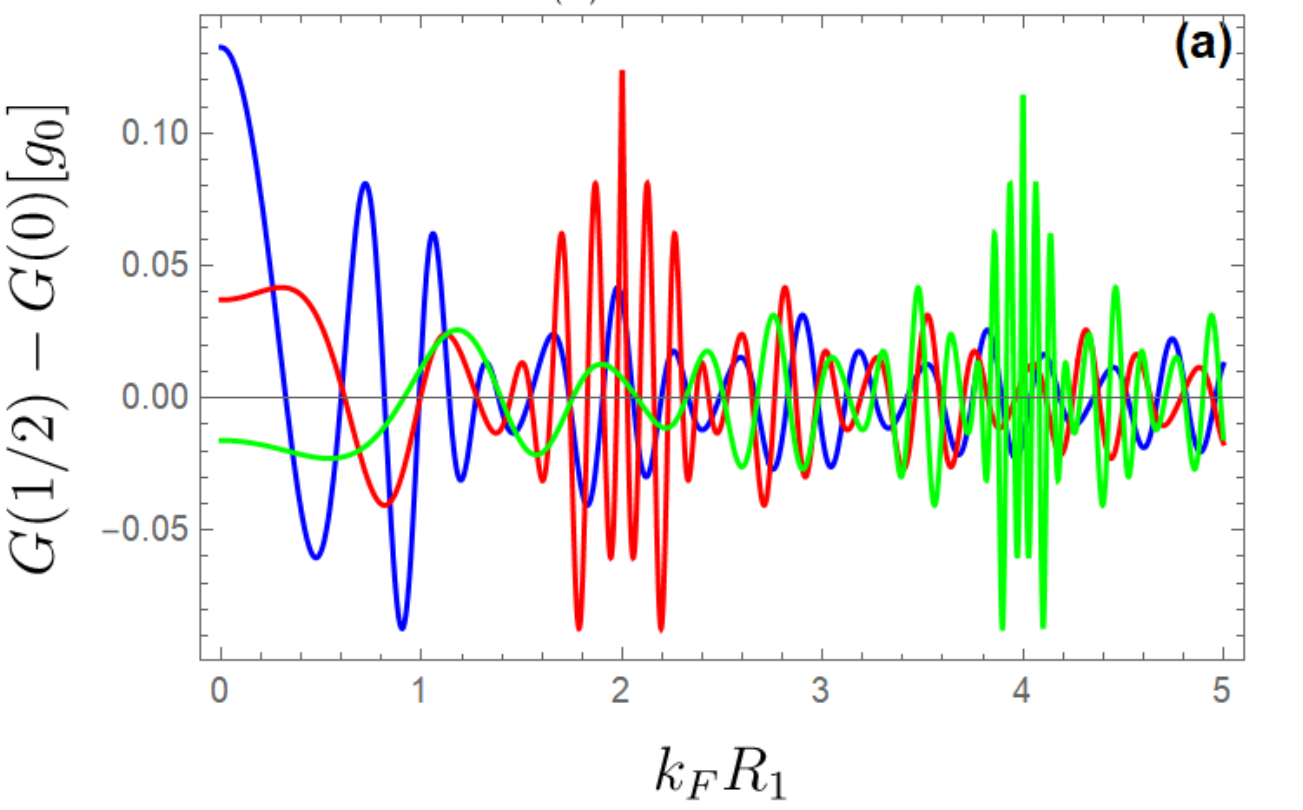}
	\includegraphics[width=0.49\linewidth]{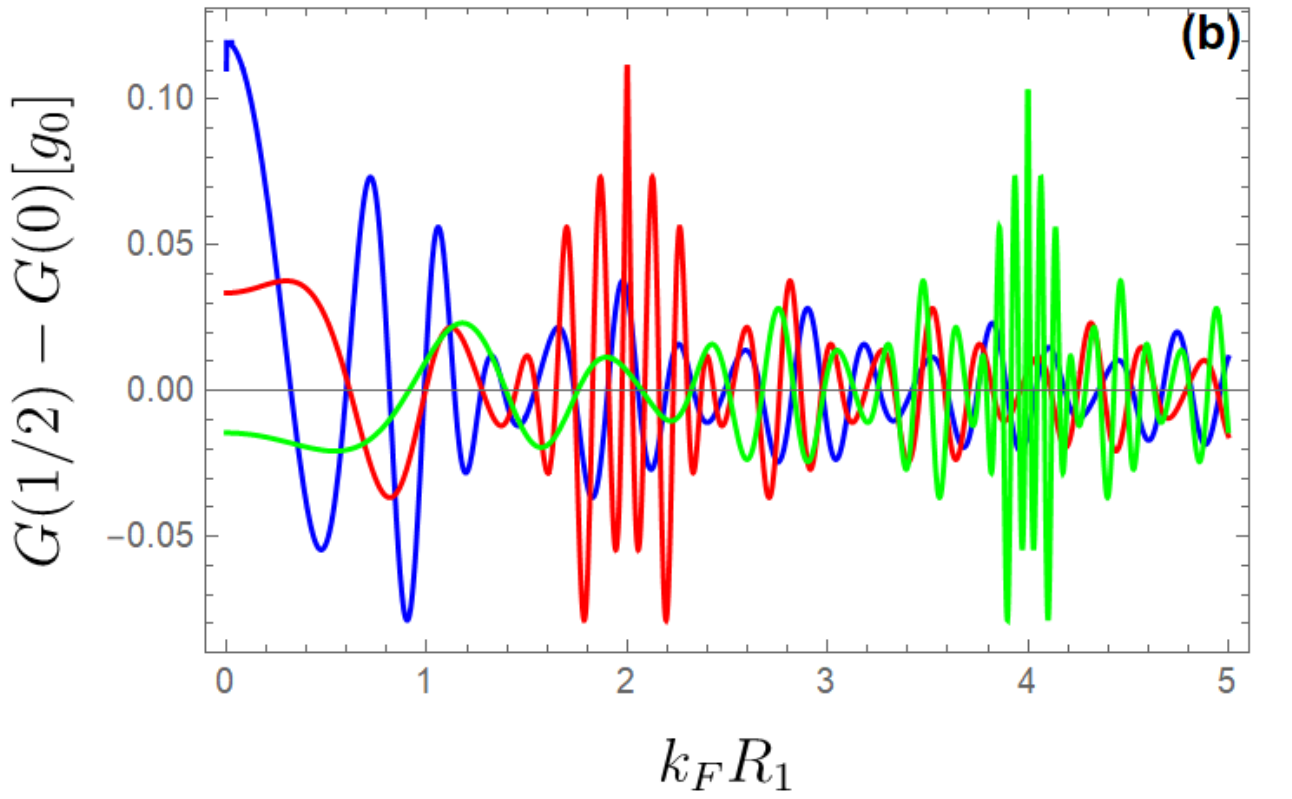}
	\includegraphics[width=0.49\linewidth]{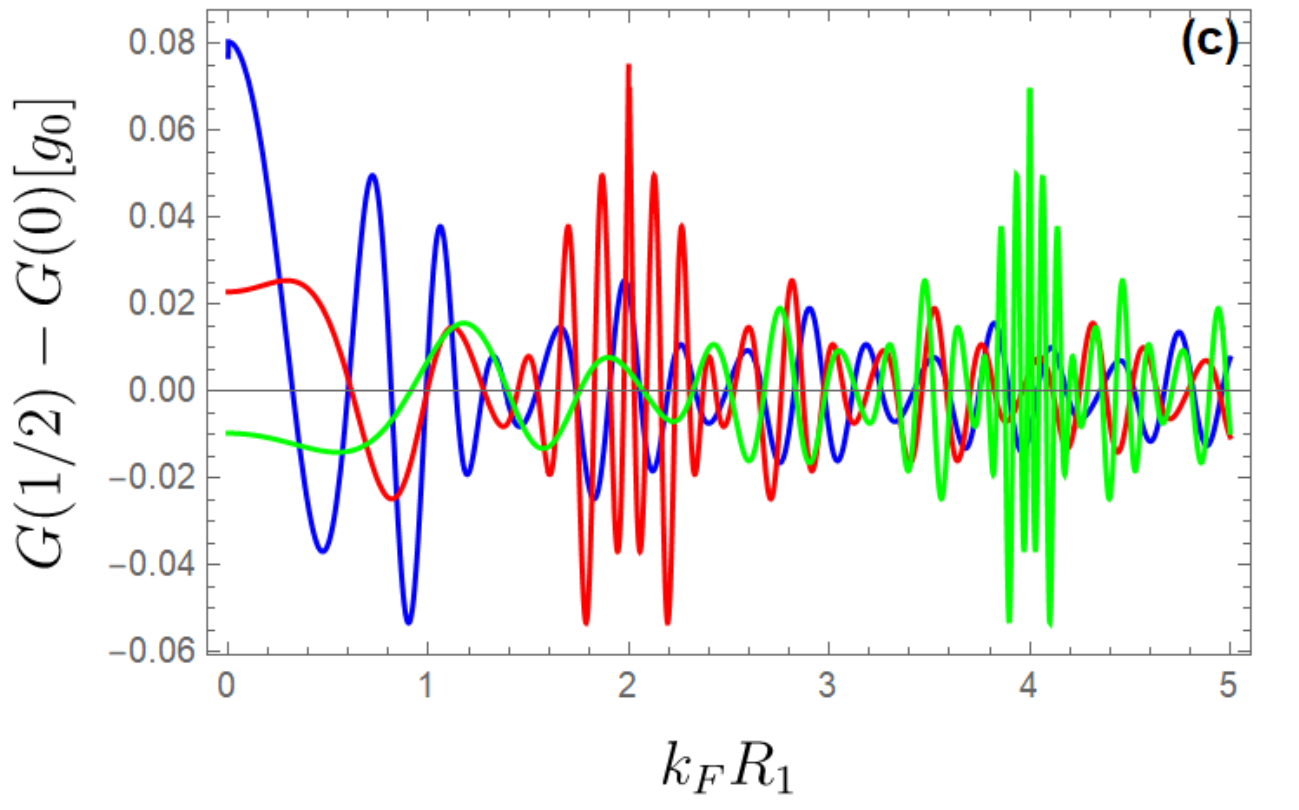}
	\includegraphics[width=0.49\linewidth]{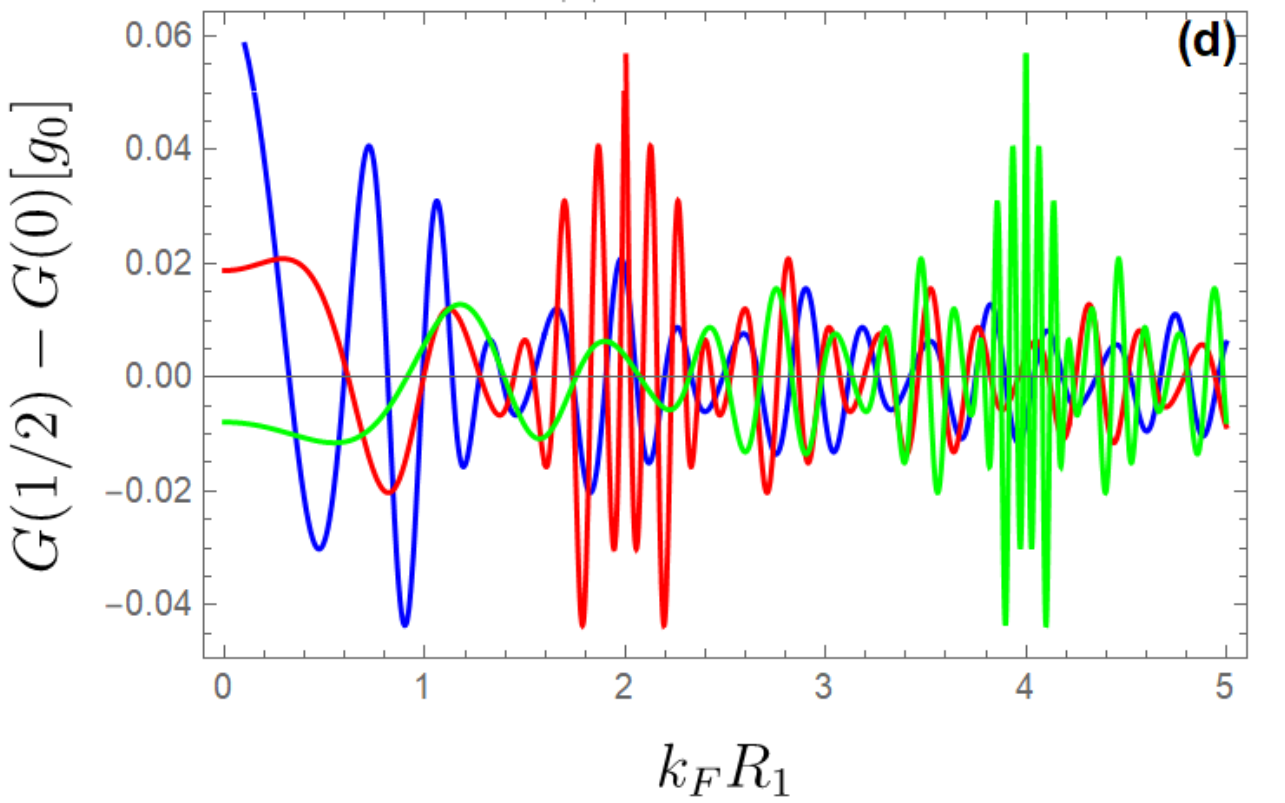}
	\includegraphics[width=0.49\linewidth]{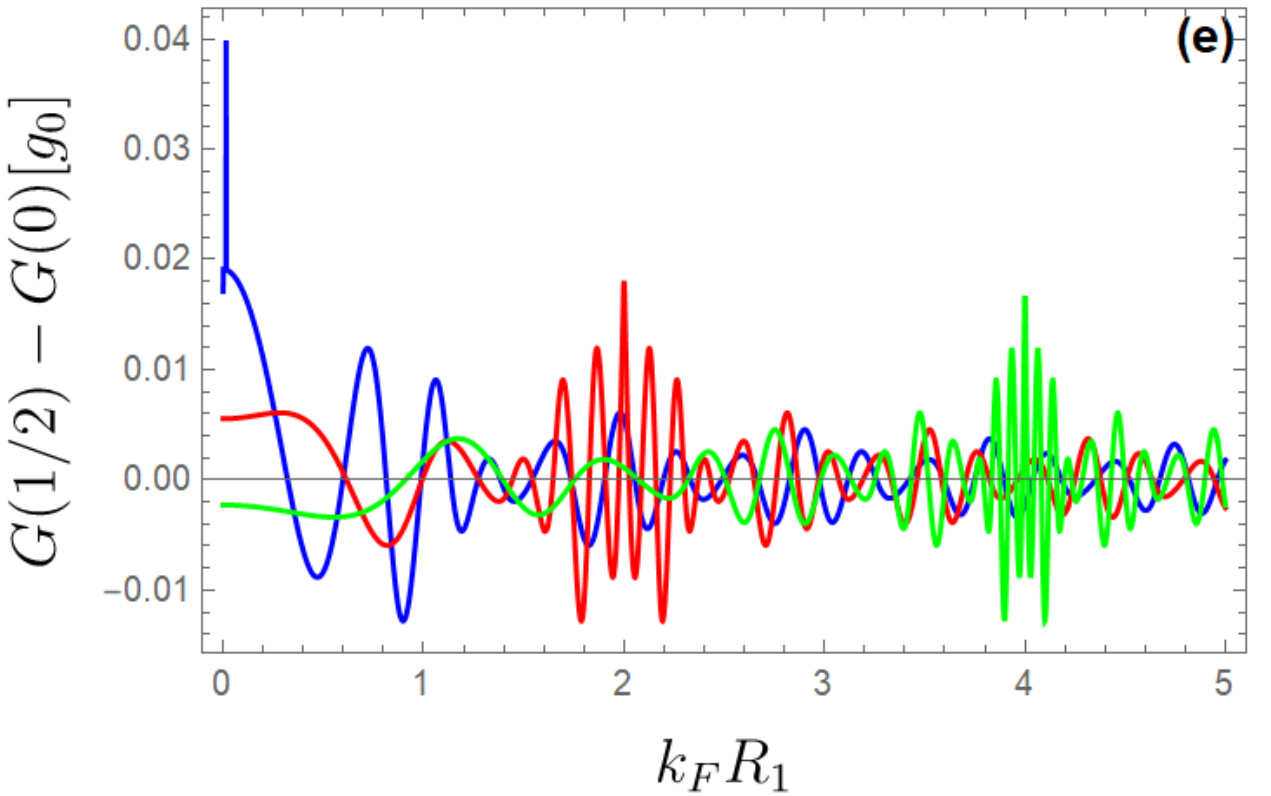}
	\includegraphics[width=0.49\linewidth]{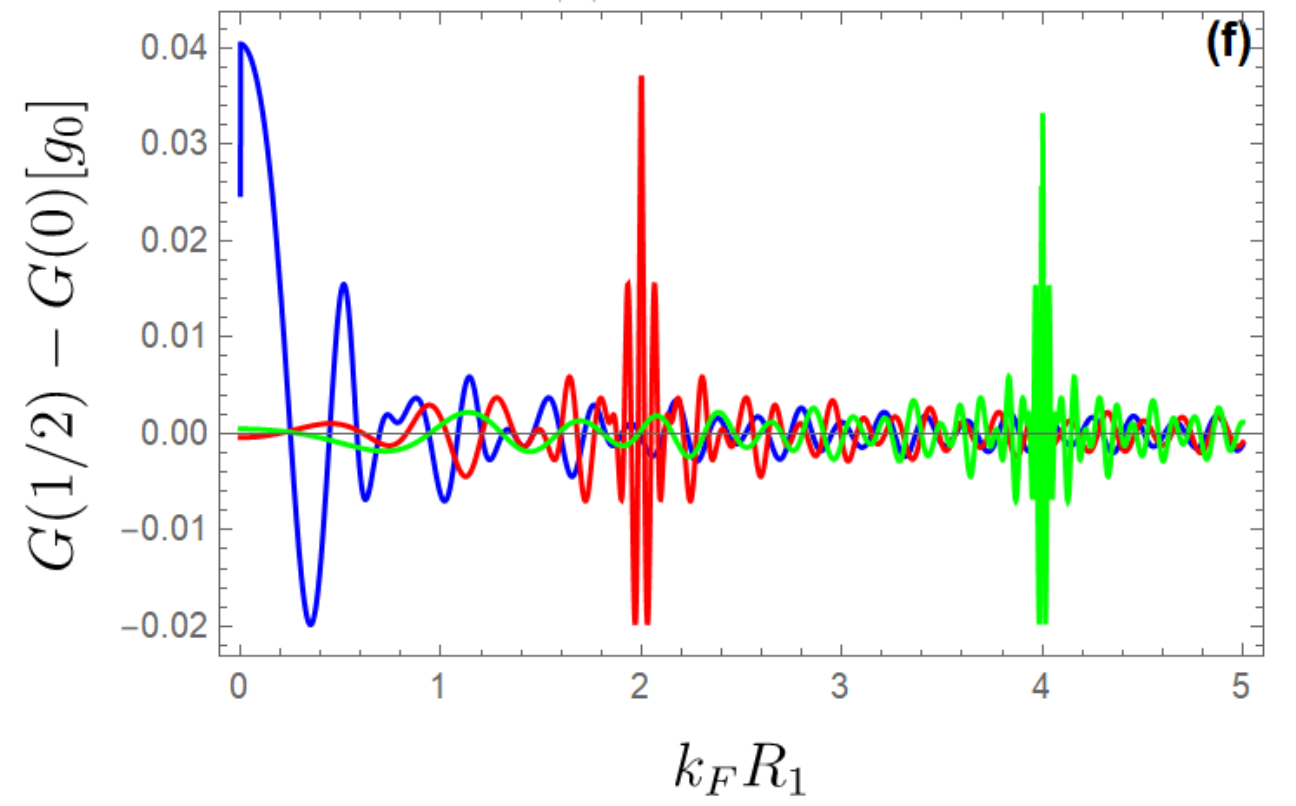}
	\includegraphics[width=0.49\linewidth]{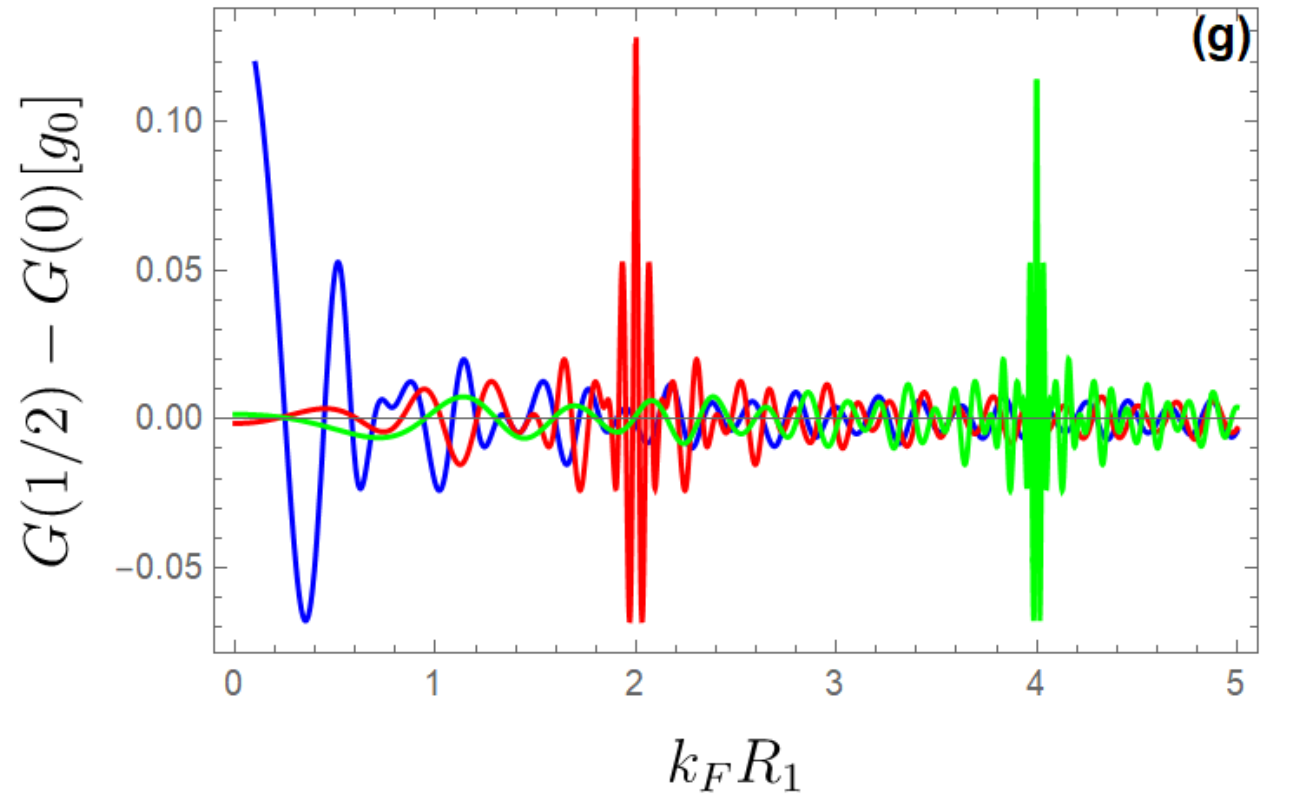}
	\includegraphics[width=0.49\linewidth]{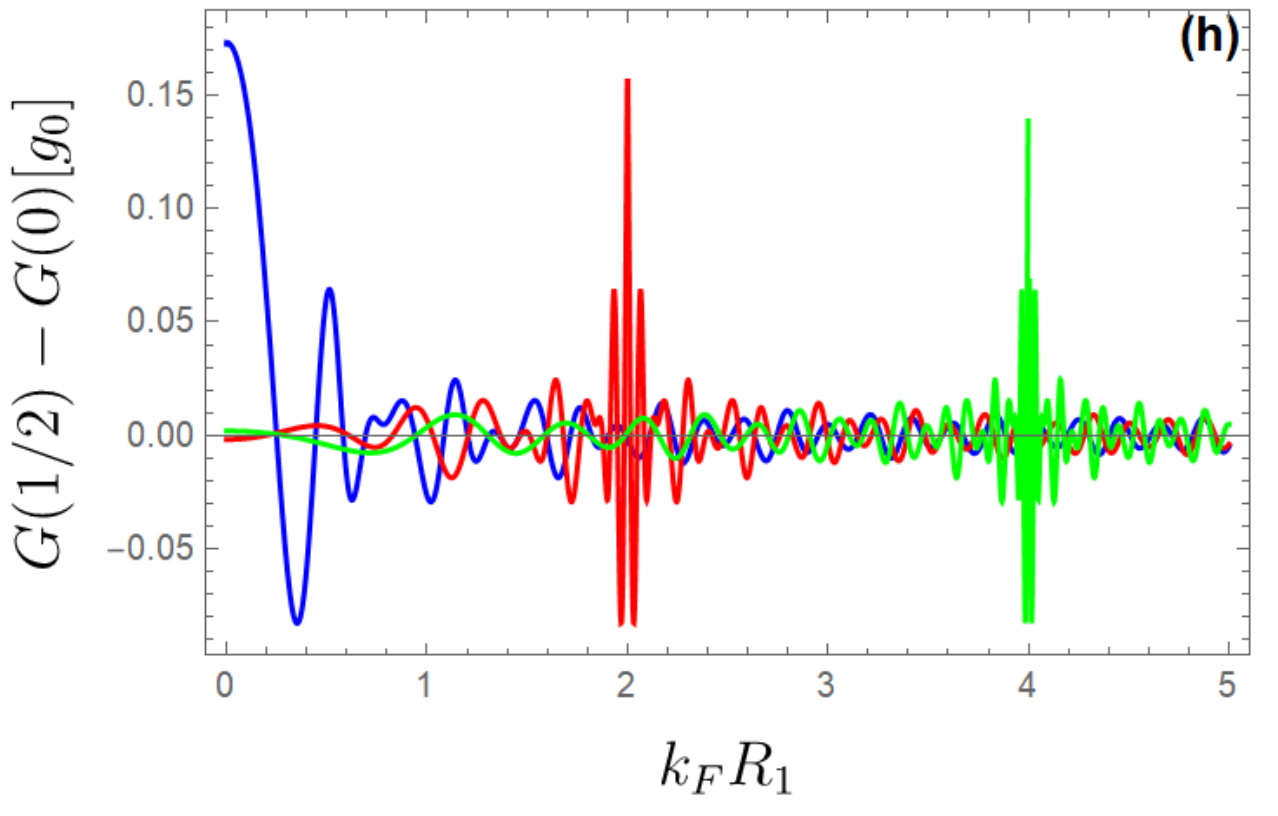}
	\includegraphics[width=0.49\linewidth]{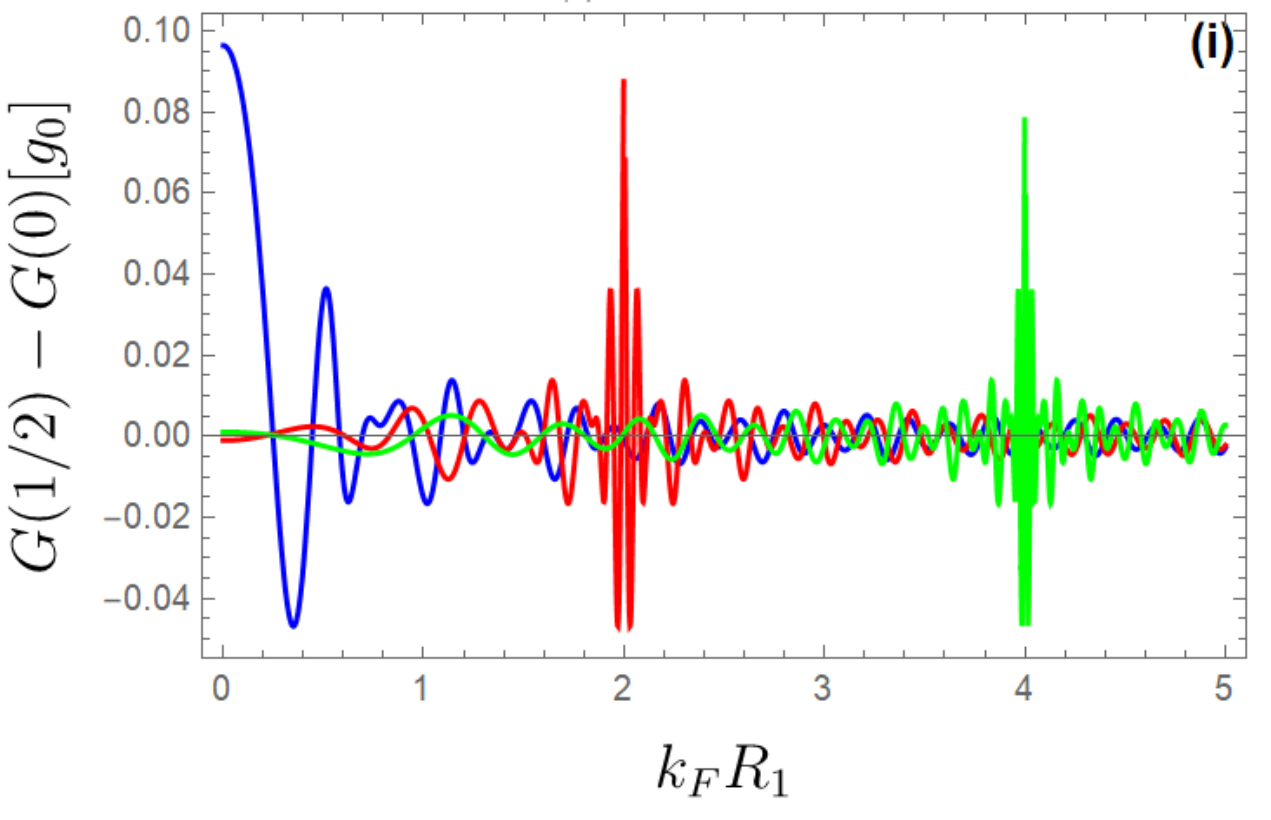}		
	\caption{(color online)  The magnitude of the conductance oscillations $ \Delta G$ 
		as a function of the doping $k_F R_1$ for 
		three values  $ R_1 \delta = 0 $ (blue line), $ 2 $ (red line) and $ 4 $ (green line). (a,b,c,d,e):  $R_2/R_1= 5$,   (f,g.h,i):  $R_2/R_1= 7.5$.  (a):  $n=0$,  (b,i):  $n=1$,  (c,h):  $n=-1$,  (d,g):  $n=2$,  (e,f):  $n=-2$. } \label{deltaconductance1}
\end{figure}

In Fig. \ref{conductance3}, we show the conductance $G[g_0]$ as a function of the magnetic flux $\Phi_i$. First we prove the approximate expressions of the transmission and the conductance for zero doping $k_F R_1 \to 0$. Indeed, for zero doping limit the transmission  for the Corbino disk in undoped graphene \eqref{eqtrans} can be simplified to
 \cite{Buttiker1985} 
\begin{equation}\label{eq1}
T^{\tau}_\nu = \frac{1}{\cosh^2[\ln(R_2/R_1)(\nu+\frac{\Phi_i}{\Omega_n})]}
\end{equation} 
 and then  the corresponding conductance \eqref{eq22} becomes 
\begin{equation}\label{conduclimit}
G(\Phi_i)=\sum^{\infty}_{j=0} G_j \cos\left(\frac{2 \pi \Phi_i}{\Omega_n}\right) 
\end{equation}
where the coefficients are given by 
\begin{align}
&G_0=\frac{2g_0}{\ln(R_2/R_1)}
\\
&G_j=\frac{4\pi^2 (-1)^j
	j g_0}{\ln(R_2/R_1)^2\sinh[\pi^2 j/\ln(R_2/R_1)]}.
\end{align}
\eqref{conduclimit} shows   the explicit dependence of  the conductance $G[g_0]$  on  the magnetic flux $\Phi_i$ and also the magnitude $\Omega_n$ as its 
period. It clearly seen that  the conductance oscillations are of Aharonov-Bohm types whose amplitude  depends on  $R_1\delta$ and $R_2/R_1$. This shows a  good agreement with our previous   results \cite{babe}. The  amplitude of $G\left(\Phi_i\right)$ oscillations   reduces by increasing  $R_1\delta$ or decreasing   $R_2/R_1$. We notice that a perfectly periodic functional dependence of $G[g_0]$ on $\Phi_i$ with an average value $G_0$ equal to the pseudo-diffusion conductance such  that the greatest value corresponds to octagon defect $n=-2$ and the smallest one is for square defect $n=2$. These symmetry properties have the consequence that the conductance  \eqref{conduclimit} is symmetric if the flux is reversed, i.e. $G(-\Phi_i)=G(\Phi_i)$.
We study the  magnitude of conductance oscillations $\Delta G$. It is defined as the difference between $G(1/2)$ and $G(0)$ 
\begin{equation}\label{eq25}
\Delta G= G(1/2)-G(0).
\end{equation}
 Fig. \ref{deltaconductance1} shows the magnitude of the conductance oscillations $\Delta G$ as a function of the doping $ k_F R_1 $ for $n=0, \pm1, \pm2$,  $R_2/R_1=5, 7.5$ and  $R_1 \delta = 0, 2, 4 $. There is  appearance of resonance peaks when  $k_F R_1$ is close to  $ R_1 \delta $ as noticed in \cite{babe}. We point out that the frequency of these resonance peaks  depends only on  $ R_1 \delta $ knowing that it increases with its  increase. It is interesting to notice that $n$ changes the amplitude of the resonance peak for $k_F R_1=R_1 \delta$. A sharp resonance is obtained in the case of  octagon defect $n=-2$.
 
\section{Conclusion}
We have studied gapped graphene in the shape of a quantum ring of inner radius $R_1$ and outer radius $R_2$ subjected to a magnetic flux $\Phi_i$ with a topological defect created using a procedure known as the Volterra process \cite{cludio94}. Taking into account the advantage of the geometry of the Corbino disk in graphene,  we have solved the corresponding stationary Dirac equation and obtained analytically the energy spectrum solutions in the three regions. We have determined the transmission probability of an electron crossing the Corbino disk in graphene as well as  the conductance $G$ and Fano factor $\mathcal{F}$. 

Our numerical results were exposed in terms of the radii ratio $R_2/R_1$, magnetic flux $\Phi_i$, energy gap $ R_1 \delta$  and wedge disclination
  $n$.
We have demonstrated the influence of  $n$ on the transmission probability. For positive $ n $   the transmission probability decreases  with growing of
 $n$, while for $ n $ negative it increases compared to gapped-graphene with $n=0$.
  We have showed that  $n$ modifies the period of Fano factor oscillations and allows to intensive peaks at some points.
 	Additionally, it was found that
 the conductance of the Corbino disk (as a function of magnetic flux piercing the disk) presents periodic oscillations of the Aharonov-Bohm type where its period depends on  $n$.  
We have seen that the energy gap, the radii ratio and $n$ can change the characteristics of the oscillations of $G(\Phi_i)$ and the resonances of $\Delta G$ when the doping $k_F R_1$ is close to the value of $R_1 \delta$.

\end{document}